# Energies of Two-State Systems

*Uri Levy*

Weizmann Institute of Science, Rehovot 7610001, Israel
E-mail: uri.levy@weizmann.ac.il

**References**

**[1]** Di Mauro, Marco, and Adele Naddeo, *Introducing quantum mechanics in high schools: A proposal based on Heisenberg's Umdeutung*, In Physical Sciences Forum, vol. 2, no. 1, p. 8. Multidisciplinary Digital Publishing Institute, 2021.

**[2]** Univ. of Washington, *Inversion Spectrum of Ammonia*, PHYS 432, (2004). http://courses.washington.edu/phys432/NH3/ammonia_inversion.pdf.

**[3]** Lavine, James P, *Time-Dependent Quantum Mechanics of Two-Level Systems*, World Scientific Publishing Co. Pte. Ltd. (2019).

**[4]** Paul C. Lauterbur – *Nobel Lecture* (December 8, 2003). NobelPrize.org. Nobel Prize Outreach AB 2021. Tue. 12 Oct 2021.
https://www.nobelprize.org/prizes/medicine/2003/lauterbur/lecture/

**[5]** Schmitt, Melanie, Andreas Potthast, David E. Sosnovik, Jonathan R. Polimeni, Graham C. Wiggins, Christina Triantafyllou, and Lawrence L. Wald, *A 128-channel receive-only cardiac coil for highly accelerated cardiac MRI at 3 Tesla*, Magnetic Resonance in Medicine: An Official Journal of the International Society for Magnetic Resonance in Medicine 59, no. 6 (2008): 1431-1439.

**[6]** Quentin Herreros, *Very low field magnetic resonance imaging*, Other [cond-mat.other]. Université René Descartes - Paris V, (2013).

**[7]** Collins, Christopher M., and Zhangwei Wang, *Calculation of radiofrequency electromagnetic fields and their effects in MRI of human subjects*, Magnetic resonance in medicine 65, no. 5 (2011): 1470-1482.

**[8]** R. Gross, A. Marx, F. Deppe, and K. Fedorov, *Control of quantum two-level systems*, Walther-Meisner-Institute (2001-2020) (BADW), https://www.wmi.badw.de/fileadmin/WMI/Lecturenotes/Applied_Superconductivity/AS2020_Chapter06_3_Slides.pdf.

**[9]** Sammet, Steffen, *Magnetic resonance safety*, Abdominal Radiology 41, no. 3 (2016): 444-451.

**[10]** P. Davidovits and R. Novick, *The optically pumped rubidium maser*, in Proceedings of the IEEE, vol. 54, no. 2, pp. 155-170, Feb. 1966, doi: 10.1109/PROC.1966.4628.

**[11]** Slawomir Bilicki, *Strontium optical lattice clocks : clock comparisons for timescales and fundamental physics applications*. Physics [physics]. Université Pierre et Marie Curie - Paris VI, (2017). English.

[12] Meynadier, F., P. Delva, C. le Poncin-Lafitte, C. Guerlin, and P. Wolf, *Atomic clock ensemble in space (ACES) data analysis*, Classical and Quantum Gravity 35, no. 3 (2018): 035018.

[13] Schmittberger, Bonnie L., and David R. Scherer, *A review of contemporary atomic frequency standards*, arXiv preprint arXiv:2004.09987 (2020).

[14] J. P. Gordon, *The Maser*, Scientific American, Dec. 1958, page 42.

[15] Feynman, Richard P., Robert B. Leighton, and Matthew Sands, *The Feynman lectures on physics, Vol. III: The new millennium edition: Quantum Mechanics*. Hachette UK, (2015).

[16] Kleppner, Daniel, H. Mark Goldenberg, and Norman F. Ramsey, *Properties of the hydrogen maser*, Applied Optics 1, no. 1 (1962): 55-60.

[17] Ramsey, Norman F., *The atomic hydrogen maser*, Metrologia 1, no. 1 (1965): 7.

[18] Rigamonti, Attilio, and Pietro Carretta, *Structure of matter*, Springer-Verlag Italia, 2007.

[19] Goujon, D., P. Rochat, P. Mosset, D. Boving, A. Perri, J. Rochat, N. Ramanan et al. *Development of the space active hydrogen maser for the aces mission*, In EFTF-2010 24th European Frequency and Time Forum, pp. 1-6. IEEE, 2010.

[20] Suzuki, Masatsugu Sei, *Ammonia Maser*, Masatsugu Sei Suzuki Department of Physics, SUNY at Binghamton (Date: November 08, 2020)." (2020).

[21] Fabre, C., S. Haroche, J.M. Raimond, P. Goy, M. Gross, and L. Moi, *Rydberg atoms and radiation in a resonant cavity II. Experiments*, Le Journal de Physique Colloques 43, no. C2 (1982): C2-275.

[22] Zagoskin, A. M., Sahel Ashhab, J. R. Johansson, and Franco Nori, *Quantum two-level systems in Josephson junctions as naturally formed qubits*, Physical review letters 97, no. 7 (2006): 077001.

[23] Holevo, Alexander S., *Statistical structure of quantum theory*, Vol. 67. Springer Science & Business Media, (2003).

[24] Stenholm, Stig, and Kalle-Antti Suominen, *Quantum approach to informatics*, John Wiley & Sons, 2005.

[25] Nielsen, Michael A; Chuang, Isaac L., 1968- (2010), *Quantum Computation and Quantum Information* (New ed., 10th anniversary ed.), Cambridge University Press, ISBN 978-1-107-00217-3

[26] Haroche, Serge, *Nobel Lecture: Controlling photons in a box and exploring the quantum to classical boundary*, Reviews of Modern Physics 85, no. 3 (2013): 1083.

[27] Devoret, Michel H., Andreas Wallraff, and John M. Martinis. *Superconducting qubits: A short review*, arXiv preprint cond-mat/0411174V1 (2004).

[28] J. Q. You and Franco Nori, *Superconducting Circuits and Quantum Information*, Physics Today 58, 11, 42 (2005).

[29] Ravid Shaniv, *Spectral noise analysis of a narrow line-width laser using a single trapped ion*, MSc thesis (2015).

**[30]**   Yotam Shapira, *Robust entanglement of trapped ion qubits*, MSc thesis (2018).

**[31]**   Bruzewicz, Colin D., John Chiaverini, Robert McConnell, and Jeremy M. Sage, *Trapped-ion quantum computing: Progress and challenges*, Applied Physics Reviews 6, no. 2 (2019): 021314.

**[32]**   Mantei, D., Jens Förstner, S. Gordon, Y. A. Leier, A. K. Rai, Dirk Reuter, A. D. Wieck, and Artur Zrenner, *Robust Population Inversion by Polarization Selective Pulsed Excitation*, Scientific reports 5, no. 1 (2015): 1-8.

**[33]**   Hargart, F., K. Roy-Choudhury, T. John, S. L. Portalupi, C. Schneider, Sven Höfling, M. Kamp, S. Hughes, and P. Michler, *Probing different regimes of strong field light–matter interaction with semiconductor quantum dots and few cavity photons*, New Journal of Physics 18, no. 12 (2016): 123031.

**[34]**   Chatterjee, Anasua, Paul Stevenson, Silvano De Franceschi, Andrea Morello, Nathalie P. de Leon, and Ferdinand Kuemmeth, *Semiconductor qubits in practice*, Nature Reviews Physics 3, no. 3 (2021): 157-177.

**[35]**   Vandersypen, Lieven MK, and Mark A. Eriksson, *Quantum computing with semiconductor spins*, Physics Today 72 (2019): 8-38.

**[36]**   Albino, Andrea, Stefano Benci, Lorenzo Tesi, Matteo Atzori, Renato Torre, Stefano Sanvito, Roberta Sessoli, and Alessandro Lunghi, *First-Principles Investigation of Spin–Phonon Coupling in Vanadium-Based Molecular Spin Quantum Bits*, Inorganic chemistry 58, no. 15 (2019): 10260-10268.

**[37]**   Najafian, Kaveh, Ziv Meir, and Stefan Willitsch, *From megahertz to terahertz qubits encoded in molecular ions: theoretical analysis of dipole-forbidden spectroscopic transitions in $N_2^+$*, Physical Chemistry Chemical Physics 22, no. 40 (2020): 23083-23098.

**[38]**   Reinhardt, Ori, Chen Mechel, Morgan Lynch, and Ido Kaminer, *Free-Electron Qubits*, Annalen der Physik 533, no. 2 (2021): 2000254.

**[39]**   DiVincenzo, David P., *The physical implementation of quantum computation*, Fortschritte der Physik: Progress of Physics 48, no. 9-11 (2000): 771-783.

**[40]**   Chow, Weng W., Stephan W. Koch, and Murray III Sargent, *Semiconductor-laser physics*, Springer Science & Business Media, (2012).

**[41]**   Shirley, Jon H, *Solution of the Schrödinger equation with a Hamiltonian periodic in time*, Physical Review 138, no. 4B (1965): B979.

**[42]**   Frimmer, Martin, and Lukas Novotny, *The classical Bloch equations*, American Journal of Physics 82, no. 10 (2014): 947-954.

**[43]**   Spreeuw, R. J. C., N. J. Van Druten, M. W. Beijersbergen, E. R. Eliel, and J. P. Woerdman, *Classical realization of a strongly driven two-level system*, Physical review letters 65, no. 21 (1990): 2642.

**[44]**   Bouwmeester, D., N. H. Dekker, FE V. Dorsselaer, C. A. Schrama, P. M. Visser, and J. P. Woerdman, *Observation of Landau-Zener dynamics in classical optical systems*, Physical Review A 51, no. 1 (1995): 646.

**[45]**   Faust, Thomas, Johannes Rieger, Maximilian J. Seitner, Jörg Peter Kotthaus, and Eva Maria Weig, *Coherent control of a classical nanomechanical two-level system*, Nature Physics 9, no. 8 (2013): 485-488.

[46]   Okamoto, Hajime, Adrien Gourgout, Chia-Yuan Chang, Koji Onomitsu, Imran Mahboob, Edward Yi Chang, and Hiroshi Yamaguchi, *Coherent phonon manipulation in coupled mechanical resonators*, Nature Physics 9, no. 8 (2013): 480-484.

[47]   Sakurai, J. J., and J. Napolitano, *Modern Quantum Mechanics. 2-nd edition*, Person New International edition (2014).

[48]   Meystre, Pierre, and Murray Sargent, *Elements of quantum optics*, Springer Science & Business Media, 2007.

[49]   Cohen-Tannoudji, Claude, Bernard Diu, and Franck Laloë, *Quantum Mechanics*, Volume 1: Basic Concepts, Tools, and Applications. John Wiley & Sons, (2019).

[50]   B.W. Shore, *The Theory of Coherent Atomic Excitations* (Wiley, New York, 1990).

[51]   Andrei Tokmakoff, *2.3: Two-Level Systems*, Chemistry LibreTexts (2020).

[52]   Wikipedia – Rabi cycle. https://en.wikipedia.org/wiki/Rabi_cycle

[53]   Problems in Mathematics, *Express the Eigenvalues of a 2 by 2 Matrix in Terms of the Trace and Determinant*, https://yutsumura.com/express-the-eigenvalues-of-a-2-by-2-matrix-in-terms-of-the-trace-and-determinant/

[54]   Wharton, K. B., and D. Koch, *Unit quaternions and the Bloch sphere*, Journal of Physics A: Mathematical and Theoretical 48, no. 23 (2015): 235302.

[55]   Allen, Leslie, and Joseph H. Eberly, *Optical resonance and two-level atoms*, Vol. 28. Courier Corporation, 1975.

[56]   Bonani, Fabio D, *The Jaynes-Cummings Model*, A· A1, no. e2 (2020): 2mec.

[57]   R. P. Feynman, F. L. Vernon, Jr., and R. W. Hellwarth, J. Appl Phys. 28, 49 (1957).

[58]   Shore, Bruce W., and Peter L. Knight, *The Jaynes-Cummings model*, Journal of Modern Optics 40, no. 7 (1993): 1195-1238.

[59]   Steck, Daniel A, *Quantum and atom optics*, (2007). (Rev. 2021)

[60]   Feynman, Richard P., Robert B. Leighton, and Matthew Sands, *The Feynman lectures on physics, Vol. I: The new millennium edition: Quantum Mechanics*. Hachette UK, (2010).

[61]   N.B. Delone, V.P. Krainov, *Atoms in Strong Light Fields*, Springer Ser. Chern. Phys. Vol. 28 (Springer, Berlin-Heidelberg 1985)

[62]   Bayfield, James E., *Quantum evolution: An introduction to time-dependent quantum mechanics*, John Wiley & Sons, Inc. (1999).

[63]   Cohen-Tannoudji, Claude N, *The Autler-Townes effect revisited*, In Amazing Light, pp. 109-123. Springer, New York, NY, (1996).

[64]   Delone, Nikolai B., and Vladimir Pavlovich Kraĭnov, *Multiphoton Processes in Atoms*, © Springer-Verlag Berlin Heidelberg 1994.

[65]   Zel'Dovich, Ya B. *Scattering and emission of a quantum system in a strong electromagnetic wave*, Soviet Physics Uspekhi 16, no. 3 (1973): 427.

**[66]** Ter-Mikhaelyan, Mikhail L, *Simple atomic systems in resonant laser fields*, Physics-Uspekhi 40, no. 12 (1997): 1195.

**[67]** Luo, Xiaobing, Qiongtao Xie, and Biao Wu, *Quasi-energies and Floquet states of two weakly coupled Bose-Einstein condensates under periodic driving*, Physical Review A 77, no. 5 (2008): 053601.

**[68]** Shimano, Ryo, and Makoto Kuwata-Gonokami, *Observation of Autler-Townes splitting of biexcitons in CuCl*, Physical review letters 72, no. 4 (1994): 530.

**[69]** Bai, Jiandong, Jieying Wang, Shuo Liu, Jun He, and Junmin Wang, *Autler–Townes doublet in single-photon Rydberg spectra of cesium atomic vapor with a 319 nm UV laser*, Applied Physics B 125, no. 3 (2019): 33.

**[70]** Mollow, B. R, *Power spectrum of light scattered by two-level systems*, Physical Review 188, no. 5 (1969).

**[71]** Eichhorn, Marc, and Markus Pollnau, *Spectroscopic foundations of lasers: Spontaneous emission into a resonator mode*, IEEE Journal of Selected Topics in Quantum Electronics 21, no. 1 (2014): 486-501.

**[72]** Schuda, F., C. R. Stroud Jr, and M. Hercher, *Observation of the resonant Stark effect at optical frequencies*, Journal of Physics B: Atomic and Molecular Physics 7, no. 7 (1974): L198.

**[73]** Wu, F. Y., Grove, R. E. & Ezekiel, S., *Investigation of the spectrum of resonance fluorescence by a monochromatic field*, Phys. Rev. Lett. 35, 1426_1429 (1975).

**[74]** Hartig, W., W. Rasmussen, R. Schieder, and H. Walther, *Study of the frequency distribution of the fluorescent light induced by monochromatic radiation*, Z. Phys. A278, no. 3 (1976): 205-210.

**[75]** Vamivakas, A. Nick, Yong Zhao, Chao-Yang Lu, and Mete Atatüre, *Spin-resolved quantum-dot resonance fluorescence*, Nature Physics 5, no. 3 (2009): 198-202.

**[76]** Ortiz-Gutiérrez, Luis, Raul Celistrino Teixeira, Aurélien Eloy, Dilleys Ferreira da Silva, Robin Kaiser, Romain Bachelard, and Mathilde Fouché, *Mollow triplet in cold atoms*, New Journal of Physics 21, no. 9 (2019): 093019.

**[77]** Hu-Berlin, *Quantized Interaction of Light and Matter*, https://www.physik.hu-berlin.de/de/nano/lehre/copy_of_quantenoptik09/Chapter8

**[78]** Weisstein, Eric W. "Eigenvalue." From *MathWorld*--A Wolfram Web Resource. https://mathworld.wolfram.com/Eigenvalue.html.

**[79]** *Laser Physics* by Murray Sargent, III, Marlan O. Scully, and Willis E. Lamb, Jr. Reviewer. American Journal of Physics 44, no. 7 (1976): 715-716.

**[80]** Weisstein, Eric W. "Hermitian Matrix." From *MathWorld*--A Wolfram Web Resource. https://mathworld.wolfram.com/HermitianMatrix.html.

**[81]** Wikipedia, *Two-state quantum system*, https://en.wikipedia.org/wiki/Two-state_quantum_system.

**[82]** Russell Anderson, *Two-Level System: Rabi Oscillations* – Yumpu (2008) https://www.yumpu.com/en/document/read/27392779/two-level-system-rabi-oscillations.

| | |
|---|---|
| **[83]** | Chemeurope encyclopedia, *"Rabi frequency"* page. https://www.chemeurope.com/en/encyclopedia/Rabi_frequency.html |
| **[84]** | Cardoso, G. C., and J. W. R. Tabosa, *Four-wave mixing in dressed cold cesium atoms*, Optics communications 185, no. 4-6 (2000): 353-358. |
| **[85]** | Cohen-Tannoudji, Claude, Jacques Dupont-Roc, and Gilbert Grynberg, *Atom-photon interactions: basic processes and applications*. (1998). |
| **[86]** | Fast way to calculate Eigen of 2x2 matrix using a formula, math.stackexchange |
| **[87]** | Problems in Mathematics, *How to Diagonalize a Matrix. Step by Step Explanation*, https://yutsumura.com/how-to-diagonalize-a-matrix-step-by-step-explanation/ |
| **[88]** | Weisstein, Eric W. "Matrix Diagonalization." From *MathWorld*--A Wolfram Web Resource. https://mathworld.wolfram.com/MatrixDiagonalization.html. |
| **[89]** | Horn, Roger A., and Charles R. Johnson. *Matrix analysis*. Cambridge university press, (2012). |
| **[90]** | Birne Binegar, *System of First Order ODEs with Constant Coefficients*, (2008) https://math.okstate.edu/people/binegar/4233/4233-l02.pdf. |
| **[91]** | Landsman, Nicolaas P, *Born rule and its interpretation*, In Compendium of quantum physics, pp. 64-70. Springer, Berlin, Heidelberg, (2009). |
| **[92]** | https://chem.libretexts.org/, *3.5: Schrödinger and Heisenberg Representations* (2020). |
| **[93]** | Paul Davidovits, in Physics in Biology and Medicine (Fifth Edition), 2019, *Larmor frequency,* https://www.sciencedirect.com/topics/computer-science/larmor-frequency. |
| **[94]** | Williams, H. Thomas, *Discrete Quantum Mechanics*, Morgan & Claypool Publishers, (2015). |
| **[95]** | Hiruy, Taddese Mengistu, *Time Dependent Quantum Mechanical Approach: Case Studies Of Ammonia Molecule*, Ph.D. diss., Addis Ababa University, (2012). |
| **[96]** | Tserkis, S. T., Ch C. Moustakidis, S. E. Massen, and C. P. Panos, *Quantum tunneling and information entropy in a double square well potential: The ammonia molecule*, Physics Letters A 378, no. 5-6 (2014): 497-504. |
| **[97]** | Yang, Ciann-Dong, and Shiang-Yi Han, *Tunneling Quantum Dynamics in Ammonia*, International Journal of Molecular Sciences 22, no. 15 (2021): 8282. |
| **[98]** | Frank Rioux, 188: The Ammonia Inversion and the Maser, Chemistry LibreTexts, (2020). |
| **[99]** | Pracna, P., V. Špirko, and W. P. Kraemer, *Electric dipole moment function of ammonia*, Journal of Molecular Spectroscopy 136, no. 2 (1989): 317-332. |
| **[100]** | Park, Youngwook, Hani Kang, Robert W. Field, and Heon Kang, *The frequency-domain infrared spectrum of ammonia encodes changes in molecular dynamics caused by a DC electric field*, Proceedings of the National Academy of Sciences 116, no. 47 (2019): 23444-23447. |

**[101]** A. Yariv, *Optical Electronics*, Saunders College Publishing, Philadelphia, (1991), 4th ed., pp. 519 – 524.

**[102]** A.W. Snyder, J.D. Love, Optical Waveguide Theory, Chapman and Hall, London, UK (1983).

**[103]** Syms, R. R. A., *Improved coupled-mode theory for codirectionally and contradirectionally coupled waveguide arrays*, JOSA A 8, no. 7 (1991): 1062-1069.

**[104]** Haus, Hermann, *Waves and fields in optoelectronics*, PRENTICE-HALL, INC., ENGLEWOOD CLIFFS, NJ 07632, USA, 1984, 402 (1984).

**[105]** Huang, Wei-Ping, *Coupled-mode theory for optical waveguides: an overview*, JOSA A 11, no. 3 (1994): 963-983.

**[106]** Levy, Uri, and Yaron Silberberg, *Electrical-Field Distributions in Waveguide Arrays-Exact and Approximate*, arXiv preprint arXiv:1401.0642 (2014).

**[107]** Humphrey, M. A., D. F. Phillips, and R. L. Walsworth, *Double-resonance frequency shift in a hydrogen maser*, Physical Review A 62, no. 6 (2000): 063405.

**[108]** Olsen, Thomas, Toke Lynæs Larsen, Carlos Leonardo Garrido Alzar, and Eugene Polzik, *Double resonance spectroscopy on the cesium atomic clock transition*, QUANTOP Lab., Niels Bohr Instit., Univ, Copenhagen (2004).

**[109]** Tannor, David J. *Introduction to quantum mechanics: a time-dependent perspective*, University Science Books (2007).

**[110]** DanielSank, *What is the "interaction picture" or "rotating frame" in quantum mechanics?* (2015).

**[111]** Orlando, T. P., *Two-level system with static and dynamic coupling* (2004).

**[112]** R. Paschotta, article on 'Rabi oscillations' in the *Encyclopedia of Laser Physics and Technology*, 1. edition October 2008, Wiley-VCH, ISBN 978-3-527-40828-3. https://www.rp-photonics.com/rabi_oscillations.html.

**[113]** BIPM - The International System of Units (2019) https://www.bipm.org/utils/common/pdf/si-brochure/SI-Brochure-9.pdf

**[114]** Ahmed, Mushtaq, Daniel V. Magalhães, Aida Bebeachibuli, Stella T. Müller, Renato F. Alves, Tiago A. Ortega, John Weiner, and Vanderlei S. Bagnato, *The Brazilian time and frequency atomic standards program*, Anais da Academia Brasileira de Ciências 80 (2008): 217-252.

**[115]** Preskill, John, *Quantum computing in the NISQ era and beyond*, Quantum 2 (2018): 79.

**[116]** Matityahu, Shlomi, Hartmut Schmidt, Alexander Bilmes, Alexander Shnirman, Georg Weiss, Alexey V. Ustinov, Moshe Schechter, and Jürgen Lisenfeld, *Dynamical decoupling of quantum two-level systems by coherent multiple Landau–Zener transitions*, npj Quantum Information 5, no. 1 (2019): 1-7.

**Abstract**

Energies of quantum states are given by the arguments of phase-evolution exponentials. It follows then that an analysis of the energies of a two-state system (TSS) can revolve around phase-emphasized description of states' probability amplitudes *in the Schrödinger picture*.

Here, studying energies of TSSs semi-classically, we suggest an energy-revealing format in which the time-dependence of the probability *amplitudes* is expressed by *phase-evolution factors only*. With this fresh energy-studying approach, we first revisit non-driven TSSs, write the conditions for setting a system (in general) in a stationary state, and identify the associated (single) definite energy. Then, more importantly, we revisit driven-TSSs, identify the two stationary states and prove the existence of two quasi-energies associated with each stationary state. Resulting from our phase-keeping framework, we display a "breathing-spiral" type precession-mode of a strongly-driven spin one-half TSS. Further, we exemplify our findings through a set of two coupling-probe computer simulations. For specific examples, we compute and list the energies of several concrete TSSs under typical operating conditions.

## 1. Introduction

*Overview*. The celebrated quantum mechanical two-state system (often referred-to also as a *two-level system - TLS*) is a favorite quantum-mechanical pedagogical "pillar" **[1]**,**[2]**. More importantly, concrete physical TSSs are in the heart of several proliferate modern applications **[3]**. Notably – magnetic resonance imaging (MRI) **[4]**-**[9]**, strontium, cesium, rubidium, aluminum, silver,…, microwave clocks and optical frequency standards **[10]**-**[13]**, ammonia and hydrogen masers **[14]**-**[20]**, Rydberg masers **[21]**, Josephson junctions **[15]**,**[22]**,**[18]**. Various types of qubits **[23]**-**[26]** – superconducting **[27]**,**[28]**, trapped-ion **[29]**-**[31]**, semiconductor **[32]**-**[34]**, QD electron spin **[35]**, molecular **[36]**,**[37]**, free electron **[38]**, which constitute the building block of any quantum information processing, all operate essentially as TSSs **[39]**. Moreover - the theory of two-level systems have been used to model semiconductor *laser* media **[40]**.

Driven two-level systems are usually associated with the realm of quantum physics. However, *classical approximations* of driven two-level systems were theoretically demonstrated **[41]**,**[42]** and even *classical implementations* were experimentally constructed and explored: photon polarizations **[43]**,**[44]**,**[15]**, propagation constants of two counter propagating optical modes in a ring resonator **[43]**, nanoelectromechanical system **[45]**, coupled driven mechanical resonators **[46]**. The purely quantum effect of spontaneous emission however, is not derived in the studies of classical systems **[43]**,**[42]**.

Five Nobel Prizes in physics were awarded to a number of distinguished scientists for studies related in some form to time-dependent two-state systems. The first four **[47]** to - Rabi (1944); Bloch and Purcell (1952); Townes, Basov, and Prochorov (1964); and Kastler (1966). The fifth Nobel prize was awarded relatively recently to Haroche and Wineland (2012).

*Energies*. The wavefunction $(\psi(\boldsymbol{r},t))$ of a quantum-mechanical system can be written as a superposition of space functions $(u_n(\boldsymbol{r}))$ each multiplied by a phase-evolution factor $\left(exp(-ĵ \cdot \omega_n \cdot t)\right)$ where $ĵ \equiv \sqrt{-1}$ and $\omega_n$ is a constant "phase rotational speed" ([48] eq. 3.13):

$$\psi(r,t) = \sum_n C_n(t) \cdot u_n(r) \cdot e^{-ĵ \cdot \omega_n \cdot t}$$

(1)

For problems described by a Hamiltonian ($\mathcal{H}$) satisfying the eigenvalue equation $\mathcal{H} \cdot u_n(r) = \hbar \cdot \omega_n \cdot u_n(r)$, the expansion coefficients $(C_n(t))$ are *time-independent* [48]. In that case - system's eigen-energies $(\mathcal{E}_n)$ each is given according to the Planck-Einstein relations [49] by the constant phase rotational speed $(\omega_n)$ of the corresponding phase-evolution factor, multiplied by the reduced Plank constant $(\hbar)$ as:

$$\mathcal{E}_n = \hbar \cdot \omega_n$$

(2)

If an interaction energy is added to the Hamiltonian, then the expansion coefficients $(C_n(t))$ become *time-dependent* [48]).

Now, given a time-dependent interaction, if the expansion coefficients for a wavefunction of the full Hamiltonian (including the interaction energy), can be expressed as a sum of *phase-evolution factors only*:

$$C_n(t) = \mathcal{A} \cdot e^{-ĵ \cdot \omega_{n,\mathcal{A}} \cdot t} + \mathcal{B} \cdot e^{-ĵ \cdot \omega_{n,\mathcal{B}} \cdot t} + \cdots$$

(3)

where $\mathcal{A}, \mathcal{B}, ...$ are time-independent constants, then the set of phase rotational speeds $\left((\omega_n + \omega_{n,\mathcal{A}}); (\omega_n + \omega_{n,\mathcal{B}}); ...\right)$ can be identified with a set of "quasi-energies" $(\mathcal{E}_{n,\mathcal{A}}; \mathcal{E}_{n,\mathcal{B}}; ...)$ of the $n$-th component of the wavefunction[1]:

$$\mathcal{E}_{n,\mathcal{A}} = \hbar \cdot (\omega_n + \omega_{n,\mathcal{A}}) \,;\, \mathcal{E}_{n,\mathcal{B}} = \hbar \cdot (\omega_n + \omega_{n,\mathcal{B}}) \,;\, ...$$

(4)

It follows then that state's energies or quasi-energies are intimately related to constant phase rotational speeds $(\omega_{n,\mathcal{A}}, \omega_{n,\mathcal{B}}, ...)$ in the arguments of phase-evolution factors $\left(e^{-ĵ \cdot \omega_{n,\mathcal{A}} \cdot t}, e^{-ĵ \cdot \omega_{n,\mathcal{B}} \cdot t}, ...\right)$ describing the expansion coefficients $(C_n(t))$ of the wavefunction.

Our work here is focused on energies of TSSs and therefore focused on phase-evolution factors describing the expansion coefficients of the associated wave functions. With this fresh energy-studying approach we first systematically revisit TSSs characterized by a *time-independent* Hamiltonians (TIHs). Next we proceed to the major and more energy-intriguing part of our study – TSSs characterized by *time-dependent* Hamiltonians (TDHs). Our suggested new angle of energy-study by phase-evolution factors thus brings a useful insight into the subject of energies of two state systems.

*Time-independent Hamiltonian.* Considering TIH-TSSs, still as part of the introduction, all types of TSSs can be characterized through a one representative four-parameter time-independent Hamiltonian [50],([15] Eq. 8.43 with small

---

[1] See [50] §4.2 for quasi-energies associated with each state of a two-state atom under *periodic excitation*, and in particular cf. two such quasi-energies associated with each state of a two-state atom given the near-resonance-valid RWA (Eq. 4.2-18).

modifications),[52],[51]. Written in a canonical set [23] of two basis states (spanning a two-dimensional Hilbert space), the TSS representative Hamiltonian reads (in units of reduced Planck constant, $\hbar$):

$$(\mathcal{H}_{TSS}) = (\mathcal{H}_{\omega 0}) + (\mathcal{H}_{12})$$

$$(\mathcal{H}_{\omega 0}) \equiv \omega_0 \cdot (I) \; ; \; (\mathcal{H}_{12}) \equiv \begin{pmatrix} -\omega_{11} & |\omega_D| \cdot e^{-\mathbb{j} \cdot \phi_D} \\ |\omega_D| \cdot e^{+\mathbb{j} \cdot \phi_D} & +\omega_{11} \end{pmatrix} \; ; \; (\omega_0, \omega_{11}, |\omega_D|, \phi_D) \in \mathbb{R}$$

(5)

where $(I)$ is a 2 × 2 identity matrix. Note that $(\mathcal{H}_{12})$ is Hermitian and traceless, thus associated with two equal-magnitude, opposite-sign, real eigenvalues [53].

Of the three Hamiltonian angular-frequency parameters $(\omega_0, \omega_{11}, |\omega_D|)$ and the single phase parameter $(\phi_D)$, the angular-frequency parameters $\omega_0$ would appear in the system description as a constant global phase-evolution factor $exp(-\mathbb{j} \cdot \omega_0 \cdot t)$. Such global phase-evolution factor does not, in itself, have any physical meaning [54], and its value is typically set arbitrarily, often to zero [55],([15] Ch. 11),[56]. The phase parameter $(\phi_D)$ of the off-diagonal "coupling" matrix element is often set to zero or to $\pi$. We are left then with two key TSS angular frequency parameters - $(\omega_{11}, |\omega_D|)$.

Indeed the two key parameters of the representative Hamiltonian $(\omega_{11}, |\omega_D|)$ enter into two angular frequency parameters - $\Omega_P, \Omega_N$ – defined as:

$$\Omega_P = \sqrt{\omega_{11}^2 + |\omega_D|^2} \; ; \; \Omega_N = -\Omega_P$$

(6)

The $\Omega_P$ angular frequency parameter of Eq. (6), very frequently appearing throughout our entire study, is related to the *generalized Rabi flopping frequency* - $\Omega_{GR}$ (elaboration and references in the text below) by a simple factor of two ($\Omega_{GR} = 2 \cdot \Omega_P$).

Writing the probability amplitudes $\big(C_1(t), C_2(t)\big)$ of the two states as a column vector $\big(\boldsymbol{C}(t)\big)$ the Schrödinger equation for the time evolution of the probability amplitudes reads:

$$\frac{d\boldsymbol{C}(t)}{dt} = -\mathbb{j} \cdot (\mathcal{H}_{TSS}) \cdot \boldsymbol{C}(t) \; ; \; \boldsymbol{C}(t) \equiv \begin{pmatrix} C_1(t) \\ C_2(t) \end{pmatrix}$$

(7)

Analytic expressions for the probability amplitudes of the TIH-TSS, solving Eq. (7), can be computed by the common method of matrix diagonalization.

The author of ([15] Ch. 9), worked out "special solutions" for the free ammonia TSS, identified these special solutions as *stationary states* or *states of definite energy*, and associated the single, constant, phase rotational speed of each such stationary state with its energy. The energy gap or "doublet split" (in units of $\hbar$) between the two energy levels of a free ammonia molecule is turned out to be the generalized Rabi flopping frequency $(\Omega_{GR})$. In fact, as we show below, the spectral doublet split of *all* TSSs (both time-dependent and time-independent) is $\Omega_{GR}$. Many of TSS publications however, are focused on expressions for state-occupation probabilities ($|C_1(t)|^2$ and $|C_2(t)|^2$) in which phase evolution information is partly lost. In these papers, TSS energies are either not central or not discussed at all.

In our study, with TSS energies in mind, we present the analytic expressions for the probability amplitudes in a rather unique way – we have lumped the state's parameters, including the initial conditions, into four *time-independent* constants $(\mathcal{A}, \mathcal{B}, \mathcal{C}, \mathcal{D})$ multiplying phase-evolution factors each written as an exponent of a single fixed phase rotational speed. With such a presentation, definite energies of stationary states are readily observed (cf. Eqs. **(3)** and **(4)**).

In addition to presenting and discussing the general solution of Schrödinger equation for the probability amplitudes, we have characterized the following concrete TIH-TSSs under typical operating parameters.

(In parenthesis – gap between the two energy levels under typical operating parameters (in micro-electronvolts).)

- A proton in a constant magnetic field (1 $\mu eV$).
- A free ammonia molecule (98.4 $\mu eV$).
- An ammonia molecule in a constant electric field (98.4 $\mu eV$ + extremely small value).
- Coupled two optical waveguides (difference between two wavenumbers $(\beta)$ – equal waveguides: 1.3 $mm^{-1}$ ; unequal waveguides: 2.5 $mm^{-1}$).

*Time-dependent Hamiltonian.* Considering TDH-TSSs, still as part of the introduction, if the TSS is under a near-resonance sinusoidal perturbation (typically related to a perturbing oscillating electric or magnetic field), then *following the adoption of the rotating wave approximation* (RWA) **[58]**, a closed-form analytic solution of Schrödinger equation for the probability amplitudes is attainable **[41]**,(**[15]** Ch. 9),**[48]**.

Discussions of non-steadily excited TSS-dynamics such as the dynamics of a linear-chirped-pulse excited TSS (leading to the Landau-Zener formula **[50]**,**[44]**,**[59]**) are outside the scope of our present work.

As elucidated above already, if a state of a quantum mechanical system is an energy eigenstate then its time evolution is described by a single phase-evolution factor **[50]** with constant – say $\omega$ – rotational speed of the phase (phase rotation in the complex plane) - $e^{-j\cdot\omega\cdot t}$. *The state's energy $(E)$ is proportional to the phase rotational speed $(\omega)$ through the Planck-Einstein relations - $E = \hbar \cdot \omega$* [2] **[49]**.

Far more intricate, and in terms of energies possibly more interesting, is the scenario of a TDH-TSS. The solution of Schrödinger equation for the probability amplitudes goes usually through transforming the Schrödinger picture into a rotating frame to get, after adopting the RWA, a solvable time-independent Hamiltonian **[61]**,**[62]**. It is readily verified that the state-occupation *probabilities* are indifferent to this rotation transformation. Considering probabilities, phase information is lost in the Schrödinger-picture into a rotating-frame transformation. To recover the phase-evolution factors, a transformation back into the Schrödinger picture is necessary. Since many already-published papers are focused on state probabilities, back rotation transformation is not executed. Here, in a dedicated section, we do execute the back rotation transformation (applied to the state's probability amplitudes in the rotating frame).

---

[2] Interestingly, an amazing resemblance exists between the characterization of the motion of a vibrating linear mechanical system by a superposition of a set of time-varying exponentials, and the quantum mechanical characterization of vibrating "objects" by a superposition of a set of energy (i.e. frequency) exponentials **[60]**.

This is done, phase-evolution factors are restored. Presenting the expressions for the probability amplitudes in our suggested $(\mathcal{A}, \mathcal{B}, \mathcal{C}, \mathcal{D})$ format, two energy effects resulting from the presence of a level-coupling near-resonance sinusoidal field are observed: a. The original energy levels of the unperturbed system are "attracted" to each other or "repelled" **[63]** away from each other, depending on weather the coupling field is red-detuned or blue-detuned respectively[3] **[61]**. b. Each pre-perturbation energy level is split (symmetrically on resonance **[64]**) into a pair of "quasi-energy" levels **[65]**,**[61]**,**[50]**,**[66]**,**[67]**. The created doublet frequencies are spectrally $\Omega_{GR}$ apart. Following our suggested presentation of the TDH-TSS's probability amplitudes, the understanding of the origin of these *known* energy levels "push/pull" and split effects gets a fresh angle.

We also present a "breathing-spiral" type precession-mode of a strongly-driven spin one-half TSS (cf. section **3.2.2** below). Computing the expressions for these "transverse precession" curves consists of adding the two "axial" probability *amplitudes* $(C_1(t), C_2(t))$ (prior to squaring) and thus mandates their (the axial amplitudes') full phase description.

In addition to presenting and discussing the general solutions of Schrödinger equation for the probability amplitudes of a driven TSS, we have characterized the following three concrete TDH-TSSs under typical operating parameters.

(In parenthesis – gap between the two energy levels for the *unperturbed* system and the doublet-splitting of each level for the typically-driven TSS *on resonance* (in micro-electronvolts)).

- A proton under a constant axial magnetic field and an harmonic transverse magnetic field ($1\ \mu eV$ ; $2.6 \cdot 10^{-7}\ \mu eV$).
- A cesium atom driven near resonance by an external harmonic magnetic field ($38\ \mu eV$ ; $2.1 \cdot 10^{-4}\ \mu eV$).
- An ammonia molecule in an harmonic electric field ($98.4\ \mu eV$ ; $3.7 \cdot 10^{-6}\ \mu eV$).

*Coupling-probe computer simulations.* A significant section of our present contribution is dedicated to a set of two semiclassical **[50]** coupling-probe computer simulations. In these simulations three distinct quantum-mechanical energy states of a free atom and two classical sinusoidal electric fields are considered. The three energy states are designated ($|g>, |e>, and\ |r>$). The two lower levels ($|g>, |e>$), considered to be the TSS, are *coupled* by a strong near-resonance sinusoidal electric field. Transitions between states ($|g>, |r>$) in one simulation scenario and between ($|e>, |r>$) in another simulation scenario are *probed* by a weak sinusoidal electric field.

Our simulation results convincingly reveal the energy states of a TDH-TSS (the coupled two lower-energy states). Each of the two simulation scenario predicts an Autler-Townes doublet transitions in a probe experiment **[68]**,**[63]**,**[69]** and *together* the two simulations predict a four-level-originated Mollow triplet spectrum **[70]**,**[71]** (a center line and two side lines $\Omega_{GR}$ apart) in a resonance and near resonance fluorescence (or "scattering") experiments **[72]**-**[76]**. {In the Autler–Townes doublet realization by the authors of **[69]**, the coupled two states (the TSS) are $|g> and\ |r>$ probed by $|g>to\ |e>$ transitions}.

---

[3] Not to be confused with the ac-Stark effect which is also observed in our calculations.

Probing deeper, we present in the simulation-section, curves for the *time-evolution* of state-occupation probabilities for each of the three states in question ($|g>, |e>, and |r>$). The dynamics curves show how the initially empty upper level - $|r>$ (initially zero occupation probability) gets gradually populated, robbing the two lower levels until a maximum occupation (probability) is reached, then gets completely empty again and so repeatedly on. In the cases where the TSS is initially launched into one of its definite-energy states, the occupation probability of the upper level ($|r>$), gets *periodically* (in time) to a *maximum of unity* while the occupation probabilities of the two lower levels ($|g>, |e>$ i.e. the TSS), go *simultaneously* to zero.

The two coupling-probe computer simulations then, not only demonstrate once more the existence of two pairs of quasi-energy levels of a sinusoidally-coupled TSS, but also bring to light the *dynamics* of the occupation probabilities.

To summarize the fresh-angles and some key points of our work as described in the introduction:

- Energies of quantum TSSs are studied here semi-classically by presenting the expansion coefficients of the wave function (the probability amplitudes) as *phase-evolution factors only*, using an ($\mathcal{A}, \mathcal{B}, \mathcal{C}, \mathcal{D}$) format.

- Revisiting TIH-TSS, based on describing the time-dependence of the expansion coefficients by phase-evolution factors only, we write the conditions for setting a system (in general) in a stationary state, and identify the associated (single) definite energy.

- Revisiting sinusoidally-perturbed TDH-TSS, expressing the time-dependence of the expansion coefficients by phase-evolution factors only (after performing transformation from the rotating frame back to the Schrödinger picture), we write the conditions for setting a system (in general) in a stationary state (one of two), and identify the associated quasi-energies (two for each stationary state).

- Resulting from our framework we display, a "breathing-spiral" type precession-mode of a strongly-driven spin one-half TSS (cf. section **3.2.2** below).

- We demonstrate the existence of four quasi-energies of TDH-TSSs by running coupling-probe experiment simulations. We also show the dynamics of occupation of each of the three levels involved, and discuss Lorentzian line widths (vs. probe detuning) of the upper level in each of the different "experiments".

- In way of examples, in addition to the general solution, we present calculated values of state energies for several concrete TIH-TSSs and TDH-TSSs under typical operating conditions.

## 2. Time-independent Hamiltonian – two energy levels

A two-level system is essentially a quantum mechanical system because in classical physics energy is a continuous variable **[61]**. In this two-part section we first present the general solution to Eq. **(7)** above, assuming a time-independent TSS Hamiltonian (Eq. **(5)** above). Then, based on the general solution, we look at a number of representative concrete examples at typical operating parameters:

*Concrete TIH-TSSs (two energy levels)*

a. A proton in a constant magnetic field

b. A free ammonia molecule
  c. An ammonia molecule in a constant electric field
  d. Coupled two optical waveguides

In both the general solution part and the concrete examples part we focus our attention on the energies of the TSSs.

### 2.1. General solution

In this "general solution" section we consider a general TIH, and through its eigenvalues and eigenvectors (or, in the case of light-matter interaction - "dressed states" **[50]**,**[77]**,**[59]**) write a general solution of the Schrödinger equation for the probability amplitudes. We write two solution versions – one is the standard, somewhat obscure matrix multiplication solution, and the other is the energy-revealing $(\mathcal{A}, \mathcal{B}, \mathcal{C}, \mathcal{D})$ analytic solution. We show, through two stationary states, the existence of two definite energies where each energy level is equal (in angular frequency units) to a distinct phase rotational speed.

On passing we look closer into the properties of eigenvalues and eigenvectors, define "modulation depth", and devote a significant subsection to the different routes of computing the average energy of a TIH-TSS.

#### 2.1.1. Basis states and the state vector

Given a set of two canonical basis states

$$|1> \equiv \begin{pmatrix} 1 \\ 0 \end{pmatrix} ; |2> \equiv \begin{pmatrix} 0 \\ 1 \end{pmatrix}$$

**(8)**

any TSS state vector can be written as a linear superposition -

$$|\psi_{1,2}(t)> = C_1(t) \cdot |1> + C_2(t) \cdot |2>$$

**(9)**

The numbers (probability amplitudes) $(C_1(t), C_2(t))$ can be regarded as the components of the state vector $|\psi_{1,2}(t)>$ **[50]**. The evolution of these time-dependent numbers $(C_1(t), C_2(t))$ is the solution of Eq. **(11)** below.

#### 2.1.2. Time-independent Hamiltonian

For clarity, to designate a TIH, we add a subscript "0" to the TSS Hamiltonian of Eq. **(5)** above:

$$(\mathcal{H}_{TSS,0}) = (\mathcal{H}_{\omega 0}) + (\mathcal{H}_{12,0})$$

$$(\mathcal{H}_{\omega 0}) \equiv \omega_0 \cdot (I) ; (\mathcal{H}_{12,0}) \equiv \begin{pmatrix} -\omega_{11} & |\omega_D| \cdot e^{-\mathbb{j} \cdot \phi_D} \\ |\omega_D| \cdot e^{+\mathbb{j} \cdot \phi_D} & +\omega_{11} \end{pmatrix} ; (\omega_0, \omega_{11}, |\omega_D|, \phi_D) \in \mathbb{R}$$

**(10)**

The Schrödinger equation for the time evolution of the probability amplitudes now reads:

$$\frac{d}{dt}\boldsymbol{C}(t) = -\mathbb{j} \cdot \left(\mathcal{H}_{TSS,0}\right) \cdot \boldsymbol{C}(t) \quad ; \quad \boldsymbol{C}(t) \equiv \begin{pmatrix} C_1(t) \\ C_2(t) \end{pmatrix}$$

(11)

Two comments are on order here:

I. The "energies" in the Hamiltonian matrix elements (Eq. **(10)**) are given in frequency units **[8]** (Planck-Einstein relations implied, or $\hbar$ is simply vanishes from the equation of motion since all energies are related to frequencies **[44]**). In the text below we use "frequency" in its actual meaning (in relation to electromagnetic fields for example) and regularly write "frequency" to mean "energy", knowing that the reader will effortlessly assign the correct meaning to this single word.

II. The $\omega_{11}$ parameter of the $\left(\mathcal{H}_{12,0}\right)$ Hamiltonian (Eq. **(10)**) is *positive*, making the $\left(\mathcal{H}_{12,0}\right)[1,1]$ matrix element negative. To insert a positive value into the $\left(\mathcal{H}_{12,0}\right)[1,1]$ matrix element, define a negative $\omega_{11}$ parameter.

### 2.1.3. Eigenvalues and eigenvectors

*Eigenvalues*. The eigenvalues $(\lambda_P, \lambda_N)$ of the *traceless* $\left(\mathcal{H}_{12,0}\right)$ Hamiltonian (Eq. **(10)**) are readily verified to be **[78]**,**[53]** -

$$\lambda_P = \Omega_P \quad ; \quad \lambda_N = \Omega_N \quad ; \quad (\lambda_N = -\lambda_P)$$

(12)

where $\left(\Omega_P(\omega_{11}, |\omega_D|); \Omega_N(\omega_{11}, |\omega_D|)\right)$ are given by Eq. **(6)**.

The eigenvalues $(\Lambda_P, \Lambda_N)$ of the full $\left(\mathcal{H}_{TSS,0}\right)$ Hamiltonian (Eq. **(10)**) are then the sums

$$\Lambda_P = \omega_0 + \lambda_P = \omega_0 + \Omega_P$$
$$\Lambda_N = \omega_0 + \lambda_N = \omega_0 + \Omega_N$$

(13)

Two other closely related angular frequencies very often appear in the scientific literature - $\Omega_R$ and $\Omega_{GR}$. In the context of the TSS's considered here, these two angular frequencies are defined as

$$\Omega_R = 2 \cdot \left(\mathcal{H}_{12,0}\right)[2,1] = 2 \cdot |\omega_D| \cdot e^{+\mathbb{j} \cdot \phi_D} \quad ; \quad \Omega_{GR} = \lambda_P - \lambda_N = 2 \cdot \Omega_P$$

(14)

The angular frequency $\Omega_R$ of Eq. **(14)** above is the *Rabi flopping frequency* (**[48]** Eq. 3.128). In general, the *Rabi flopping frequency* is a complex number. The real and positive angular frequency parameter $\Omega_{GR}$ of Eq. **(14)** is the *generalized Rabi flopping frequency* **[83]**,(**[48]** Eq. (3.132)),(**[55]** Eq. 3.12),(**[42]** Eq. 20),**[84]**,(**[59]** Eq. 5.54),(**[85]** Part III). The authors of (**[79]** Eq. 61) refer to $\Omega_{GR}$ as *Rabi* "flopping frequency" (without "generalized") defined explicitly as the difference of the eigenvalues of $\left(\mathcal{H}_{12,0}\right)$ of Eq. **(10)** (($\lambda_P - \lambda_N$) cf. Eq. **(14)** above). In the literature however there is some inconsistency[4].

---

[4] In **[50]** the dipole element *Rabi frequency* is in general a complex number. In **[40]** the Rabi (flopping) frequency is defined as a complex number. In the "Two-state quantum system" page of Wikipedia the Rabi (flopping) frequency is very explicitly stated as a complex number **[81]**.

Often the Hamiltonian $(\mathcal{H}_{12,0})$ of Eq. **(10)** is preceded by a factor of $\frac{1}{2}$ which is likely a source of confusion.

The eigenvalues $(\lambda_P, \lambda_N)$ of the $(\mathcal{H}_{12,0})$ Hermitian, zero trace Hamiltonian matrix (Eq. **(10)**) are both real **[80]** and are equal in their absolute value. Their value varies with the two key TSS parameters - $(\omega_{11}, |\omega_D|)$ as shown by the panels of ***Figure 1***.

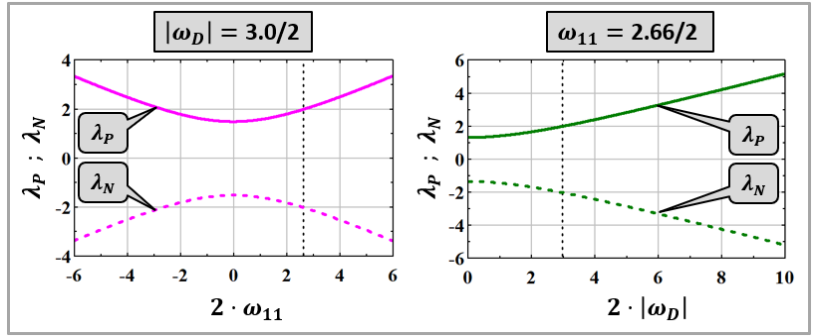

***Figure 1:*** *The eigenvalues ($\lambda_P = \Omega_P$ ; $\lambda_N = -\Omega_P$ cf. Eq. **(12)**) of the $(\mathcal{H}_{12,0})$ Hermitian, zero-trace Hamiltonian matrix (Eq. **(10)**). Left: Vs. $\omega_{11}$. Right: Vs. $|\omega_D|$. The dashed vertical line in each of the panels is set at specific respective parameter-value that is used in the coupling-probe simulations (cf. section **4**).*

*Eigenvectors.* Expressions for the eigenvectors of $(\mathcal{H}_{TSS,0})$ of Eq. **(10)** are given through an angle $\theta_R\big((\omega_{11}, |\omega_D|)\big)$:

$$\theta_R(\omega_{11}, |\omega_D|) = \frac{1}{2} \cdot acos\left(\frac{\omega_{11}}{\Omega_P(\omega_{11}, |\omega_D|)}\right) \quad ; \quad 0 \leq \theta_R \leq \frac{\pi}{2}$$

**(15)**

---

The authors of (**[63]** Eq. 3),**[49]**,**[66]**,**[73]** all refer to the dipole-element-**only** as *Rabi frequency*. In **[48]**, the *real number* $|\Omega_D|$ of the coupling matrix-element is called the *Rabi flopping frequency*. In analyzing level damping, or in introducing the vector model of density matrix, the authors of **[48]** assume that the coupling matrix-element ($|\omega_D| \cdot e^{-j \cdot \phi_D}$) is real. In **[82]** the Rabi (flopping) frequency is a complex number but *can be made real by choosing the phase of the driving field to be zero*. Similarly, in **[57]**,**[55]**, the Rabi (flopping) frequency *can be made real* for some transitions ($\Delta m = 0$) by adjusting the phase of the two states involved. In **[8]** the coupling matrix-element $(\Omega_{eg})$ is *usually assumed* to be real *without loss of generality*.

In our general TSS analysis here we explicitly and consistently take the Rabi flopping frequency to be a complex number $\big(|\Omega_D| \cdot e^{-j \cdot \phi_D}\big)$ **[50]**.

The generalized Rabi flopping frequency $(\Omega_{GR})$ is a real positive number. In **[50]**,**[62]** $\Omega_{GR}$ is the *Rabi flopping frequency*. If $(\mathcal{H}_{12,0})[2,1]$ is real ($\phi_D = 0$) *and* $(\mathcal{H}_{12,0})[1,1] \equiv -\omega_{11} = 0$ then the value of $\Omega_R$ and the value of $\Omega_{GR}$ coincide ($\Omega_{GR} = |\Omega_R|$).

The authors of (**[8]** – pg. 17) refer to $\Omega_{GR}$ as *Rabi frequency*. The author of (**[71]** Eq. 4.31c) refers to $\Omega_{GR}$ as *Rabi frequency of population inversion*. The authors of **[33]**,**[75]** refer to $\Omega_{GR}$ as *effective Rabi frequency*. Regarding the generalized Rabi flopping frequency then, there is some inconsistency in the language but there is no confusion in the literature as to the mathematical-expression description.

Curves of the angle $\theta_R$ (Eq. (15)) and of the related trigonometric functions are plotted in *Figure 2* as a function of the diagonal-element's value ($\omega_{11}$).

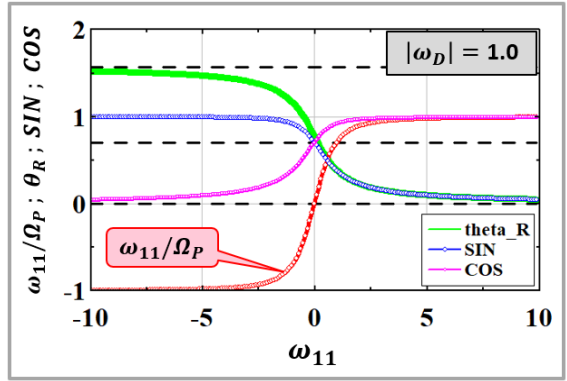

*Figure 2:* Curves related to the angle $\theta_R$ (Eq. (15)). The argument of the acos function of Eq. (15) (red) takes values between $-1$ an 1. The angle $\theta_R$ (green) is thus between zero and $\pi/2$. The trigonometric functions (blue and magenta) are positive across the entire range (any $\omega_{11}$ value). The horizontal black lines designate vertical-axis values of $(0; 1/\sqrt{2}; \pi/2)$.

The orthonormal eigenvectors ($\xi_P; \xi_N$) corresponding to the eigenvalues ($\lambda_P; \lambda_N$) of the *traceless* Hamiltonian matrix ($\mathcal{H}_{12,0}$) of Eq. (10) [86] (also the eigenvectors of the complete ($\mathcal{H}_{TSS,0}$) Hamiltonian matrix) are expressed as:

$$\xi_P = \begin{pmatrix} \sin\theta_R \\ \cos\theta_R \cdot e^{+\mathrm{j}\cdot\phi_D} \end{pmatrix} ; \quad \xi_N = \begin{pmatrix} \cos\theta_R \\ -\sin\theta_R \cdot e^{+\mathrm{j}\cdot\phi_D} \end{pmatrix}$$

(16)

### 2.1.4. Dependence of eigenvectors' components on TSS parameters

We show below that *eigenvectors' components are the initial probability amplitudes of definite-energy states*. As such, it is instructive to probe further into their characteristics. Specifically – dependence of eigenvectors' components (($\xi_{P1}, \xi_{P2}$); ($\xi_{N1}, \xi_{N2}$)) on the key TSS parameters ($\omega_{11}, |\omega_D|$). In our study here, we refer to an eigenvector as "Symm" if its components are both real and positive (not necessarily equal) and as "Asym" if its components are both real and of opposite signs.

The panels of *Figure 3* and of *Figure 4* show the dependence of the eigenvectors components on $\omega_{11}$ (*Figure 3*) and on $|\omega_D|$ (*Figure 4*), both for $\phi_D = 0$. If $\phi_D = \pi$ then Symm ↔ Asym, i.e. $\xi_P \to$ Asym ; $\xi_N \to$ Symm.

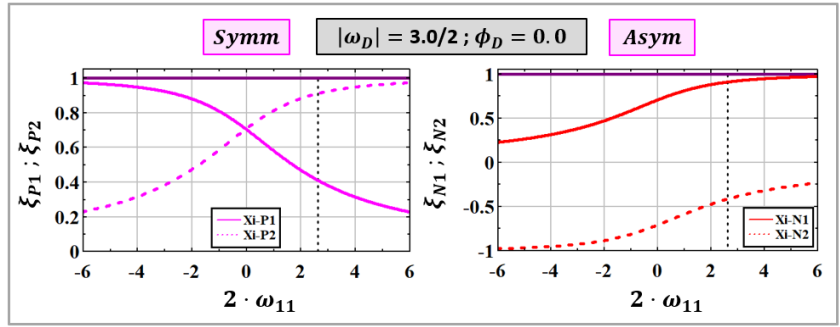

*Figure 3:* *Components of the eigenvectors (Eq. (16)) of the $(\mathcal{H}_{12,0})$ Hamiltonian matrix (Eq. (10)) vs. $\omega_{11}$. Left: The "P" eigenvector components. Right: The "N" eigenvector components. The dashed vertical line in each of the panels and in each of the four panels of* ***Figure 4*** *below is set at specific parameter-value that is used in the coupling-probe simulations (cf. section 4). The curves are plotted for $\phi_D = 0$. If $\phi_D = \pi$ then $\xi_P \to Asym$ ; $\xi_N \to Symm$.*

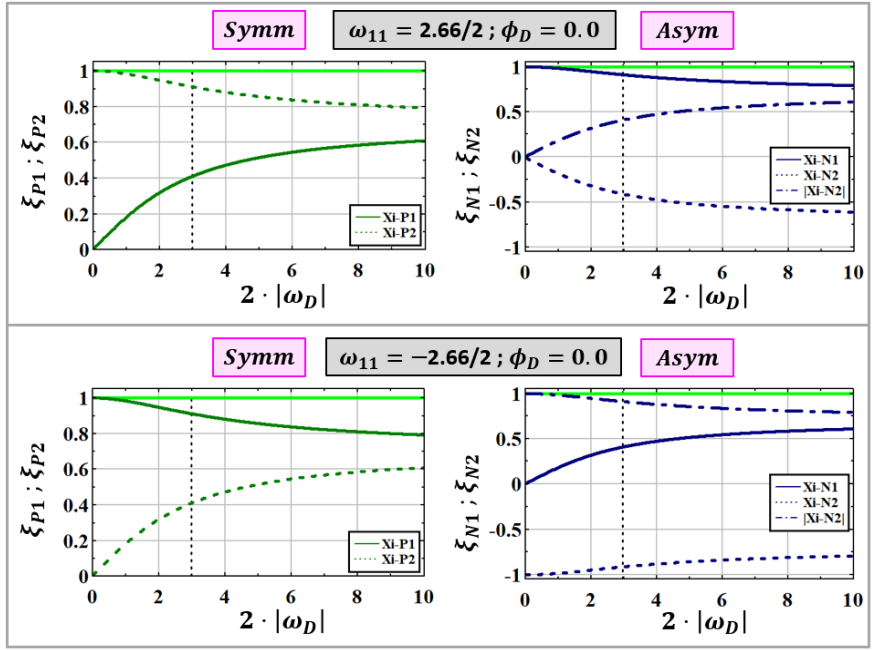

*Figure 4:* *Components of the eigenvectors (Eq. (16)) of the $(\mathcal{H}_{12,0})$ Hamiltonian matrix (Eq. (10)) vs. $|\omega_D|$. Left column: The "P" eigenvector components. Right column: The "N" eigenvector components. As shown, the P eigenvector is consistently Symm according to our definition (see the text above). the N eigenvector is consistently Asym. The curves are plotted for $\phi_D = 0$. If $\phi_D = \pi$ then $\xi_P \to Asym$ ; $\xi_N \to Symm$.*

### 2.1.5. Matrix multiplication solution

The Schrödinger equation for the time evolution of the probability amplitudes $(C_1(t), C_2(t))$ is Eq (11) above (repeated here):

$$\frac{d\mathbf{C}(t)}{dt} = -\mathring{\jmath} \cdot (\mathcal{H}_{TSS,0}) \cdot \mathbf{C}(t) \; ; \; \mathbf{C}(t) \equiv \begin{pmatrix} C_1(t) \\ C_2(t) \end{pmatrix}$$

(17)

with $(\mathcal{H}_{TSS,0})$ given by Eq. (10).

To write the matrix-multiplication solution of Eq. (17) we need to first define an eigenvectors matrix $(\xi)$ as [87]:

$$(\xi) = (\xi_P, \xi_N)$$

(18)

where $\xi_P, \xi_N$ are the *column* eigenvectors (Eq. (16)) of the $(\mathcal{H}_{TSS,0})$ Hamiltonian matrix.

Note that the element $\xi_{P2}$ of the eigenvector $\boldsymbol{\xi}_P$ is the element $\xi_{21}$ of the matrix $(\xi)$. And similarly for the element $\xi_{N1}$ of the eigenvector $\boldsymbol{\xi}_N$ and the element $\xi_{12}$ of the matrix $(\xi)$ (cf. ***Figure 5***).

$$(\xi) \equiv \left[ \begin{pmatrix} \xi_{P1} \\ \xi_{P2} \end{pmatrix} \begin{pmatrix} \xi_{N1} \\ \xi_{N2} \end{pmatrix} \right] = \begin{bmatrix} \xi_{11} & \xi_{12} \\ \xi_{21} & \xi_{22} \end{bmatrix}$$

***Figure 5:*** *The eigenvectors matrix (Eq. **(18)**). The eigenvectors of the $(\mathcal{H}_{TSS,0})$ Hamiltonian matrix are placed as columns next to each other to form the eigenvectors matrix $(\xi)$.*

Note also that the eigenvectors matrix $(\boldsymbol{\xi})$ is a unitary matrix since $(\boldsymbol{\xi})^\dagger = (\boldsymbol{\xi})^{-1}$.

The solution of Eq. **(17)** for the components of the amplitude vector $(\boldsymbol{C}(t))$ in terms of the eigenvalues of Eq. **(12)**, the eigenvectors of Eq. **(16)**, and the initial conditions $(\boldsymbol{C}(0))$ is **[88],[89],[90]**:

$$\begin{pmatrix} C_1(t) \\ C_2(t) \end{pmatrix} = \left( e^{-j \cdot \omega_0 \cdot t} \cdot (I) \right) \cdot (\xi) \cdot (I_\lambda(t)) \cdot (\xi)^{-1} \begin{pmatrix} C_1(0) \\ C_2(0) \end{pmatrix}$$

$$(I_\lambda(t)) \equiv \begin{pmatrix} e^{-j \cdot \lambda_P \cdot t} & 0 \\ 0 & e^{-j \cdot \lambda_N \cdot t} \end{pmatrix}$$

(19)

On passing, and as follows from Eq. **(19)**, let us also isolate the *time evolution operator* $(U(t))$ solving Schrödinger equation **(17)** with the Hamiltonian **(10)**:

$$(U(t)) = \left( e^{-j \cdot \omega_0 \cdot t} \cdot (I) \right) \cdot (\xi) \cdot (I_\lambda(t)) \cdot (\xi)^{-1}$$

(20)

### 2.1.6. The $(\mathcal{A}, \mathcal{B}, \mathcal{C}, \mathcal{D})$ analytic solution

Thinking of state's energies, it is convenient to write an explicit expression for the amplitudes $(\boldsymbol{C}(t))$ solving Schrödinger Eq. **(17)**, not through the time evolution operator but using a four parameter *time-independent* matrix that includes the initial conditions $(\boldsymbol{C}(0))$ as:

$$\boldsymbol{C}(t) = \begin{pmatrix} \mathcal{A} & \mathcal{B} \\ \mathcal{C} & \mathcal{D} \end{pmatrix} \begin{pmatrix} e^{-j \cdot (\cdot \omega_0 - \Omega_P) \cdot t} \\ e^{-j \cdot (\cdot \omega_0 + \Omega_P) \cdot t} \end{pmatrix}$$

$$\mathcal{A} \equiv c^2 \cdot C_1(0) - c \cdot s \cdot e^{-j \cdot \phi_D} \cdot C_2(0)$$
$$\mathcal{B} \equiv s^2 \cdot C_1(0) + c \cdot s \cdot e^{-j \cdot \phi_D} \cdot C_2(0)$$
$$\mathcal{C} \equiv -c \cdot s \cdot e^{+j \cdot \phi_D} \cdot C_1(0) + s^2 \cdot C_2(0)$$
$$\mathcal{D} \equiv c \cdot s \cdot e^{+j \cdot \phi_D} \cdot C_1(0) + c^2 \cdot C_2(0)$$

$$c \equiv \cos \theta_R \ ; \ s \equiv \sin \theta_R$$

(21)

Explicitly:

$$C_1(t) = [\mathcal{A} \cdot e^{+\mathrm{j} \cdot \Omega_P \cdot t} + \mathcal{B} \cdot e^{-\mathrm{j} \cdot \Omega_P \cdot t}] \cdot e^{-\mathrm{j} \cdot \omega_0 \cdot t}$$
$$C_2(t) = [\mathcal{C} \cdot e^{+\mathrm{j} \cdot \Omega_P \cdot t} + \mathcal{D} \cdot e^{-\mathrm{j} \cdot \Omega_P \cdot t}] \cdot e^{-\mathrm{j} \cdot \omega_0 \cdot t}$$

(22)

Thus, the TSS state vector of Eq. **(9)** $(|\psi_{1,2}(t)>)$ can be written with time-dependence specified by only two exponential phase-evolution factors, each having a single phase rotational speeds (either $\omega_P$ or $\omega_N$):

$$(|\psi_{1,2}(t)>) = (\mathcal{A} \cdot e^{-\mathrm{j} \cdot \omega_N \cdot t} + \mathcal{B} \cdot e^{-\mathrm{j} \cdot \omega_P \cdot t}) \cdot |1>$$
$$+ (\mathcal{C} \cdot e^{-\mathrm{j} \cdot \omega_N \cdot t} + \mathcal{D} \cdot e^{-\mathrm{j} \cdot \omega_P \cdot t}) \cdot |2>$$
$$\omega_P \equiv \omega_0 + \Omega_P \; ; \quad \omega_N \equiv \omega_0 - \Omega_P$$

(23)

As explained in the introduction, a constant phase rotational speed of an exponential phase-evolution factor can be identified with energy ($\hbar = 1$):

$$\mathcal{E}_P \equiv \omega_P \; ; \quad \mathcal{E}_N \equiv \omega_N$$

(24)

Note that the $\mathcal{A}$ and $\mathcal{C}$ parameters multiply phase terms with the same phase rotational speed ($\omega_N$). The $\mathcal{B}$ and $\mathcal{D}$ parameters also multiply phase terms with the same phase rotational speed ($\omega_P$). If the initial conditions are such that $\mathcal{A}$ and $\mathcal{C}$ are both zero or that $\mathcal{B}$ and $\mathcal{D}$ are both zero, then the TSS is in a stationary state associated with a *definite-energy* level (either $\omega_P$ or $\omega_N$). Following a brief discussion on state probabilities, we will continue from here and examine the setting of the TSS into a definite-energy state.

In many of the illustrating figures posted below (including figures related to driven TSSs) we show two identical (overlayed) curves, one calculated by matrix multiplication (Eq. **(19)**) and the other according to the $(\mathcal{A}, \mathcal{B}, \mathcal{C}, \mathcal{D})$ solution (Eq. **(21)**).

*State-occupation probabilities.* In terms of the $(\mathcal{A}, \mathcal{B}, \mathcal{C}, \mathcal{D})$ parameters (Eq. **(21)**), the state-occupation probabilities $(\mathcal{P}_1(t) \equiv |C_1(t)|^2 \; ; \; \mathcal{P}_2(t) \equiv |C_2(t)|^2)$ are expressed as (cf. Eq. **(22)**):

$$\mathcal{P}_1(t) = |\mathcal{A}|^2 + |\mathcal{B}|^2 + \mathcal{A} \cdot \mathcal{B}^* \cdot e^{+\mathrm{j} \cdot 2 \cdot \Omega_P \cdot t} + \mathcal{A}^* \cdot \mathcal{B} \cdot e^{-\mathrm{j} \cdot 2 \cdot \Omega_P \cdot t}$$
$$\mathcal{P}_2(t) = |\mathcal{C}|^2 + |\mathcal{D}|^2 + \mathcal{C} \cdot \mathcal{D}^* \cdot e^{+\mathrm{j} \cdot 2 \cdot \Omega_P \cdot t} + \mathcal{C}^* \cdot \mathcal{D} \cdot e^{-\mathrm{j} \cdot 2 \cdot \Omega_P \cdot t}$$

(25)

If the off-diagonal coupling element of $(\mathcal{H}_{TSS,0})$ is real ($\phi_D = 0$ or $\phi_D = \pi$, cf. Eq. **(10)**) and the two initial amplitudes are real ($C_1(0), C_2(0) \in \mathbb{R}$), then the four parameters $(\mathcal{A}, \mathcal{B}, \mathcal{C}, \mathcal{D})$ are all real and the state-occupation probability expressions of Eq. **(25)** are conveniently simplified to:

$$\mathcal{P}_1(t)_{\phi_D = 0/\pi, C(0) \in \mathbb{R}} = \mathcal{A}^2 + \mathcal{B}^2 + 2 \cdot \mathcal{A} \cdot \mathcal{B} \cdot \cos(2 \cdot \Omega_P \cdot t)$$
$$\mathcal{P}_2(t)_{\phi_D = 0/\pi, C(0) \in \mathbb{R}} = \mathcal{C}^2 + \mathcal{D}^2 + 2 \cdot \mathcal{C} \cdot \mathcal{D} \cdot \cos(2 \cdot \Omega_P \cdot t)$$

(26)

According to Eqs. **(25)** and **(26)**, the state-occupation probabilities (of non-stationary states – cf. sections **2.1.7** and **2.1.8** below) of the TIH-TSS (Eq. **(10)**) oscillate at the generalized Rabi flopping frequency ($\Omega_{GR} = 2 \cdot \Omega_P$).

### 2.1.7. Definite-energy states of time-independent-Hamiltonian TSSs

If the two TSS probability amplitudes are initiated to be equal to the components of a time-independent TSS Hamiltonian eigenvector, then the TSS is initiated into a definite-energy state. In the present subsection we prove this key statement step by step.

The set of the two orthonormal eigenvectors or "eigenstates" ($|\xi_P>, |\xi_N>$) of the Hamiltonian ($\mathcal{H}_{TSS,0}$) is a special set of *stationary* or *definite-energy* basis states. In the basis of *definite-energy* states ($|\xi_P>, |\xi_N>$), the Hamiltonian ($\mathcal{H}_{(\xi)}$) is diagonal:

$$(\mathcal{H}_{(\xi)}) = (\xi)^{-1} \cdot (\mathcal{H}_{TSS,0}) \cdot (\xi) = \omega_0 \cdot (I) + \begin{pmatrix} \lambda_P & 0 \\ 0 & \lambda_N \end{pmatrix}$$

(27)

The state vector in the basis of *definite-energy* states is thus

$$|\psi_{(\xi)}(t)> = D_P(0) \cdot e^{-\mathrm{j}\cdot\omega_P\cdot t} \cdot |\xi_P> + D_N(0) \cdot e^{-\mathrm{j}\cdot\omega_N\cdot t} \cdot |\xi_N>$$

(28)

where $\omega_P$ and $\omega_N$ are defined by Eq. (23).

Indeed, according to Eq. (28), the *magnitude* of each of the $D(t)$ amplitudes is time-independent. The explicit expression for the $D(t)$ amplitudes of Eq. (28) then reads:

$$D_P(t) = D_P(0) \cdot e^{-\mathrm{j}\cdot(\omega_0+\lambda_P)\cdot t} = D_P(0) \cdot e^{-\mathrm{j}\cdot\omega_P\cdot t}$$
$$D_N(t) = D_N(0) \cdot e^{-\mathrm{j}\cdot(\omega_0+\lambda_N)\cdot t} = D_N(0) \cdot e^{-\mathrm{j}\cdot\omega_N t}$$

(29)

A general TSS state vector, say $|\phi(t)>$, can be written in either one of these two sets of basis states:

$$|\phi(t)> = |\psi_{1,2}(t)> = C_1(t) \cdot |1> + C_2(t) \cdot |2>$$
$$= |\psi_{(\xi)}(t)> = D_P(t) \cdot |\xi_P> + D_N(t) \cdot |\xi_N>$$

(30)

The expression for $|\psi_{(\xi)}(t)>$ in Eq. (30) with the coefficients $(D_P(t), D_N(t))$ given by Eq. (29) is the TIH-TSS form of the general Eq. (1) with constant expansion coefficients $(D_P(0), D_N(0))$, as stated explicitly by Eq. (28).

Transforming from one set of probability amplitudes to the other can be determined for instance, by consulting the solution to the set of $C(t)$ amplitudes of Eq. (19). If we multiply both sides of solution (19) by the unitary $(\xi)^{-1}$ matrix we get:

$$\begin{pmatrix} D_P(t) \\ D_N(t) \end{pmatrix} \equiv (\xi)^{-1} \cdot \begin{pmatrix} C_1(t) \\ C_2(t) \end{pmatrix} = \left( e^{-\mathrm{j}\cdot\omega_0\cdot t} \cdot (I) \right) \cdot \left( I_\lambda(t) \right) \cdot (\xi)^{-1} \begin{pmatrix} C_1(0) \\ C_2(0) \end{pmatrix}$$
$$= \left( e^{-\mathrm{j}\cdot\omega_0\cdot t} \cdot (I) \right) \cdot \left( I_\lambda(t) \right) \cdot \begin{pmatrix} D_P(0) \\ D_N(0) \end{pmatrix}$$

(31)

Thus, the transformations from the amplitudes $(C(t))$ given the original set of basis states to the amplitudes $(D(t))$ given the set of definite-energy basis states and back are:

$$\begin{pmatrix} D_P(t) \\ D_N(t) \end{pmatrix} = (\xi)^{-1} \cdot \begin{pmatrix} C_1(t) \\ C_2(t) \end{pmatrix} \;;\; \begin{pmatrix} C_1(t) \\ C_2(t) \end{pmatrix} = (\xi) \cdot \begin{pmatrix} D_P(t) \\ D_N(t) \end{pmatrix}$$

(32)

Eq. (32) holds for all times, including $t = 0$. Then, to initiate the TSS into the definite-energy $|\psi_{(\xi),P}(t)>$ state, set $D_P(0) = 1\;;\; D_N(0) = 0$ to find:

$$\begin{pmatrix} C_1(0) \\ C_2(0) \end{pmatrix}_P = \begin{pmatrix} \xi_{P1} \\ \xi_{P2} \end{pmatrix}$$

$$\begin{pmatrix} C_1(t) \\ C_2(t) \end{pmatrix}_P = \left( e^{-j \cdot \omega_0 \cdot t} \cdot (I) \right) \cdot (\xi) \cdot \left( I_\lambda(t) \right) \cdot (\xi)^{-1} \begin{pmatrix} \xi_{P1} \\ \xi_{P2} \end{pmatrix} = e^{-j \cdot (\omega_0 + \Omega_P) \cdot t} \cdot \begin{pmatrix} \xi_{P1} \\ \xi_{P2} \end{pmatrix}$$

$$|\psi_{(\xi),P}(t)> = e^{-j \cdot (\omega_0 + \Omega_P) \cdot t} \cdot |\xi_P> = C_1(t)_P \cdot |1> + C_2(t)_P \cdot |2>$$

(33)

Similarly, set $D_P(0) = 0\;;\; D_N(0) = 1$ in order to initiate the TSS into the $|\psi_{(\xi),N}(t)>$ definite-energy state:

$$\begin{pmatrix} C_1(0) \\ C_2(0) \end{pmatrix}_N = \begin{pmatrix} \xi_{N1} \\ \xi_{N2} \end{pmatrix}$$

$$\begin{pmatrix} C_1(t) \\ C_2(t) \end{pmatrix}_N = \left( e^{-j \cdot \omega_0 \cdot t} \cdot (I) \right) \cdot (\xi) \cdot \left( I_\lambda(t) \right) \cdot (\xi)^{-1} \begin{pmatrix} \xi_{N1} \\ \xi_{N2} \end{pmatrix} = e^{-j \cdot (\omega_0 - \Omega_P) \cdot t} \cdot \begin{pmatrix} \xi_{N1} \\ \xi_{N2} \end{pmatrix}$$

$$|\psi_{(\xi),N}(t)> = e^{-j \cdot (\omega_0 - \Omega_P) \cdot t} \cdot |\xi_N> = C_1(t)_N \cdot |1> + C_2(t)_N \cdot |2>$$

(34)

Once the TSS is initiated into one of the two definite-energy states (either $D_P(0) = 1\;;\; D_N(0) = 0$ or $D_P(0) = 0\;;\; D_N(0) = 1$), the state-occupation probability for that state is unity at all times (and of course zero state-occupation probability for the other definite-energy state). *In the canonical basis representation* the state-occupation probabilities are time-independent $\left( (|C_1(t)|^2, |C_2(t)|^2) = \right.$ either $(|\xi_{P1}|^2, |\xi_{P2}|^2)$ or $(|\xi_{N1}|^2, |\xi_{N2}|^2) \left. \right)$.

### 2.1.8. Energies by phase rotational speeds

The only system change in time for a TSS in a definite-energy state is given by an exponential phase-evolution factor with a constant phase rotational speed $(\omega_0 + \lambda_P = \omega_0 + \Omega_P)$ or $(\omega_0 + \lambda_N = \omega_0 - \Omega_P)$. The pair of definite-energy levels $(\mathcal{E}_P; \mathcal{E}_N)$ associated with the two stationary states ($\hbar = 1$) are:

$$\mathcal{E}_P \equiv \omega_P = \omega_0 + \Omega_P \;;\; \mathcal{E}_N \equiv \omega_N = \omega_0 - \Omega_P$$

(35)

If launched into the $P$-state, then the TSS will stay in the $P$-state with definite-energy $\mathcal{E}_P$. If launched into the $N$-state, then the TSS will stay in the $N$-state with definite-energy $\mathcal{E}_N$.

Namely:

$$\begin{pmatrix} C_1(0) \\ C_2(0) \end{pmatrix}_P = \begin{pmatrix} \xi_{P1} \\ \xi_{P2} \end{pmatrix} \Rightarrow \mathcal{P}_P(t) = 1 \; ; \; \mathcal{E}_P \equiv \omega_P = \omega_0 + \Omega_P$$

$$\begin{pmatrix} C_1(0) \\ C_2(0) \end{pmatrix}_N = \begin{pmatrix} \xi_{N1} \\ \xi_{N2} \end{pmatrix} \Rightarrow \mathcal{P}_N(t) = 1 \; ; \; \mathcal{E}_N \equiv \omega_N = \omega_0 - \Omega_P$$

(36)

Energy diagram for a general non-driven TSS (Hamiltonian of Eq. (10)) is shown by the (right part of) *Figure 6*.

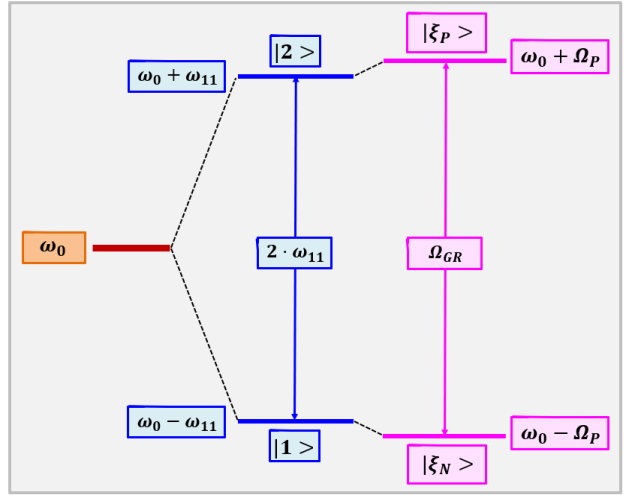

*Figure 6:* TIH-TSS energy diagrams. Blue: The unperturbed TSS (Hamiltonian of Eq. (10) with $\omega_D = 0$). Magenta: Energy diagram of a general non-driven TSS (Hamiltonian of Eq. (10)). The state $|\xi_P>$ is a "Symm" stationary state at a definite energy of $\omega_0 + \Omega_P$. The state $|\xi_N>$ is an "Asym" stationary state at a definite energy of $\omega_0 - \Omega_P$. The energy gap between the two definite energies in frequency units is the generalized Rabi flopping frequency ($\Omega_{GR} = 2 \cdot \Omega_P$).

For general initial conditions of the TSS, state-occupation probabilities ($|C_1(t)|^2$ and $|C_2(t)|^2$) *will* depend on time (cf. Eq. (22)).

In terms of the alternative ($\mathcal{A}, \mathcal{B}, \mathcal{C}, \mathcal{D}$) state description, let's consult Eq. (23), repeated here for convenience -

$$|\psi_{1,2}(t)> = (\mathcal{A} \cdot e^{-j \cdot \omega_N \cdot t} + \mathcal{B} \cdot e^{-j \cdot \omega_P \cdot t}) \cdot |1> \\ + (\mathcal{C} \cdot e^{-j \cdot \omega_N \cdot t} + \mathcal{D} \cdot e^{-j \cdot \omega_P \cdot t}) \cdot |2>$$

(37)

It is a matter of straightforward evaluation to verify that if $C_1(0) = \xi_{P1}$ and $C_2(0) = \xi_{P2}$ then *both* $\mathcal{A} = 0$ *and* $\mathcal{C} = 0$ whereas $\mathcal{B} = \xi_{P1}$ and $\mathcal{D} = \xi_{P2}$. Again the result is time evolution description by a *single* phase-evolution factor with a constant phase rotational speed ($\omega_P$) and we are back to the definite-energy state of Eq. (33).
Similarly - if $C_1(0) = \xi_{N1}$ and $C_2(0) = \xi_{N2}$ then *both* $\mathcal{B} = 0$ *and* $\mathcal{D} = 0$ whereas $\mathcal{A} = \xi_{N1}$ and $\mathcal{C} = \xi_{N2}$ and we are back to the definite-energy state of Eq. (34).
To conclude the last two definite-energy subsections – a TSS described by a TIH (typically time-independent off-diagonal perturbation elements) is characterized by two definite-energy states with energy levels given by Eq. (35).

Graphical illustrations in terms of phase rotational speeds are shown by the panels of *Figure 7*.

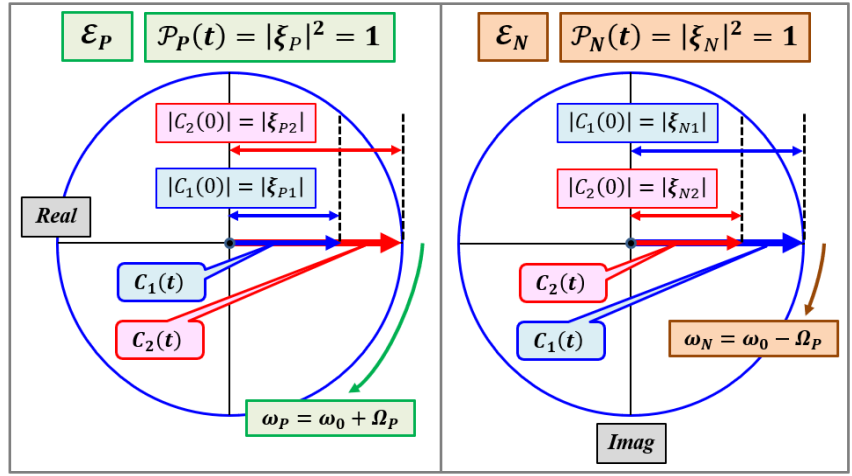

*Figure 7:* *The two energies of a time-independent-Hamiltonian TSS in terms of phase rotational speeds. Each of the panels shows (schematically) the two constant-magnitude probability amplitudes rotate in the complex plane "**together**" at the same rotational speed. Phase rotational speed is identified with system energy [15] through the Planck-Einstein relations.* $\{(|\xi_P|^2 \equiv |\xi_{P1}|^2 + |\xi_{P2}|^2)\,;\,(|\xi_N|^2 \equiv |\xi_{N1}|^2 + |\xi_{N2}|^2)\}$.

Another manifestation of the two definite-energy states of a TIH-TSS is offered by the panels of *Figure 8*.

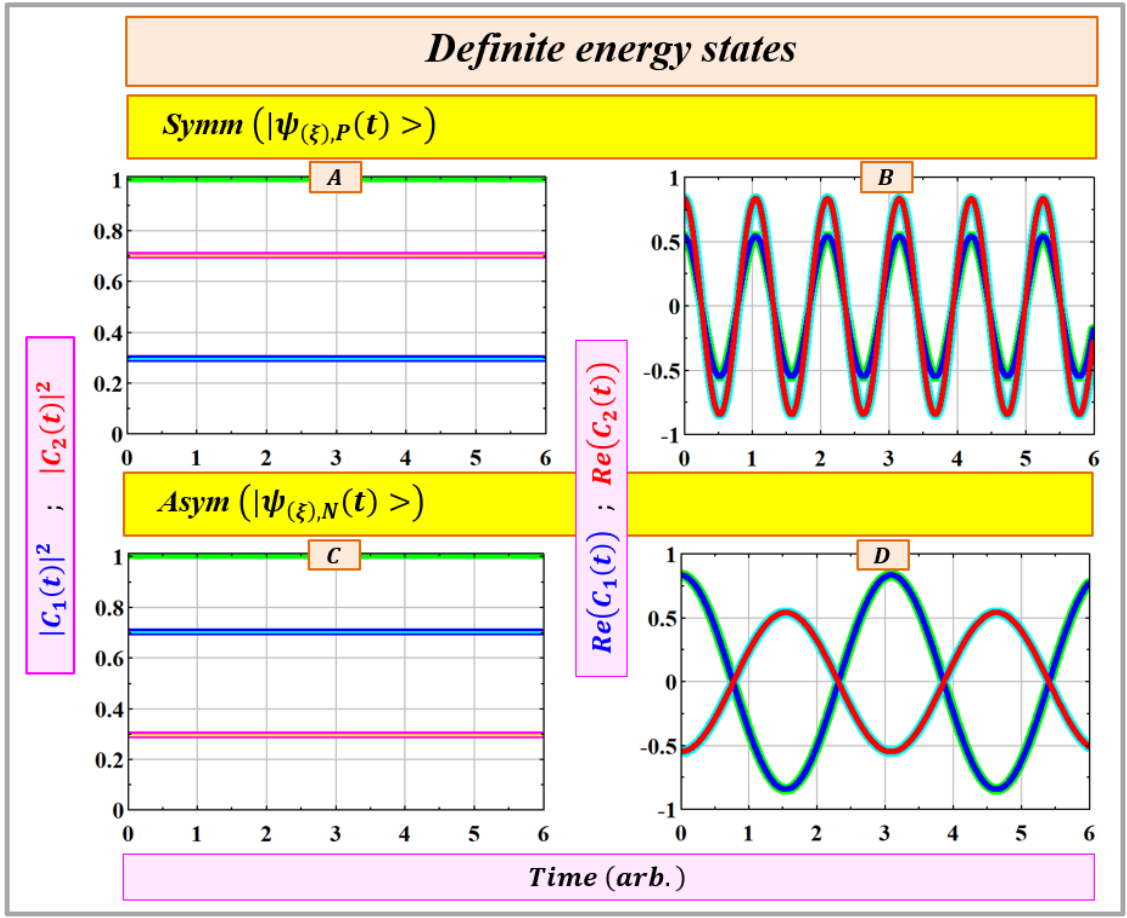

*Figure 8: Definite-energy states of a TIH-TSS. Left column (A,C): Stationary probabilities. A: System launched into the Symm $|\psi_{(\xi),P}(t)>$ eigenstate ($|C_1(0)| = |\xi_{P1}|, |C_2(0)| = |\xi_{P2}|$). C: System launched into the Asym $|\psi_{(\xi),N}(t)>$ eigenstate ($|C_1(0)| = |\xi_{N1}|, |C_2(0)| = |\xi_{N2}|$, cf. Eqs. **(33)** and **(34)**). Right column (B,D): Real part of the probability amplitudes showing two pairs of **identical** phase rotational speeds. B: Time evolution of the probability amplitudes $(C_1(t), C_2(t))$ for the TSS launched into the Symm stationary $|\psi_{(\xi),P}(t)>$ eigenstate. The Real part of the time evolution of the probability amplitudes shows identical phase rotational speeds $(\omega_0 + \Omega_P)$ for the two amplitudes. D: Same with the Asym stationary $|\psi_{(\xi),N}(t)>$ eigenstate (two identical phase rotational speeds of $(\omega_0 - \Omega_P)$).*

### 2.1.9. Average energy of a TIH-TSS

Most straight-forward calculation of a state's expected average energy ($\langle \mathcal{E} \rangle_{av}$) (many random-time measurements) is the weighted sum **[15],[91]** –

$$\langle \mathcal{E} \rangle_{av} = \sum_i \mathcal{P}_i \cdot \mathcal{E}_i$$

(38)

For the TIH-TSS considered here, in the eigenvectors basis representation (Eq. **(30)** and Eq. **(35)**), the TSS's expected average energy is the weighted sum –

$$\langle \mathcal{E} \rangle_{av} = |D_P(t)|^2 \cdot \omega_P + |D_N(t)|^2 \cdot \omega_N = |D_P(0)|^2 \cdot \omega_P + |D_N(0)|^2 \cdot \omega_N$$

(39)

As expected, given a non-driven fixed-parameters TSS, the expectation value of its energy is time-independent, determined only by the initial conditions.

Clearly, the TSS's measured energy values and the expected average energy values are between $\omega_N$ and $\omega_P$. The span of measured energy values is $\omega_P - \omega_N = \Omega_{GR}$ (cf. *Figure 9* below). We see here too (Eq. **(39)**) that if the TSS is in the $|\xi_P>/|\xi_N>$ state then its expected average energy ($\langle \mathcal{E} \rangle_{av}$) is $\omega_P/\omega_N$ (cf. *Figure 9* below).

Given a TIH - ($\mathcal{H}$), the average energy of a state $|\psi(t)>$ can also be written as **[80], [92]**:

$$\langle \mathcal{E} \rangle_{av} = <\psi(t)|(\mathcal{H})|\psi(t)>$$

(40)

Equation **(40)** ties expectation value of an operator (here Hamiltonian) to a system's wavefunction. For the TSS at hand, and in the canonical basis states,

$$\langle \mathcal{E} \rangle_{av} = \mathbf{C}(t)^\dagger \cdot (\mathcal{H}_{TSS,0}) \cdot \mathbf{C}(t) = \mathbf{C}(0)^\dagger \cdot (\mathcal{H}_{TSS,0}) \cdot \mathbf{C}(0)$$

(41)

For a pure state $|\psi(t)>$ and a TIH - ($\mathcal{H}$), the average energy of the state can also be calculated through a density matrix (or "operator") $(\rho(t))$ **[23]**:

$$\langle \mathcal{E} \rangle_{av} = <\psi(t)|(\mathcal{H})|\psi(t)> = tr(\rho(t) \cdot (\mathcal{H})) \quad ; \quad \rho(t) = |\psi(t)><\psi(t)|$$

(42)

For the TIH-TSS considered here, in the canonical basis representation, the explicit expression for the density matrix $(\rho_{12}(t))$ is -

$$\rho_{12}(t) = \mathbf{C}(t) \cdot \mathbf{C}(t)^\dagger = \begin{pmatrix} C_1(t) \\ C_2(t) \end{pmatrix} \cdot (C_1^*(t), C_2^*(t)) = \begin{pmatrix} |C_1(t)|^2 & C_1(t) \cdot C_2^*(t) \\ C_1^*(t) \cdot C_2(t) & |C_2(t)|^2 \end{pmatrix}$$
(43)

Multiplying by $(\mathcal{H}_{TSS,0})$, we can then write another explicit expression for the TSS's expected average energy:

$$\begin{aligned}\langle \mathcal{E} \rangle_{av} &= tr\left(\rho_{12}(t) \cdot (\mathcal{H}_{TSS,0})\right) = \\ &= |C_1(t)|^2 \cdot (\omega_0 - \omega_{11}) + |C_2(t)|^2 \cdot (\omega_0 + \omega_{11}) + 2 \\ &\quad \cdot Re\left[C_1(t) \cdot C_2^*(t) \cdot (\mathcal{H}_{TSS,0})[2,1]\right] \\ &= |C_1(0)|^2 \cdot (\omega_0 - \omega_{11}) + |C_2(0)|^2 \cdot (\omega_0 + \omega_{11}) + 2 \\ &\quad \cdot Re\left[C_1(0) \cdot C_2^*(0) \cdot |\Omega_D| \cdot e^{+\mathrm{j} \cdot \phi_D}\right]\end{aligned}$$
(44)

Density matrix $\rho_{(\xi)}(t)$ can be written in the eigenvectors basis representation. The equation for the TSS's expected average energy then reads:

$$\langle \mathcal{E} \rangle_{av} = tr\left(\rho_{(\xi)}(t) \cdot (\mathcal{H}_{TSS,0})\right) = tr\left(\rho_{(\xi)}(0) \cdot (\mathcal{H}_{TSS,0})\right)$$
(45)

Identical curves of TSS's expected average energy calculated by the four different expressions recorded above (Eqs. **(39), (41), (44), (45)**) are shown by the two panels of ***Figure 9***.

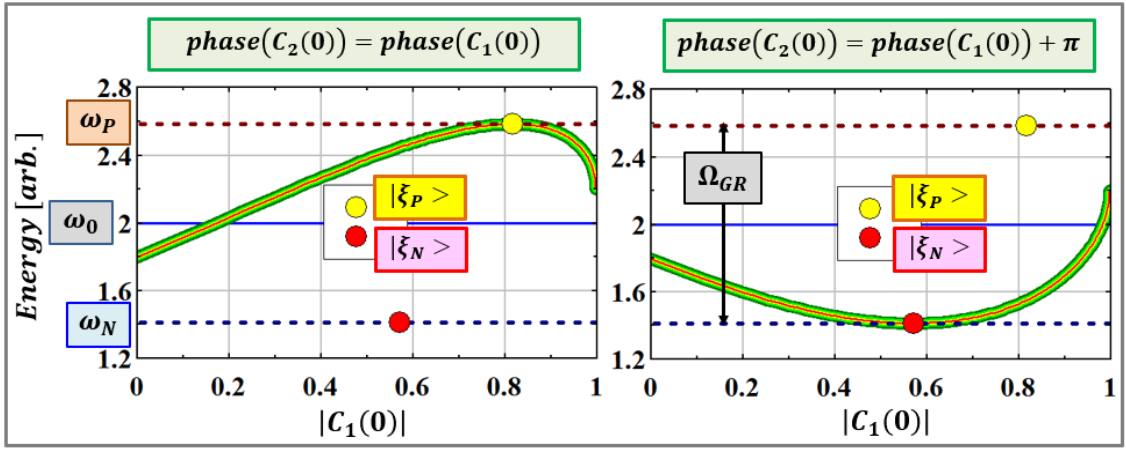

***Figure 9:*** *Expected average energy ($\langle \mathcal{E} \rangle_{av}$) of a TIH-TSS (Hamiltonian of Eq. **(10)**). Four identical curves are shown in each panel calculated according to four different recipes (Eqs. **(39), (41), (44), (45)**). TSS average energy varies with the TSS initial conditions ($|C_1(0)|$ and a slave $|C_2(0)|$, up to a phase-evolution factor). In general, one of two energy values can be recorded in a single energy measurement. If the TSS is launched into an eigenstate (cf. Eqs. **(33),(34)** ) only a single ("definite") energy value would be consistently recorded (one of the two red/yellow circles). If $phase(C_1(0)) = phase(C_2(0))$ then upon continues sweep of the $|C_1(0)|$ value, the TSS will "visit" the definite energy of eigenstate $|\xi_P\rangle$ (yellow circle – Symm eigenvector). If $phase(C_1(0)) = phase(C_2(0)) + \pi$ then upon continues sweep of the $|C_1(0)|$ value, the TSS will "visit" the definite energy of eigenstate $|\xi_N\rangle$ (red circle - Asym eigenvector). The span of possible energy measurement outcomes is the generalized Rabi flopping frequency ($\omega_P - \omega_N = \Omega_{GR}$, Planck-Einstein relations implied).*

## 2.1.10. Modulation depth

Modulation depth (degree of population-inversion completeness [55]) is defined as the maximum occupation probability of state $|2>$ when the system is prepared in state $|1>$ ($C_1(0) = 1$) [59]. The modulation depth ($m$) is computed against the diagonal element $\left((\mathcal{H}_{12,0})[1,1]\right)$ of the Hamiltonian matrix ($m(\omega_{11}) = max_t\{|C_2(\omega_{11};t)|^2\}$). Consulting Eqs. (15) for the definition of $\theta_R$, and consulting Eq. (25) for the state-occupation probabilities, the curve of the modulation depth $\left(m(\omega_{11})\right)$ turns out to be of a Lorentzian shape with half-width-half-max of $|\omega_D|$ [62]:

$$m(\omega_{11}) = \frac{|\omega_D|^2}{\Omega_P^2} = \frac{|\omega_D|^2}{\omega_{11}^2 + |\omega_D|^2}$$

(46)

Three $m(\omega_{11})$ curves for three different values of $|\omega_D|$ are plotted in *Figure 10*.

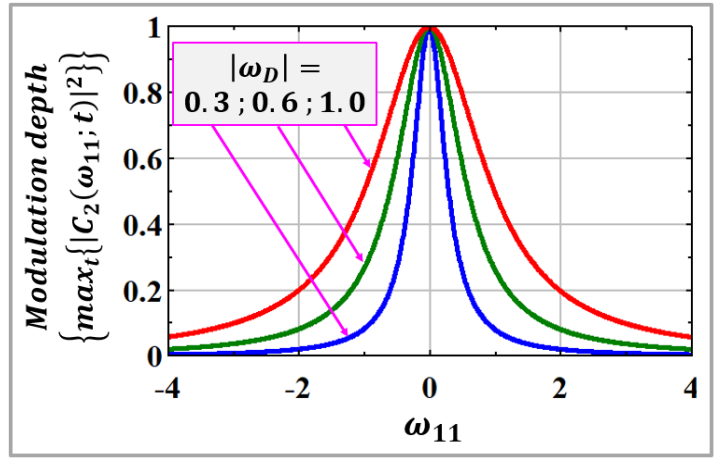

*Figure 10:* *Modulation depth of a TSS's state-occupation probability (Eq. (46)). The curves are of a Lorentzian shape. The width of the "line" increases with increasing absolute value of the off-diagonal element ($|\omega_D|$) of the TSS Hamiltonian $\left((\mathcal{H}_{12,0})[1,2]\right)$. In the computer-simulations section we discuss different-widths Lorentzian-line-shaped populations of the upper-level (above the TSS energy levels) vs. probe detuning frequency.*

At $\omega_{11} = 0$, the modulation depth is at a maximum of unity and the frequency of oscillations of the state-occupation probabilities is minimal (for a given $|\omega_D|$ value). As the value of $|\omega_{11}|$ increases, modulation depth decreases and oscillations' frequency increases.

We get back to Lorentzian lines in the computer simulations section to find, somewhat surprising at first encounter, third-level Lorentzian-line-shaped populations of different widths.

With the dynamics and some characteristics of a general abstract TSS at hand we proceed to examine several concrete systems – three quantum mechanical TSSs and one classical TSS (coupled optical waveguides).

## 2.2. A proton in a constant magnetic field

We start with the most basic TSS - a proton in a constant magnetic field in the "z" direction [$\boldsymbol{B} = (0,0,B)$]. To treat a concrete system we consider a spin one-half proton but the analysis holds for *any* spin one-half object. The Hamiltonian for the system is diagonal – the simplest TSS Hamiltonian possible, there is no coupling between the two levels, the eigenvalues are the "internal" energy diagonal elements **[15]**, and the eigenvectors coincide with the canonical basis.

### 2.2.1. Basis states, Hamiltonian, eigenvalues

The Hamiltonian is given in the basis state $|up>$ and $|dn>$:

$$|up> = \begin{pmatrix} 1 \\ 0 \end{pmatrix} \; ; \; |dn> = \begin{pmatrix} 0 \\ 1 \end{pmatrix}$$

(47)

The state vector $|\psi_{spin}(t)>$ is then

$$|\psi_{spin}(t)> = C_{up}(t) \cdot |up> + C_{dn}(t) \cdot |dn>$$

(48)

The Hamiltonian for a proton in a constant magnetic field is diagonal **[15]**

$$(\mathcal{H}_{spin05}) = \begin{pmatrix} -\omega_{pr} & 0 \\ 0 & \omega_{pr} \end{pmatrix}$$

(49)

The constant magnetic field takes-on a typical MRI value **[93],[6],[5]** -

$$B = 3 \, Tesla \, (MRI \, typical)$$

(50)

The angular frequency $(\omega_{pr})$ is the product of the proton's magnetic moment $(\mu_P)$ and the constant magnetic field (that actually *defines* the "z" axis **[15]**) -

$$\omega_{pr} \equiv \mu_P \cdot B \left(\frac{rad}{sec}\right) = 2.67 \cdot 10^8 \frac{radians}{second * Tesla} \cdot 3 \, Tesla \cong 8 \cdot 10^8 \frac{radians}{second}$$

(51)

The eigenvalues $(\lambda_P, \lambda_N)$ are (cf. Eq. **(27)** with $\omega_0 = 0$):

$$\lambda_P = \omega_{pr} \; ; \; \lambda_N = -\omega_{pr}$$

(52)

The energy range $(\lambda_P - \lambda_N = 2 \cdot \omega_{pr})$ translates to

$$2 \cdot \omega_{pr} \to 1.056 \, \mu eV$$

(53)

### 2.2.2. The $(\mathcal{A}, \mathcal{B}, \mathcal{C}, \mathcal{D})$ solution and definite-energy states

*Eigenvectors.* For the Hamiltonian **(49)** we find $\theta_R = 0 \Rightarrow s = 0, c = 1$ (cf. Eq. **(15)**) and the eigenvectors (Eq. **(16)**) are:

$$\xi_P = \begin{pmatrix} 0 \\ 1 \end{pmatrix} \; ; \; \xi_N = \begin{pmatrix} 1 \\ 0 \end{pmatrix}$$
(54)

*Definite-energy states*. To get the $|\psi_{(\xi),P}(t)>$ definite-energy state we set (Eq. (33)):

$$\begin{pmatrix} C_{up}(0) \\ C_{dn}(0) \end{pmatrix}_P = \begin{pmatrix} \xi_{P1} \\ \xi_{P2} \end{pmatrix} = \begin{pmatrix} 0 \\ 1 \end{pmatrix}$$
(55)

to get (Eq. (33)):

$$|\psi_{(\xi),P}(t)> = e^{-j\cdot(+\omega_{pr})\cdot t} \cdot |\xi_P> = e^{-j\cdot(+\omega_{pr})\cdot t} \cdot [0 \cdot |up> + 1 \cdot |dn>]$$
$$= e^{-j\cdot(+\omega_{pr})\cdot t} \cdot |dn>$$
(56)

Similarly, to get the $|\psi_{(\xi),N}(t)>$ state we set (Eq. (34)):

$$\begin{pmatrix} C_{up}(0) \\ C_{dn}(0) \end{pmatrix}_N = \begin{pmatrix} \xi_{N1} \\ \xi_{N2} \end{pmatrix} = \begin{pmatrix} 1 \\ 0 \end{pmatrix}$$
(57)

to get (Eq. (34)):

$$|\psi_{(\xi),N}(t)> = e^{-j\cdot(-\omega_{pr})\cdot t} \cdot |\xi_P> = e^{-j\cdot(-\omega_{pr})\cdot t} \cdot [1 \cdot |1> + 0 \cdot |2>]$$
$$= e^{-j\cdot(-\omega_{pr})\cdot t} \cdot |1>$$
(58)

*The $(\mathcal{A}, \mathcal{B}, \mathcal{C}, \mathcal{D})$ solution*.

For the Hamiltonian (49) we find $\theta_R = 0 \Rightarrow s = 0, c = 1$ (cf. Eq. (15)) and

$$\begin{aligned} \mathcal{A} &\equiv C_{up}(0) \\ \mathcal{B} &\equiv 0 \\ \mathcal{C} &\equiv 0 \\ \mathcal{D} &\equiv C_{dn}(0) \end{aligned}$$
(59)

For the spin one-half TSS $\omega_0 = 0$ and positive eigenvalue $\Lambda_P$ of $(\mathcal{H}_{spin05})$ is $\Lambda_P = \omega_{pr}$. According to the general solution (21):

$$\begin{pmatrix} C_{up}(t) \\ C_{dn}(t) \end{pmatrix} = \begin{pmatrix} C_{up}(0) & 0 \\ 0 & C_{dn}(0) \end{pmatrix} \begin{pmatrix} e^{-j\cdot(-\omega_{pr})\cdot t} \\ e^{-j\cdot(+\omega_{pr})\cdot t} \end{pmatrix}$$
(60)

After multiplication:

$$\begin{aligned} C_{up}(t) &= C_{up}(0) \cdot e^{-j\cdot(-\omega_{pr})\cdot t} \\ C_{dn}(t) &= C_{dn}(0) \cdot e^{-j\cdot(+\omega_{pr})\cdot t} \end{aligned}$$
(61)

And a general state function reads:

$$|\psi_{spin}(t)> = C_{up}(0) \cdot e^{-j\cdot(-\omega_{pr})\cdot t} \cdot |up> + C_{dn}(0) \cdot e^{-j\cdot(+\omega_{pr})\cdot t} \cdot |dn>$$
(62)

### 2.2.3. Energy levels

As follows from Eq. **(52)**, the energy levels $(\mathcal{E}_P, \mathcal{E}_N)$ (cf. Eq. **(24)**) of a spin one-half object in a constant magnetic field are -

$$\mathcal{E}_P \equiv \omega_{pr} \; ; \; \mathcal{E}_N \equiv -\omega_{pr}$$

(63)

### 2.2.4. Probability of the spin one-half particle to be in the $+x$ and or $-x$ state

The amplitude $A_+(t)$ to be in the $(+x)$ state at time $t$ is given by **[15]**:

$$A_+(t) = \frac{1}{\sqrt{2}} \cdot C_{up}(t) + \frac{1}{\sqrt{2}} \cdot C_{dn}(t)$$

(64)

Similarly for the amplitude $A_-(t)$ to be in the $(-x)$ state at time $t$:

$$A_-(t) = \frac{1}{\sqrt{2}} \cdot C_{up}(t) - \frac{1}{\sqrt{2}} \cdot C_{dn}(t)$$

(65)

The probability $\left(\mathcal{P}_{(\pm x)}(t)\right)$ of the spin one-half particle to be in the $(\pm x)$ state, is then given by:

$$\mathcal{P}_{(\pm x)}(t) = |A_\pm(t)|^2 = \frac{1}{2} \cdot [1 \pm 2 \cdot C_{up}(0) \cdot C_{dn}(0) \cdot \cos(2 \cdot \omega_{pr} \cdot t)]$$

(66)

Thus, if a spin one-half particle in a constant magnetic field in the $+z$ direction is initiated with its spin in the $+x$ direction $\left(C_{up}(0) = C_{dn}(0) = \frac{1}{\sqrt{2}}\right)$, then according to Eq. **(66)** the probability of the spin one-half particle to be in the $+x$ direction varies in time sinusoidally (with extreme values of 0 and 1) **[94]**.

The precession angular frequency $(2 \cdot \omega_{pr})$ is the *full* energy gap (in angular frequency units) generated by the interaction of the spin one-half particle (through its magnetic moment) with the applied constant magnetic field.

Note however that if $\left(C_{up}(0), C_{dn}(0)\right) = (1,0)$ or $\left(C_{up}(0), C_{dn}(0)\right) = (0,1)$ (i.e. the TSS is in one of its stationary states – Eq. **(54)**) then *precession ceases* and the probability of the spin one-half particle to be in the $+x$ direction or in the $-x$ direction is a constant one half $\left(\mathcal{P}_{(\pm x)}(t) = \frac{1}{2}\right)$.

Interestingly, precession frequency $(2 \cdot \omega_{pr})$ equals the generalized Rabi flopping frequency $(\Omega_{GR})$ since for Hamiltonian's zero-valued-off-diagonal elements $\Omega_{GR} = 2 \cdot \omega_{pr}$. It turns out that if interaction energy due to a *constant transverse* magnetic field (typically designated $B_x$) is added to the spin one-half TSS Hamiltonian as off-diagonal elements, then the "precession" frequency of the spin one-half particle (some periodic motion, not the familiar circular motion) is higher and is still, consistently, the generalized Rabi flopping frequency (cf. section **3.2.2**).

### 2.2.5. Graphical illustrations

The panels of *Figure 11* illustrate some characteristics of the "proton in a constant magnetic field" characteristics.

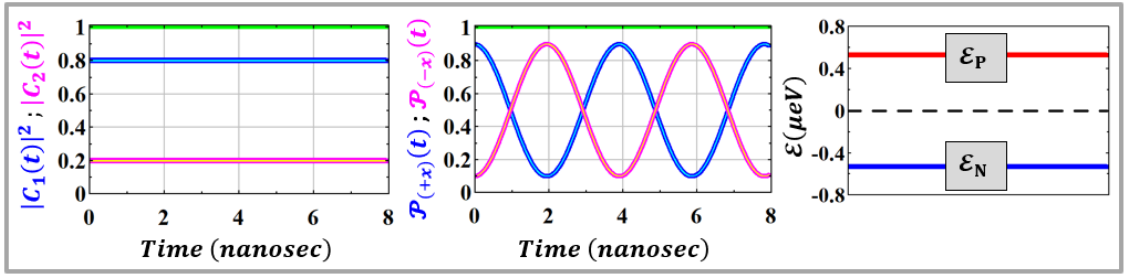

*Figure 11: A proton in a constant magnetic field – TSS characteristics. Left: Occupation probabilities of the basis states (values set arbitrarily). Each of the basis states ($|up>, |dn>$) is a state of definite energy. As such, its occupation probability is time-independent. Center: Probability $\left(\mathcal{P}_{(\pm x)}(t)\right)$ of the (spin one-half) proton to be in the $\pm x$ direction (Eq. (66)). Overall, the proton spin is precessing around the $+z$ axis at a frequency of $2 \cdot \omega_{pr}$ (Eq. (66)) [15]. In a constant magnetic field of 3 Tesla the proton's spin precession period is about 4 nanoseconds. Right: The two energies of the TSS under consideration. Energy gap is about 1 $\mu eV$.*

### 2.3. A free ammonia molecule

An ammonia molecule consists of three hydrogen atoms in a plane and a single nitrogen atom "above" (or "below") the plane. Consider a molecule rotating about an axis passing through the nitrogen, perpendicular to the hydrogen plane. The "above" and "below" configurations, apparently of equal energy, are taken as a set ($|1>, |2>$) of two canonical basis states. (According to [95], due to zero-point rotational motion the "above" and "below" configurations are not of exactly equal energy levels, to be ignored here). An energy barrier exists between the two states but there is some probability that the nitrogen atom will penetrate the energy barrier [95]. Thus, the two basis states are tunneling-coupled [96],[97]. The coupling of the two basis states is expressed as *negative* off-diagonal energy terms in the ammonia TSS Hamiltonian [15],[20]. The two eigenstates of the molecule's Hamiltonian are symmetric and antisymmetric *superpositions* of the basis states, characterized by different energy levels (that are equal to the corresponding eigenvalues – larger value for the antisymmetric state [98], as is the case for the modes of the first two energy levels of an harmonic oscillator [98]).

Analysis of the dynamics of an ammonia molecule appears in a number of already-published studies [14],[96],[94],[98],[97].

### 2.3.1. Basis states, Hamiltonian, eigenvalues

The basis taken for the ammonia molecule is the set $|1>$ and $|2>$ of the canonical basis states:

$$|1> = \begin{pmatrix} 1 \\ 0 \end{pmatrix} \; ; \; |2> = \begin{pmatrix} 0 \\ 1 \end{pmatrix}$$

(67)

The state vector of a free ammonia molecule $\left(|\psi_{free}(t)>\right)$ is then

$$|\psi_{free}(t)> = C_{1,free}(t) \cdot |1> + C_{2,free}(t) \cdot |2>$$

(68)

The Hamiltonian for the free ammonia molecule $\left((\mathcal{H}_{free})\right)$ is the sum –

$$(\mathcal{H}_{free}) = \omega_0 \cdot (I) + (H_{12,f})$$

$$(H_{12,f}) = \begin{pmatrix} 0 & \omega_b \cdot e^{-\jmath \cdot \phi_b} \\ \omega_b \cdot e^{+\jmath \cdot \phi_b} & 0 \end{pmatrix} \; ; \; 0 < \omega_b \in \mathbb{R}$$

$\omega_0 = 3 \cdot 10^{11}$ (set arbitrarily) ; $\omega_b = \frac{1}{2} \cdot (2 \cdot \pi) \cdot 23.786 \cdot 10^9 \left(\frac{rad}{sec}\right)$ ; $\phi_b = \pi$

(69)

The eigenvalues of $(\mathcal{H}_{free})$ are:

$$\Lambda_{P,free} = \omega_0 + \Omega_P$$
$$\Lambda_{N,free} = \omega_0 - \Omega_P$$
$$\Omega_P \equiv \lambda_P = \omega_b$$

(70)

{In **[15],[20]** $\hbar \cdot \omega_0 = E_0$ and $\hbar \cdot \omega_b = A$. The microwave frequency of $23.786 \cdot 10^9 \; Hz$ in Eq. **(69)** is taken from **[20],[97]**.}.

### 2.3.2. The $(\mathcal{A}, \mathcal{B}, \mathcal{C}, \mathcal{D})$ solution and definite-energy states

*Eigenvectors*. For $\omega_{11} = 0$ and $\phi_b = \pi$ we get $\theta_R = \frac{\pi}{4}$ (Eq. **(15)**) $\Rightarrow s = c = 1/\sqrt{2}$ and the eigenvectors (Eq. **(16)**) are:

$$\xi_P = \frac{1}{\sqrt{2}} \cdot \begin{pmatrix} -1 \\ 1 \end{pmatrix} \; ; \; \xi_N = \frac{1}{\sqrt{2}} \cdot \begin{pmatrix} 1 \\ 1 \end{pmatrix}$$

(71)

*Definite-energy states*. To get the $|\psi_{(\xi),P}(t)>$ definite-energy state we set (Eq. **(33)**):

$$\begin{pmatrix} C_1(0) \\ C_2(0) \end{pmatrix}_P = \begin{pmatrix} \xi_{P1} \\ \xi_{P2} \end{pmatrix} = \frac{1}{\sqrt{2}} \cdot \begin{pmatrix} -1 \\ 1 \end{pmatrix}$$

(72)

to get (Eq. **(33)**):

$$|\psi_{(\xi),P}(t)> = e^{-\jmath \cdot (\omega_0 + \omega_b) \cdot t} \cdot |\xi_P> = e^{-\jmath \cdot (\omega_0 + \omega_b) \cdot t} \cdot \left[\left(-\frac{1}{\sqrt{2}}\right) \cdot |1> + \frac{1}{\sqrt{2}} \cdot |2>\right]$$

(73)

Similarly, to get the $|\psi_{(\xi),N}(t)>$ state we set (Eq. **(34)**):

$$\begin{pmatrix} C_1(0) \\ C_2(0) \end{pmatrix}_N = \begin{pmatrix} \xi_{N1} \\ \xi_{N2} \end{pmatrix} = \frac{1}{\sqrt{2}} \cdot \begin{pmatrix} 1 \\ 1 \end{pmatrix}$$

(74)

to get (Eq. **(34)**)

$$|\psi_{(\xi),N}(t)> = e^{-j\cdot(\omega_0-\omega_b)\cdot t} \cdot |\xi_N> = e^{-j\cdot(\omega_0-\omega_b)\cdot t} \cdot \left[\frac{1}{\sqrt{2}}\cdot |1> + \frac{1}{\sqrt{2}}\cdot |2>\right]$$

(75)

Equations **(73)** and **(75)** indicate that the $|\psi_{(\xi),P}(t)>$ state is an antisymmetric ("$a$") superposition of the original $|1>; |2>$ states whereas $|\psi_{(\xi),N}(t)>$ state is a symmetric ("$s$") superposition of the original $|1>; |2>$ states. Designating $|\psi_{(\xi),P}(t)> \equiv |\phi_a(t)>$ and $|\psi_{(\xi),N}(t)> \equiv |\phi_s(t)>$ we rewrite the eigenstates as:

$$|\phi_a(t)> = e^{-j\cdot(\omega_0+\omega_b)\cdot t} \cdot |\xi_P> \; ; \; \{D_a(t) = 1\cdot e^{-j\cdot(\omega_0+\omega_b)\cdot t} \; ; \; D_s(t) = 0\}$$
$$|\phi_s(t)> = e^{-j\cdot(\omega_0-\omega_b)\cdot t} \cdot |\xi_N> \; ; \; \{D_a(t) = 0 \; ; \; D_s(t) = 1\cdot e^{-j\cdot(\omega_0-\omega_b)\cdot t}\}$$

(76)

*The $(\mathcal{A}, \mathcal{B}, \mathcal{C}, \mathcal{D})$ solution.*

The general solution for the probability amplitudes of the free ammonia molecule is (Eq. **(22)**):

$$C_{1,free}(t) = [\mathcal{A}\cdot e^{+j\cdot\omega_b\cdot t} + \mathcal{B}\cdot e^{-j\cdot\omega_b\cdot t}]\cdot e^{-j\cdot\omega_0\cdot t}$$
$$C_{2,free}(t) = [\mathcal{C}\cdot e^{+j\cdot\omega_b\cdot t} + \mathcal{D}\cdot e^{-j\cdot\omega_b\cdot t}]\cdot e^{-j\cdot\omega_0\cdot t}$$

(77)

With $\phi_D = \pi$ ; $c = s = \frac{1}{\sqrt{2}}$, the $(\mathcal{A}, \mathcal{B}, \mathcal{C}, \mathcal{D})$ coefficients are (Eq. **(21)**):

$$\mathcal{A} \equiv \frac{1}{2}\cdot[C_{1,free}(0) + C_{2,free}(0)] \; ; \; \mathcal{B} \equiv \frac{1}{2}\cdot[C_{1,free}(0) - C_{2,free}(0)]$$
$$\mathcal{C} \equiv \frac{1}{2}\cdot[C_{1,free}(0) + C_{2,free}(0)] \; ; \; \mathcal{D} \equiv \frac{1}{2}\cdot[-C_{1,free}(0) + C_{2,free}(0)]$$

(78)

For the $|\psi_{(\xi),P}(t)>$ state $\left(C_{1,free}(0) = -\frac{1}{\sqrt{2}} \; ; \; C_{2,free}(0) = \frac{1}{\sqrt{2}}\right)_P$:

$$C_{1,free}(t)_P = \left[0\cdot e^{+j\cdot\omega_b\cdot t} - \frac{1}{\sqrt{2}}\cdot e^{-j\cdot\omega_b\cdot t}\right]\cdot e^{-j\cdot\omega_0\cdot t}$$
$$C_{2,free}(t)_P = \left[0\cdot e^{+j\cdot\omega_b\cdot t} + \frac{1}{\sqrt{2}}\cdot e^{-j\cdot\omega_b\cdot t}\right]\cdot e^{-j\cdot\omega_0\cdot t}$$

(79)

The $|\phi_a(t)>$ definite-energy state turns out to be the antisymmetric superposition (of the $|1>, |2>$ basis states) with the *higher-level* energy value:

$$|\phi_a(t)> \equiv |\psi_{(\xi),P}(t)> = \left(C_{1,free}(t)\right)_P\cdot |1> + \left(C_{2,free}(t)\right)_P\cdot |2>$$
$$= \left[-\frac{1}{\sqrt{2}}\cdot |1> + \frac{1}{\sqrt{2}}\cdot |2>\right]\cdot e^{-j\cdot(\omega_0+\omega_b)\cdot t}$$

(80)

For the $|\psi_{(\xi),N}(t)>$ state $\left(C_{1,free}(0) = \frac{1}{\sqrt{2}}\,;\, C_{2,free}(0) = \frac{1}{\sqrt{2}}\right)_N$:

$$C_{1,free}(t)_N = \left[\frac{1}{\sqrt{2}} \cdot e^{+j\cdot\omega_b\cdot t} + 0 \cdot e^{-j\cdot\omega_b\cdot t}\right] \cdot e^{-j\cdot\omega_0\cdot t}$$

$$C_{2,free}(t)_N = \left[\frac{1}{\sqrt{2}} \cdot e^{+j\cdot\omega_b\cdot t} + 0 \cdot e^{-j\cdot\omega_b\cdot t}\right] \cdot e^{-j\cdot\omega_0\cdot t}$$

(81)

The $|\phi_N(t)>$ definite-energy state then turns out to be the symmetric superposition (of the $|1>,|2>$ basis states) with the *lower-level* energy value:

$$|\phi_s(t)> \equiv |\psi_{(\xi),N}(t)> = \left(C_{1,free}(t)\right)_N \cdot |1> + \left(C_{2,free}(t)\right)_N \cdot |2>$$
$$= \left[\frac{1}{\sqrt{2}} \cdot |1> + \frac{1}{\sqrt{2}} \cdot |2>\right] \cdot e^{-j\cdot(\omega_0 - \omega_b)\cdot t}$$

(82)

Equations ((80) and (82)) are in agreement with equations ((73) and (75)) respectively.

### 2.3.3. Energy levels

From (76) we get for the two energy levels $(E_a, E_s)$:

$$\mathcal{E}_a \equiv \mathcal{E}_P = \omega_0 + \omega_b\,;\quad \mathcal{E}_s \equiv \mathcal{E}_N = \omega_0 - \omega_b$$

(83)

For the free ammonia, the strictly quantum-energy doublet formation, is the result of the molecule's flip-flop effect [15].

### 2.3.4. Graphical illustrations

The two panels of *Figure 12* and the six panels of *Figure 13* illustrate some of the characteristics of a free ammonia molecule, with focus on its energy doublet.

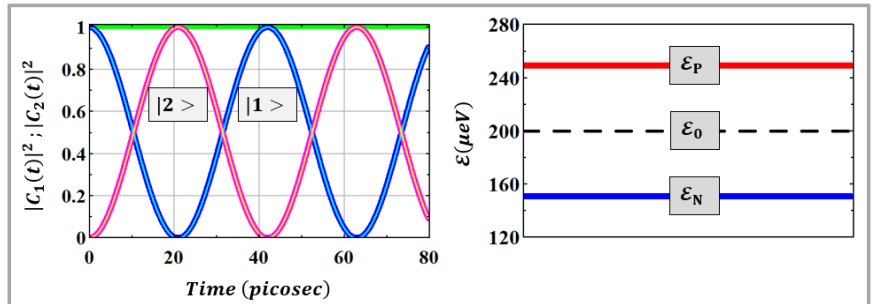

*Figure 12: Oscillations of the state-occupation probabilities (left) and the energy doublet of the symmetric/antisymmetric superposition eigenstates of a free ammonia molecule (right). An ammonia molecule at the upper-energy antisymmetric-superposition eigenstate (Eq. (80) and the $\mathcal{E}_P$ red line on the right panel) will spontaneously decay into the symmetric lower-energy superposition eigenstate, emitting microwave radiation with a period of about 40 picoseconds (see the oscillations period on the left panel). As is well known, the global phase rotational speed $(\omega_0)$ has no effect on state-occupation probabilities (cf. Eqs. (25),(26)). The corresponding energy-level splitting is about $100\ \mu eV$ (the shown value of $\mathcal{E}_0 = \hbar \cdot \omega_0 = 200\ \mu eV$ is arbitrary).*

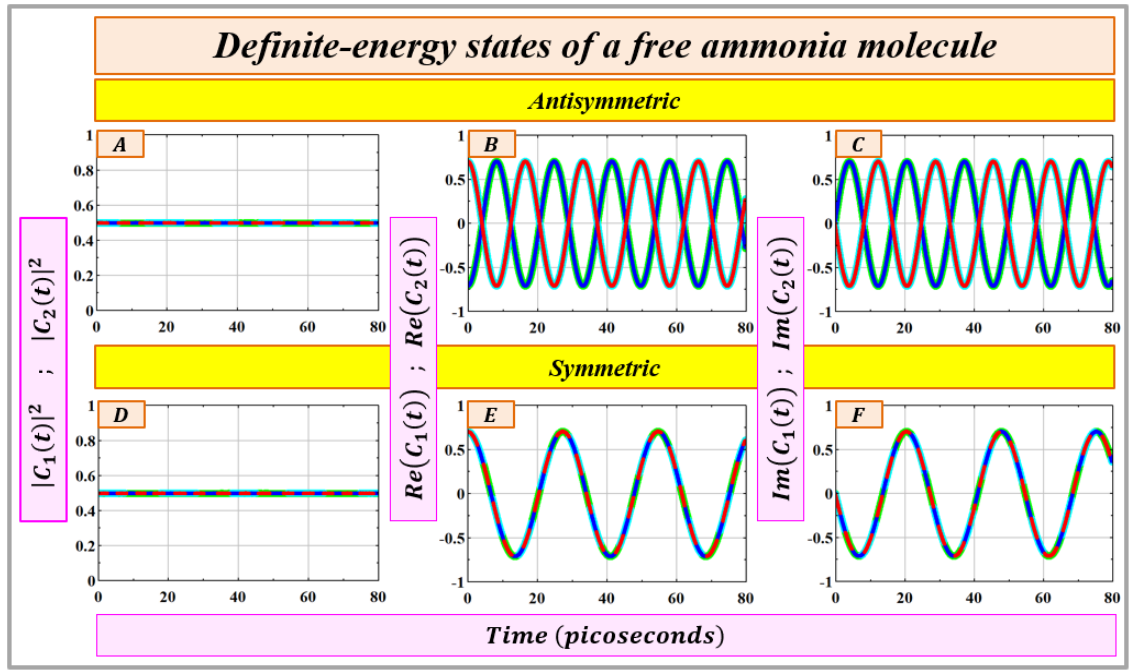

*Figure 13:* *State occupation probabilities and probability amplitudes of a free ammonia molecule in its definite-energy (stationary) states. Top row: The antisymmetric definite-energy state $\left(C_1(t) = -\frac{1}{\sqrt{2}} \cdot e^{-j \cdot (\omega_0 + \omega_b) \cdot t}; C_2(t) = +\frac{1}{\sqrt{2}} \cdot e^{-j \cdot (\omega_0 + \omega_b) \cdot t}\right)$. Bottom row: The symmetric definite-energy state $\left(C_1(t) = +\frac{1}{\sqrt{2}} \cdot e^{-j \cdot (\omega_0 - \omega_b) \cdot t}; C_2(t) = +\frac{1}{\sqrt{2}} \cdot e^{-j \cdot (\omega_0 - \omega_b) \cdot t}\right)$. Left column (A and D): State-occupation probabilities. For both definite-energy states, the occupation-**probabilities** of the canonical basis states $(|C_1(t)|^2, |C_2(t)|^2)$ are 0.5. States' energy information in the form of phase rotational speeds is lost. Center column (B and E): Real part of the probability **amplitudes** - $C_1(t), C_2(t)$ (in each panel). Comparing the amplitude's oscillation frequencies (B vs. E), the difference in phase rotational speeds and thus in energy levels becomes convincingly clear. Similarly for the right column (C and F) showing the imaginary part of the probability **amplitudes**.*

## 2.4. An ammonia molecule in a constant electric field

Another quantum mechanical TIH-TSS is an ammonia molecule in a constant electric field. The electric field ($\boldsymbol{E}$) is assumed to be perpendicular to the plane of the hydrogens. Due to the permanent dipole moment ($\boldsymbol{\mu_E}$) of the ammonia molecule, the external field will *raise* the energy of, say, the |1> basis state by $+\mu_E \cdot E$ and will *reduce* the energy of the |2> basis state by $-\mu_E \cdot E$. Namely – in the presence of the constant electric field, the "internal" energies of the two states |1> and |2> will shift from $\omega_0$ to $\omega_0 \pm \omega_E$ with $\omega_E \equiv \mu_E \cdot E/\hbar$. Compared to the energy gap of the free ammonia molecule then, the energy gap of the molecule in the constant electric field will *increase* (extremely slightly for a typical amplitude of the electric field [20]).

### 2.4.1. Basis states, Hamiltonian, eigenvalues

The basis states for the ammonia molecule under constant electric field are taken again to be the canonical $|1>$ and $|2>$:

$$|1> = \begin{pmatrix} 1 \\ 0 \end{pmatrix} \; ; \; |2> = \begin{pmatrix} 0 \\ 1 \end{pmatrix}$$

(84)

The state vector $|\psi_E(t)>$ is then

$$|\psi_E(t)> = C_{1,E}(t) \cdot |1> + C_{2,E}(t) \cdot |2>$$

(85)

The interaction of the field with the permanent electric dipole moment of the molecule effects the energy of the two basis states and the Hamiltonian for the ammonia molecule in a static electric field thus becomes [15],[18]:

$$(\mathcal{H}_E) = \omega_0 \cdot (I) + (H_{12,E})$$

$$(H_{12,E}) = \begin{pmatrix} \omega_E & \omega_D \cdot e^{-j \cdot \phi_D} \\ \omega_D \cdot e^{+j \cdot \phi_D} & -\omega_E \end{pmatrix}$$

$$\omega_E = \mu_E \cdot E_0 = 2.82 \cdot 10^3 \left(\frac{rad}{sec}\right)$$

$$\omega_0 = 3 \cdot 10^{11} \text{ (arbitrary)} \; ; \; \omega_D = \frac{1}{2} \cdot (2 \cdot \pi) \cdot 23.786 \cdot 10^9 \left(\frac{rad}{sec}\right) \; ; \; \phi_D = \pi$$

(86)

Note that the interaction of the molecule with the electric field changes the energies on the *diagonal* elements of the $(H_{12,E})$ Hamiltonian. And we have assumed [15] that the off-diagonal element for the tunneling-coupling energy is unaffected. The parameter $\mu_E = 4.9098 \cdot 10^{-30} \; coulomb \cdot meter$ is the permanent electric dipole moment of the ammonia molecule [99]. The value of $E_0 = 2.36 \cdot 10^{-2} \frac{Volt}{meter}$ was estimated by [20] for the maximum amplitude of the electric field inside an ammonia maser cavity. Here we take the estimated amplitude value as the amplitude of the *constant* field for the discussed example.

*Eigenvalues*. The eigenvalues of the $(H_E)$ Hamiltonian matrix are -

$$\Lambda_{P,E} = \omega_0 + \Omega_{P,E}$$
$$\Lambda_{N,E} = \omega_0 - \Omega_{P,E}$$
$$\Omega_{P,E} \equiv \sqrt{\omega_E^2 + \omega_D^2}$$

(87)

{In [15],[20] $\hbar \cdot \omega_E = \mu_E \cdot E$ }.

### 2.4.2. The $(\mathcal{A}, \mathcal{B}, \mathcal{C}, \mathcal{D})$ solution and definite-energy states

*Eigenvectors*. Given $\phi_D = \pi$, the eigenvectors (Eq. (16)) are:

$$\xi_P = \begin{pmatrix} \sin \theta_R \\ -\cos \theta_R \end{pmatrix} \; ; \; \xi_N = \begin{pmatrix} \cos \theta_R \\ \sin \theta_R \end{pmatrix}$$

(88)

*Definite-energy states*. To get the $|\psi_{(\xi),P}(t)>$ definite-energy state we set (Eq. **(33)**):

$$\begin{pmatrix} C_1(0) \\ C_2(0) \end{pmatrix}_P = \begin{pmatrix} \xi_{P1} \\ \xi_{P2} \end{pmatrix} = \begin{pmatrix} \sin\theta_R \\ -\cos\theta_R \end{pmatrix}$$

(89)

to get (Eq. **(33)**):

$$|\psi_{(\xi),P}(t)> = e^{-\jmath\cdot(\omega_0+\Omega_{P,E})\cdot t}\cdot|\xi_P> = e^{-\jmath\cdot(\omega_0+\Omega_{P,E})\cdot t}\cdot[\sin\theta_R\cdot|1> -\cos\theta_R\cdot|2>]$$

(90)

Similarly, to get the $|\psi_{(\xi),N}(t)>$ state we set (Eq. **(34)**):

$$\begin{pmatrix} C_1(0) \\ C_2(0) \end{pmatrix}_N = \begin{pmatrix} \xi_{N1} \\ \xi_{N2} \end{pmatrix} = \begin{pmatrix} \cos\theta_R \\ \sin\theta_R \end{pmatrix}$$

(91)

to get (Eq. **(34)**):

$$|\psi_{(\xi),N}(t)> = e^{-\jmath\cdot(\omega_0-\Omega_{P,E})\cdot t}\cdot|\xi_N> = e^{-\jmath\cdot(\omega_0-\Omega_{P,E})\cdot t}\cdot[\cos\theta_R\cdot|1> +\sin\theta_R\cdot|2>]$$

(92)

Equations **(90)** and **(92)** indicate that the $|\psi_{(\xi),P}(t)>$ state is an antisymmetric ("$a$") superposition of the original $|1>;|2>$ states whereas $|\psi_{(\xi),N}(t)>$ state is a symmetric ("$s$") superposition of the original $|1>;|2>$ states. Designating $|\psi_{(\xi),P}(t)>\equiv|\phi_{a,E}(t)>$ and $|\psi_{(\xi),N}(t)>\equiv|\phi_{s,E}(t)>$ we rewrite the eigenstates as:

$$|\phi_{a,E}(t)> = e^{-\jmath\cdot(\omega_0+\Omega_{P,E})\cdot t}\cdot|\xi_P> \;;\; \{D_{a,E}(t) = 1\cdot e^{-\jmath\cdot(\omega_0+\Omega_{P,E})\cdot t}\;;\; D_{s,E}(t) = 0\}$$
$$|\phi_{s,E}(t)> = e^{-\jmath\cdot(\omega_0-\Omega_{P,E})\cdot t}\cdot|\xi_N> \;;\; \{D_{a,E}(t) = 0\;;\; D_{s,E}(t) = 1\cdot e^{-\jmath\cdot(\omega_0-\Omega_{P,E})\cdot t}\}$$

(93)

*The $(\mathcal{A},\mathcal{B},\mathcal{C},\mathcal{D})$ solution.*

The general solution for the probability amplitudes of the ammonia molecule under a constant electric field is (Eq. **(22)**):

$$C_{1,E}(t) = [\mathcal{A}\cdot e^{+\jmath\cdot\Omega_{P,E}\cdot t} + \mathcal{B}\cdot e^{-\jmath\cdot\Omega_{P,E}\cdot t}]\cdot e^{-\jmath\cdot\omega_0\cdot t}$$
$$C_{2,E}(t) = [\mathcal{C}\cdot e^{+\jmath\cdot\Omega_{P,E}\cdot t} + \mathcal{D}\cdot e^{-\jmath\cdot\Omega_{P,E}\cdot t}]\cdot e^{-\jmath\cdot\omega_0\cdot t}$$

(94)

With $\phi_D = \pi$ the $(\mathcal{A},\mathcal{B},\mathcal{C},\mathcal{D})$ coefficients are (Eq. **(21)**):

$$\mathcal{A} \equiv [c^2\cdot C_{1,E}(0) + c\cdot s\cdot C_{2,E}(0)]\;;\; \mathcal{B} \equiv [s^2\cdot C_{1,E}(0) - c\cdot s\cdot C_{2,E}(0)]$$
$$\mathcal{C} \equiv [c\cdot s\cdot C_{1,E}(0) + s^2\cdot C_{2,E}(0)]\;;\; \mathcal{D} \equiv [-c\cdot s\cdot C_{1,E}(0) + c^2\cdot C_{2,E}(0)]$$

(95)

### 2.4.3. Energy levels

From **(93)** we get for the two energy levels $(E_{a,E}, E_{s,E})$:

$$\mathcal{E}_{a,E} \equiv \mathcal{E}_{P,E} = \omega_0 + \Omega_{P,E} \;;\; \mathcal{E}_{s,E} \equiv \mathcal{E}_{N,E} = \omega_0 - \Omega_{P,E}$$

(96)

For typical constant electric fields the change in the energy levels (vs. the levels of a free ammonia molecule) is negligibly small. Changes in the emission spectrum of ammonia molecules under *uniquely strong* electric fields (up to $10^8$ V/m ) were reported in **[100]**.

### 2.4.4. Graphical illustrations

Diagrams for energy levels and probability oscillations of an ammonia molecule are shown by the panels of *Figure 14* and of *Figure 15*.

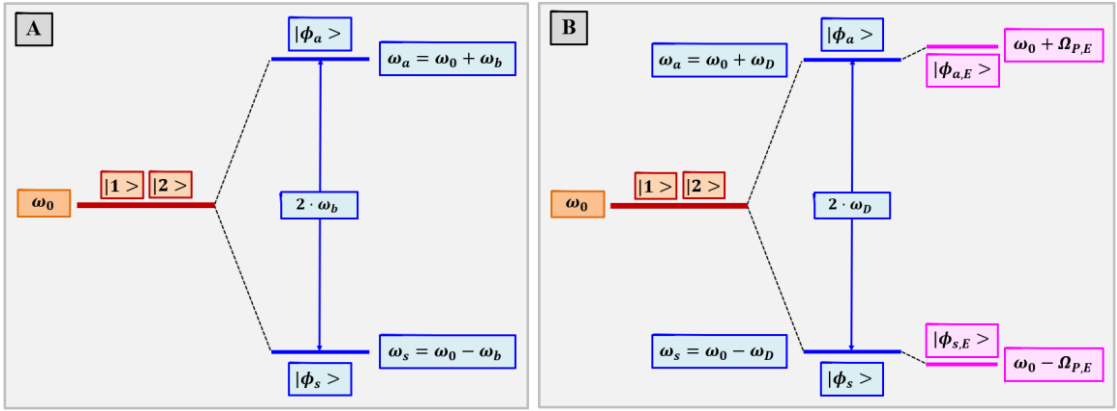

*Figure 14:* *Energy levels diagrams for an ammonia molecule. A: Free molecule. An energy doublet is formed due to the molecule's flip-flop effect. The symmetric superposition eigenstate of a free molecule Hamiltonian (bottom blue) is of a lower energy level (vs. the energy level of the antisymmetric superposition eigenstate (top blue) – Eq.* **(83)***). The energy gap is nearly* $100\ \mu eV$. *B: Ammonia molecule in a constant electric field along the flip-flop axis. In the presence of the field the state-superposition energy levels are (very slightly) pushed further away from each other – Eq.* **(96)**. *(In the shown diagram the extra level-spread is highly exaggerated).*

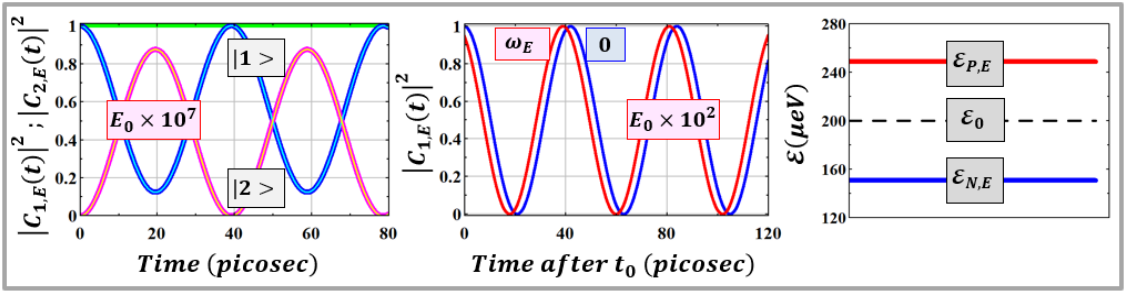

*Figure 15:* *Ammonia molecule in a constant electric field. Left: State-occupation probabilities. In a typical constant electric field the modulation depth (degree of energy exchange between the basis states) is very nearly unity* $(m(\omega_E) \leq 1)$. *To show the effect (presence of a constant electric field), the curves in the panel were calculated for a field* $\times\ 10^7$ *stronger than a typical ammonia-maser field* **[20]**. *Center: Occupation probability of the* $|1>$ *basis state. Without the field (blue) and with the field (red). In the presence of the field the period of probability oscillations is negligibly shorter. The shown red curve was calculated with a field* $10^2$ *stronger than a typical field and the time segment shown in the panel is after* $10^{10}$ *oscillation periods. Right: Energy levels of an ammonia molecule in a constant, typical-amplitude electric field. The extra level-spread due to the interaction with the constant field is invisible (compare with the right panel of* **Figure 12**).

## 2.5. Coupled two optical waveguides

A system of coupled two optical waveguides constitutes the classical analog of a quantum-mechanical TIH-TSS. Consider two "left" and "right" identical single-mode optical waveguides in close proximity. The "wavefunction" is a linear superposition of the guided optical modes with propagation-distance-dependent superposition amplitudes. Once the distance-dependence of the superposition amplitudes is known, the distance-evolution of the wavefunction is determined.

Coupling between the proximity-placed waveguides goes as follows. As a coherent electromagnetic field is coupled to – say – the left waveguide, the evanescent part of the guided mode, reaching the right waveguide, excites a wave with amplitude *trailing by ninety degrees*. In turn, the evanescent part of the growing mode amplitude on the right excites a wave trailing by ninety degrees i.e. by *one-hundred and eighty degrees* relative to the already-propagating mode on the left. The net result is a growing amplitude on the right and, through destructive interference, diminishing amplitude on the left. If the waveguides are identical, then the entire launched electromagnetic power is fully transferred to the right waveguide. At that point (in space) the next half cycle of power transfer (right to left with inverted phase) starts. The classical optical evanescent coupling effect very much resembles the quantum-mechanical tunneling effect.

The distance-evolution equation for the two superposition amplitudes, put forward by the coupled-mode theory **[101]**, is mathematically identical to the TIH-TSS Schrödinger equation.

For the coupled waveguides' TSS, the energy-equivalent phase rotational speeds are given by the wavevectors (typically designated "$\beta$") of the propagating optical modes (in units of $distance^{-1}$).

### 2.5.1. Electromagnetic field and Hamiltonian-equivalent

The electromagnetic field for the two-waveguide system, assumed to be $x$ polarized $\left(E_x(x,y,z)\right)$, is a linear superposition of the two individual guided modes with $z$ dependent coefficients (**[102]** Eq. 27-1,(**[103]** Eq. 6):

$$E_x(x,y,z) = A_L(z) \cdot U_L(x,y) + A_R(z) \cdot U_R(x,y)$$
$$A_L(z) \equiv E_L(z) \cdot e^{j \cdot \beta_L \cdot z} \; ; \quad A_R(z) \equiv E_R(z) \cdot e^{j \cdot \beta_R \cdot z}$$

**(97)**

where $U_L(x,y) \; ; U_R(x,y)$ are the (real functions) solutions of the scalar wave equation in the two transverse coordinates (**[102]** Eq. 13-8) for the *LEFT* and for the *RIGHT* waveguides, and are assumed to be normalized:

$$\langle U_i(x,y), U_i(x,y) \rangle = 1 \; ; \quad i = L, R \; .$$

**(98)**

The set of two functions in the transverse coordinates $\left(U_L(x,y) \; ; U_R(x,y)\right)$ are the classical waveguide-system analog to the set of base states ($|1>\;;\;|2>$) in the quantum mechanical treatment of two state systems.

The powers carried by each of the field terms $[P_L(z) \; ; \; P_R(z)]$ are

$$P_L(z) = P_0 \cdot A_L(z) \cdot A_L^*(z) \quad ; \quad P_R(z) = P_0 \cdot A_R(z) \cdot A_R^*(z) \ .$$
(99)

The total power-flow across the two WGs is thus ([102] Eq. 27-5)

$$P_{all}(z) = P_R(z) + P_L(z) \ .$$
(100)

Equation **(100)** for the total power-flow holds under the approximation of weakly guiding waveguides with the hidden assumption that $\langle U_L(x,y), U_R(x,y) \rangle \ll 1$.

For a lossless (or gain-less) waveguide system, assuming Eq. **(100)** holds, power conservation requires (at any distance $z$) -

$$\frac{dP_{all}(z)}{dz} = 0 \ .$$
(101)

According to the coupled-mode theory, the set of two coupled equations for the coefficients $\big(A_L(z) \ ; \ A_R(z)\big)$ reads ([102] Eq. 29-4a,b),([104] Eqs. 7.59,7.60),([105] Eqs. 3.2a,b):

$$\frac{d}{dz}\begin{pmatrix} A_L(z) \\ A_R(z) \end{pmatrix} = \mathbb{j} \cdot (\mathcal{H}_{WG}) \begin{pmatrix} A_L(z) \\ A_R(z) \end{pmatrix}$$
(102)

with the Hamiltonian-equivalent $(\mathcal{H}_{WG})$:

$$(\mathcal{H}_{WG}) = \beta_{avg} \cdot (I) + \begin{pmatrix} -\delta\beta & K_{eff} \\ K_{eff} & +\delta\beta \end{pmatrix}$$

$$\beta_{avg} \equiv \frac{\beta_L + \beta_R}{2} \quad ; \quad \delta\beta \equiv \frac{\beta_R - \beta_L}{2} \geq 0 \quad ; \quad 0 < K_{eff} \in \mathbb{R}$$
(103)

The parameter $\beta_{avg}$ is the average propagation vector for the set of two guided modes. The "effective" off-diagonal *cross-coupling* element $(K_{eff})$ is a real number.

Notes:

- The Hamiltonian-equivalent $(\mathcal{H}_{WG})$ matrix is Hermitian ($\mathbb{j} \cdot (\mathcal{H}_{WG})$ is anti-Hermitian or *skew-Hermitian*) and thus energy conservation is assured. In general the coupling constants $L$ to $R$ and $R$ to $L$ are *not* equal ($K_{L \to R} \neq K_{R \to L}$). The value for each of the two cross-coupling constants can be determined analytically ([101], Eq. 13.8-4),([102] Eq. 29-5),([105] Eq. 2.2),([106] Eqs. 43,44,45). The off-diagonal coupling constants in $(\mathcal{H}_{WG})$ of Eq. **(103)** *are taken to be equal* either under the assumption of weakly guiding and *not too dissimilar* waveguides ([102] Ch. 29) or by taking their average values ([105] Eq. 3.44).

- Yariv considered only the slowly varying part of the amplitudes $\big(E_L(z), E_R(z)\big)$ to get $(\mathcal{H}_{WG})$ of Eq. **(103)** without the $\beta_{avg} \cdot (I)$ component ([101] Eq. 13.8-4).

- Here, as in ([104] Eqs. 7.59,7.60), we have left out a small *self-coupling* coefficient that is negligible compared with the cross-coupling coefficient and can anyway be readily included by redefining the propagation constants $(\beta_L, \beta_R)$ [102].

Note that the Hamiltonian-equivalent ($\mathcal{H}_{WG}$) for the coupled optical waveguides configuration (Eq. **(103)**) is similar to the Hamiltonian for the ammonia molecule in a constant electric field (Eq. **(86)**) except that here the diagonal element and the off-diagonal coupling element switch signs ($\delta\beta \geq 0$) and in particular $(K_{eff} > 0)$.

*Eigenvalues.*

$$\Lambda_{P,WG} = \beta_{avg} + \Omega_{P,WG}$$
$$\Lambda_{N,WG} = \beta_{avg} - \Omega_{P,WG}$$
$$\Omega_{P,WG} \equiv \sqrt{(\delta\beta)^2 + K_{eff}^2}$$

**(104)**

### 2.5.2. Definite-energy states

Considering the optical waveguide system, the "energy" (in units of inverse distance) is measured by the phase rotational speeds of the wavevectors (for example $\beta_{avg} + \Omega_{P,WG}$ or $\beta_{avg} - \Omega_{P,WG}$).

*Eigenvectors.* Given $\phi_D = 0$, the eigenvectors (Eq. **(16)**) are:

$$\xi_P = \begin{pmatrix} \sin\theta_R \\ \cos\theta_R \end{pmatrix}; \quad \xi_N = \begin{pmatrix} \cos\theta_R \\ -\sin\theta_R \end{pmatrix}$$

**(105)**

According to our soft definition of symmetry (components of the same sign) $\xi_P$ is a Symm eigenvector whereas $\xi_N$ is an Asym eigenvector.

*Definite-energy states.* To get the Symm definite-energy state (eigenmode) we set (Eq. **(33)**):

$$\begin{pmatrix} A_L(0) \\ A_R(0) \end{pmatrix}_P = \begin{pmatrix} \xi_{P1} \\ \xi_{P2} \end{pmatrix} = \begin{pmatrix} \sin\theta_R \\ \cos\theta_R \end{pmatrix}$$

**(106)**

to get (Eq. **(33)**):

$$\phi_P(x,y,z) \equiv E_{symm}(x,y,z)$$
$$= e^{j\cdot(\beta_{avg}+\Omega_{P,WG})\cdot z} \cdot [\sin\theta_R \cdot U_L(x,y) + \cos\theta_R \cdot U_R(x,y)]$$

**(107)**

Similarly, to get the Asym definite-energy state (eigenmode) we set (Eq. **(34)**):

$$\begin{pmatrix} A_L(0) \\ A_R(0) \end{pmatrix}_N = \begin{pmatrix} \xi_{N1} \\ \xi_{N2} \end{pmatrix} = \begin{pmatrix} \sin\theta_R \\ -\cos\theta_R \end{pmatrix}$$

**(108)**

to get (Eq. **(34)**)

$$\phi_N(x,y,z) \equiv E_{asym}(x,y,z)$$
$$= e^{j\cdot(\beta_{avg}-\Omega_{P,WG})\cdot z} \cdot [\cos\theta_R \cdot U_L(x,y) - \sin\theta_R \cdot U_R(x,y)]$$

**(109)**

Equations **(107)** and **(109)** indicate that the $\phi_P(x,y,z)$ eigenmode is a Symm $(E_{symm}(x,y,z))$ superposition of the original $(U_L(x,y); U_R(x,y))$ states whereas $\phi_N(x, y, z)$ eigenmode is an Asym $(E_{asym}(x, y, z))$ superposition of the original $(U_L(x, y); U_R(x, y))$ states.

A general superposition of the two eigenmodes will result in periodic power transfer between the two waveguides – complete power transfer if the waveguides are identical and incomplete power transfer if the waveguides are nonidentical (cf. *Figure 16* below). The *spatial* frequency of power transfer is the generalized Rabi flopping (spatial) frequency $(\Omega_{GR,WG} = 2 \cdot \Omega_{P,WG})$.

### 2.5.3. Energy levels

From **(107)** and **(109)** we get for the two "energy" levels $(K_P, K_N)$:

$$K_P \equiv K_{symm} = \beta_{avg} + \Omega_{P,WG} \; ; \; K_N \equiv K_{asym} = \beta_{avg} - \Omega_{P,WG}$$

$$\Omega_{GR,WG} = 2 \cdot \Omega_{P,WG}$$

**(110)**

Considering the coupled two optical waveguides system, the Asym superposition is of a *lower* energy (slower phase rotational speed of $(\beta_{avg} - \Omega_{P,WG})$) and the Symm superposition is of a *higher* energy (phase rotational speed of $(\beta_{avg} + \Omega_{P,WG})$). These Symm-Asym energy levels are reversed compared with the energy levels of the ammonia molecule in a constant electric field. The source of the difference can be traced to the sign of the off-diagonal coupling element of the Hamiltonian (or Hamiltonian-equivalent) matrix (Eq. **(86)** with $\phi_D = \pi$ vs. Eq. **(103)**). The waveguides' energy gap is of course $\Omega_{GR,WG} = 2 \cdot \Omega_{P,WG}$.

### 2.5.4. Graphical illustrations

Computer simulations of power-exchange oscillations between two coupled optical waveguides are depicted in *Figure 16*. For the simulation, assumed equal/unequal system parameters are: vacuum wavelength - 0.8 $\mu m$, core-glass indices (left ; right) - (1.5018; 1.5018) and (1.5018; 1.5024), coupling coefficient $(K_{eff})$ was found to be – 0.63/0.71 $mm^{-1}$. As shown, in the case of unequal waveguides, modulation depth is less than unity and oscillation-period is shorter (vs. the oscillation-period of the equal waveguides pair).

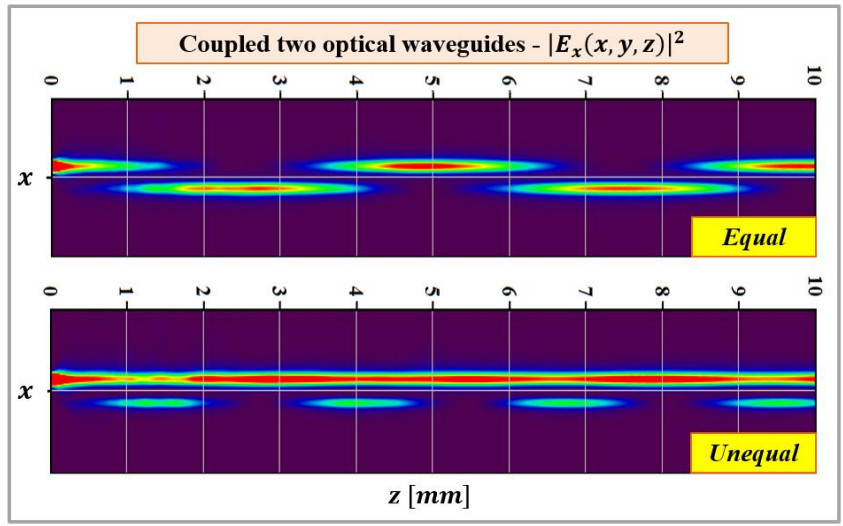

*Figure 16: Oscillations of the electrical power propagating along a TSS of coupled two optical waveguides. Top: Equal waveguides. Modulation depth is unity and oscillation period is "long". Bottom: Unequal waveguides. Modulation depth is less than unity and oscillation period is "short". "Energy" levels of the optical TSS are $K_P = \beta_{avg} + \Omega_{P,WG}$ ; $K_N = \beta_{avg} - \Omega_{P,WG}$. Energy gap is $\Omega_{GR,WG} = 2 \cdot \Omega_{P,WG}$ (cf. Eq. (110))*

So much for TIH-TSSs characterized by two stationary eigenstates, and by two associated distinct ("definite") energies as indicated by two distinct phase rotational speeds. In the next section – discussion of more involved cases of TDH-TSSs characterized by two stationary states, and by *four* quasi-energies.

### 3. Time-dependent Hamiltonian – four quasi-energy levels

If a "monochromatic" near-resonance sinusoidal term $\bigl(cos(\omega_C \cdot t)\bigr)$ is included in the time-dependent TSS Hamiltonian, then after adopting the RWA, closed-form *analytic* expressions for the probability amplitudes can be derived. In the majority of concrete TSSs such as atoms in an electric field **[48]**,**[61]**, driven protons (MRI) **[8]**, the Hydrogen maser **[107]**, the collection of atomic clocks **[108]**, or trapped-ion qubits **[30]**, the coupling sinusoidal term is introduced into the two off-diagonal terms. In these cases, the analytic solution for the probability amplitudes involves three steps – rotate (and adopt the RWA) – solve (a stationary problem) – rotate back.

An outstanding TSS is the driven ammonia molecule. The sinusoidal term for the driven permanent-dipole ammonia TSS goes into the diagonal Hamiltonian elements **[15]**. In such a case, a preparation step of basis change (from the canonical basis to the eigenstates basis of the free-molecule Hamiltonian) is required. The preparation step "translates" (later) the sinusoidal term into the off-diagonal elements.

It is readily verified, that the state-occupation *probabilities* (absolute-squared of the probability amplitudes) are *independent of the back rotation step*. For this reason, many of the published studies, being focused on state-occupation probabilities, do not perform the back-rotation extra step. However, it is through phase rotational speeds appearing in the expressions for the probability amplitudes *following back rotation* (back to the Schrödinger picture), that the four quasi-energy levels are recognized.

The author of **[50]** presents a "Floquet Hamiltonian" and shows a *single* eigenstate and associated *two* frequency eigenvalues (leading to two quasi-energies) of the Floquet Hamiltonian.

In the fully quantum-mechanical "atom + driving photons" ("dressed atom") model, the four energy levels are the energies of four distinct *energy eigenstates*. In the dressed atom model then, the prefix "quasi" assigned to the energy levels during a semi-classical treatment, is omitted all together **[50]**.

One way or the other, the four quasi-energies are definitely "physical". That is - depending on the physical conditions of the system, each of the four quasi-energy states may be populated. Their existence is manifested, for example, in detecting Autler-Townes doublet transitions in probe experiments **[84]**,**[69]** or in realizing Mollow triplet transitions in fluorescence (spontaneous emission) spectrum measurements **[72]**-**[76]**.

In the present two-part section then, we first present a solution to Schrödinger equation (Eq. **(114)** below), assuming a time-dependent TSS Hamiltonian (Eq. **(113)** below). Then, based on the general solution, we look at a number of representative concrete examples at typical operating parameters:

*Concrete TDH-TSSs (four quasi-energy levels)*

d)  A proton under a constant axial magnetic field and an harmonic transverse magnetic field
e)  A cesium atom driven near resonance by an external harmonic magnetic field
f)  An ammonia molecule in an harmonic electric field

In both the general solution part and the concrete examples part we focus our attention on the energies of the TSSs.

### 3.1. General solution

In this "general solution" section we consider the most general (sinusoidally driven) TDH and execute the three rotate-solve-rotate back steps. Again we write two solution versions – one is the standard, somewhat obscure matrix multiplication solution, and the other is the energy-revealing $(\mathcal{A}, \mathcal{B}, \mathcal{C}, \mathcal{D})$ analytic solution. We show the existence of four quasi-energies each with its own distinct phase rotational speed.

#### 3.1.1. Basis states and the state vector

Given a set of two canonical basis states

$$|1> \equiv \begin{pmatrix} 1 \\ 0 \end{pmatrix} \; ; \; |2> \equiv \begin{pmatrix} 0 \\ 1 \end{pmatrix}$$

**(111)**

any TSS state vector can be written as a linear superposition -

$$|\psi_{12,t}(t)> = C_1(t) \cdot |1> + C_2(t) \cdot |2>$$

**(112)**

The evolution of the time-dependent probability amplitudes $(C_1(t), C_2(t))$ is the solution of Schrödinger equation for these amplitudes (Eq. **(114)** below).

For clarity, to designate time-dependent TSS quantities, we add a subscript "$t$" to the state vector $(|\psi_{12,t}(t)>)$ and to the TSS Hamiltonian – see Eq. **(113)** (but keep the general designation for the probability amplitudes $(C_1(t), C_2(t))$).

#### 3.1.2. Time-dependent Hamiltonian

The general *time-dependent* TSS Hamiltonian $(\mathcal{H}_{TSS,t})$ with *harmonic coupling* off-diagonal elements reads **[41]**,(**[15]** Ch. 11),**[109]**:

$$(\mathcal{H}_{TSS,t}) = (\mathcal{H}_{\omega 0}) + (\mathcal{H}_{12,t})$$

$$(\mathcal{H}_{\omega 0}) \equiv \omega_0 \cdot (I) \; ; \; (\mathcal{H}_{12,t}) \equiv \begin{pmatrix} -\dfrac{\omega_A}{2} & \Omega_D \cdot cos(\omega_C \cdot t) \\ \Omega_D^* \cdot cos(\omega_C \cdot t) & +\dfrac{\omega_A}{2} \end{pmatrix} \; ; \; \omega_A \in \mathbb{R}$$

**(113)**

Note that the time-independent Hamiltonian is a special case of the more general TDH of Eq. **(113)** (by setting $\omega_C = 0$).

The Schrödinger equation for the amplitude vector reads:

$$\frac{d}{dt}\boldsymbol{C}(t) = -\mathrm{j}\cdot(\mathcal{H}_{TSS,t})\cdot\boldsymbol{C}(t) \quad ; \quad \boldsymbol{C}(t) \equiv \begin{pmatrix} C_1(t) \\ C_2(t) \end{pmatrix}$$

**(114)**

### 3.1.3. Rotation

To solve Eq. **(114)** we first apply a rotation transformation to the $(\mathcal{H}_{TSS,t})$ Hamiltonian **[110]**:

$$(H_{12,ROT,RWA}) = (R_\omega(t))^\dagger \cdot (\mathcal{H}_{12;t}) \cdot (R_\omega(t)) - \mathrm{j}\cdot(R_\omega(t))^\dagger \cdot \frac{d(R_\omega(t))}{dt}$$

$$(R_\omega(t)) = \begin{pmatrix} e^{+\mathrm{j}\cdot\frac{\omega_C}{2}\cdot t} & 0 \\ 0 & e^{-\mathrm{j}\cdot\frac{\omega_C}{2}\cdot t} \end{pmatrix} \quad ; \quad (R_\omega(t))^\dagger = \begin{pmatrix} e^{-\mathrm{j}\cdot\frac{\omega_C}{2}\cdot t} & 0 \\ 0 & e^{+\mathrm{j}\cdot\frac{\omega_C}{2}\cdot t} \end{pmatrix}$$

**(115)**

Working the math and adopting the RWA, the TIH in the rotating frame is given as **[3]**:

$$(H_{12,ROT,RWA}) = (\mathcal{H}_{\omega 0}) + (\mathcal{H}_{C,ROT,RWA})$$

$$(\mathcal{H}_{C,ROT,RWA}) = \begin{pmatrix} -\frac{\delta_C}{2} & \frac{\Omega_D}{2} \\ \frac{\Omega_D^*}{2} & +\frac{\delta_C}{2} \end{pmatrix} \quad ; \quad \delta_C \equiv (\omega_A - \omega_C)$$

**(116)**

### 3.1.4. Time-dependence of the rotated state vector

Designate the rotated amplitude vector as

$$\boldsymbol{C}_x(t) \equiv \begin{pmatrix} C_{1x}(t) \\ C_{2x}(t) \end{pmatrix}$$

**(117)**

The Schrödinger equation for the rotated amplitude vector reads:

$$\frac{d}{dt}\boldsymbol{C}_x(t) = -\mathrm{j}\cdot(H_{12,ROT,RWA})\cdot\boldsymbol{C}_x(t)$$

**(118)**

Identify the elements of $(\mathcal{H}_{C,ROT,RWA})$ with the elements of $(\mathcal{H}_{12,0})$ (cf. Eq. **(10)** ):

$$\begin{aligned}
(\mathcal{H}_{C,ROT,RWA})[1,1] &\equiv -\omega_{11} \\
(\mathcal{H}_{C,ROT,RWA})[1,2] &\equiv |\omega_{12}|\cdot e^{-\mathrm{j}\cdot\phi_D} \\
(\mathcal{H}_{C,ROT,RWA})[2,1] &\equiv |\omega_{12}|\cdot e^{+\mathrm{j}\cdot\phi_D} \\
(\mathcal{H}_{C,ROT,RWA})[2,2] &\equiv +\omega_{11}
\end{aligned}$$

**(119)**

Let us designate the important parameter $\Omega_P$ as $\Omega_P \to \Omega_{P,t}$ and write the expression for this parameter in terms of the entries of the rotated Hamiltonian $(\mathcal{H}_{C,ROT,RWA})$. According to Eq. **(6)**:

$$\Omega_{P,t} = \sqrt{\left(\frac{\delta_C}{2}\right)^2 + \left|\frac{\Omega_D}{2}\right|^2} \quad ; \quad \Omega_{GR,t} = 2 \cdot \Omega_{P,t}$$

(120)

Write the matrix-multiplication solution for the $\boldsymbol{C}_x(t)$ probability amplitudes:

$$\begin{pmatrix} C_{1x}(t) \\ C_{2x}(t) \end{pmatrix} = \left(e^{-\jmath \cdot \omega_0 \cdot t} \cdot (I)\right) \cdot (\xi) \cdot \left(I_\lambda(t)\right) \cdot (\xi)^{-1} \begin{pmatrix} C_{1x}(0) \\ C_{2x}(0) \end{pmatrix}$$

$$\left(I_\lambda(t)\right) = \begin{pmatrix} e^{-\jmath \cdot \lambda_P \cdot t} & 0 \\ 0 & e^{-\jmath \cdot \lambda_N \cdot t} \end{pmatrix}$$

(121)

And write the $(\mathcal{A}, \mathcal{B}, \mathcal{C}, \mathcal{D})$ analytic solution:

$$C_{1x}(t) = [\mathcal{A} \cdot e^{+\jmath \cdot \Omega_{P,t} \cdot t} + \mathcal{B} \cdot e^{-\jmath \cdot \Omega_{P,t} \cdot t}] \cdot e^{-\jmath \cdot \omega_0 \cdot t}$$
$$C_{2x}(t) = [\mathcal{C} \cdot e^{+\jmath \cdot \Omega_{P,t} \cdot t} + \mathcal{D} \cdot e^{-\jmath \cdot \Omega_{P,t} \cdot t}] \cdot e^{-\jmath \cdot \omega_0 \cdot t}$$

(122)

### 3.1.5. Back rotation

Now apply back rotation to the $\boldsymbol{C}_x(t)$ amplitude vector (in the rotating frame), to get the $\boldsymbol{C}(t) = \left(R_\omega(t)\right) \cdot \boldsymbol{C}_x(t)$ (in the Schrödinger picture). The full solution to Schrödinger equation **(114)** then reads:

$$\begin{pmatrix} C_1(t) \\ C_2(t) \end{pmatrix} = \left(R_\omega(t)\right) \cdot \left(e^{-\jmath \cdot \omega_0 \cdot t} \cdot (I)\right) \cdot (\xi) \cdot \left(I_\lambda(t)\right) \cdot (\xi)^{-1} \begin{pmatrix} C_{1x}(0) \\ C_{2x}(0) \end{pmatrix}$$

(123)

At $t = 0$ (Eq. **(123)**):

$$\begin{pmatrix} C_{1x}(0) \\ C_{2x}(0) \end{pmatrix} = \begin{pmatrix} C_1(0) \\ C_2(0) \end{pmatrix}$$

(124)

Insert **(124)** into **(123)** to arrive at the solution to Schrödinger equation **(114)** for the time evolution of the probability amplitudes given a general *time-dependent* TSS Hamiltonian $\left(\mathcal{H}_{TSS,t}\right)$:

$$\begin{pmatrix} C_1(t) \\ C_2(t) \end{pmatrix} = \left(R_\omega(t)\right) \cdot \left(e^{-\jmath \cdot \omega_0 \cdot t} \cdot (I)\right) \cdot (\xi) \cdot \left(I_\lambda(t)\right) \cdot (\xi)^{-1} \begin{pmatrix} C_1(0) \\ C_2(0) \end{pmatrix}$$

(125)

On passing, and as follows from Eq. **(125)**, let us also isolate the *time evolution operator* $\left(U_t(t)\right)$ solving Schrödinger equation **(114)** with the Hamiltonian **(113)**:

$$\left(U_t(t)\right) = \left(R_\omega(t)\right) \cdot \left(e^{-\jmath \cdot \omega_0 \cdot t} \cdot (I)\right) \cdot (\xi) \cdot \left(I_\lambda(t)\right) \cdot (\xi)^{-1}$$

(126)

The explicit expressions for the $\boldsymbol{C}(t)$ amplitudes in terms of the $(\mathcal{A}, \mathcal{B}, \mathcal{C}, \mathcal{D})$ parameters of Eq. **(21)** are spelled out as:

$$C_1(t) = [\mathcal{A} \cdot e^{+\jmath \cdot \Omega_{P,t} \cdot t} + \mathcal{B} \cdot e^{-\jmath \cdot \Omega_{P,t} \cdot t}] \cdot e^{-\jmath \cdot \left(\omega_0 - \frac{\omega_C}{2}\right) \cdot t}$$
$$C_2(t) = [\mathcal{C} \cdot e^{+\jmath \cdot \Omega_{P,t} \cdot t} + \mathcal{D} \cdot e^{-\jmath \cdot \Omega_{P,t} \cdot t}] \cdot e^{-\jmath \cdot \left(\omega_0 + \frac{\omega_C}{2}\right) \cdot t}$$

(127)

Equation **(127)** can be recast in a different way. Replacing $\omega_C$ by $\omega_A - \delta_C$ (Eq. **(116)**) and moving "$\delta_C$" left to the two terms in the square brackets, we get:

$$C_1(t) = \left[\mathcal{A} \cdot e^{\mathbb{j}\cdot\left(-\frac{\delta_C}{2}+\Omega_{P,t}\cdot t\right)} + \mathcal{B} \cdot e^{\mathbb{j}\cdot\left(-\frac{\delta_C}{2}-\Omega_{P,t}\cdot t\right)}\right] \cdot e^{-\mathbb{j}\cdot\left(\omega_0-\frac{\omega_A}{2}\right)\cdot t}$$

$$C_2(t) = \left[\mathcal{C} \cdot e^{\mathbb{j}\cdot\left(+\frac{\delta_C}{2}+\Omega_{P,t}\cdot t\right)} + \mathcal{D} \cdot e^{\mathbb{j}\cdot\left(+\frac{\delta_C}{2}-\Omega_{P,t}\cdot t\right)}\right] \cdot e^{-\mathbb{j}\cdot\left(\omega_0+\frac{\omega_A}{2}\right)\cdot t}$$

**(128)**

Consulting the expression for the probability amplitude $C_1(t)$ of Eq. **(128)**, the phase rotational speed on the extreme right $\left(\omega_0 - \frac{\omega_A}{2}\right)$ is identified as the lower eigenenergy of the unperturbed TSS. Given a red detuned sinusoidal perturbation ($\delta_C > 0$), the original lower eigenenergy we are considering is pushed *up* by $\delta_C/2$ and splits in two by $\pm\Omega_{P,t}$ for a total lower-energy doublet split of $2 \cdot \Omega_{P,t} = \Omega_{GR,t}$ (Eq. **(120)**). Consulting the expression for the second probability amplitude $C_2(t)$ of Eq. **(128)**, we see the unperturbed upper eigenenergy $\left(\omega_0 + \frac{\omega_A}{2}\right)$ pulled *down* by $\delta_C/2$ and splits in two by $\pm\Omega_{P,t}$ for a total upper-energy doublet split of $2 \cdot \Omega_{P,t} = \Omega_{GR,t}$. These are two concrete examples of the general quasi-energies statement expressed by Eq. **(4)** above.

Thus, the TDH-TSS state vector of Eq. **(112)** $\left(|\psi_{12,t}(t)>\right)$ can be written as a sum of *four* amplitudes[5], each characterized by a *distinct phase rotational speed* **[50]**:

$$|\psi_{12,t}(t)> = \left(\mathcal{A} \cdot e^{-\mathbb{j}\cdot\omega_N(L)\cdot t} + \mathcal{B} \cdot e^{-\mathbb{j}\cdot\omega_P(L)\cdot t}\right) \cdot |1>$$
$$+\left(\mathcal{C} \cdot e^{-\mathbb{j}\cdot\omega_N(H)\cdot t} + \mathcal{D} \cdot e^{-\mathbb{j}\cdot\omega_P(H)\cdot t}\right) \cdot |2>$$

**(129)**

The four distinct rotational speeds $\left(\omega_P(L), \omega_P(H), \omega_N(L), \omega_N(H)\right)$ are listed by Eq. **(130)** below in relations to stationary states and quasi-energies.

State-occupation probabilities $(|C_1(t)|^2, |C_2(t)|^2)$ for the driven TSS are given by Eqs. **(25)** and **(26)** *with $\Omega_P$ replaced by $\Omega_{P,t}$* of Eq. **(120)**.

Note that the state-occupation *probabilities* $(|C_1(t)|^2, |C_2(t)|^2)$ are indifferent to the rotation transformation. In other words - state-occupation probabilities in the Schrödinger picture and in the rotating frame are identical. In terms of modulation depth (Eq. **(46)**) given the Hamiltonian $\left(\mathcal{H}_{C,ROT,RWA}\right)$ of Eq. **(116)** we see (cf. section **2.1.10**) that modulation depth of unity (complete population inversion) is achieved *if and only if the sinusoidal driving field is at resonance* ($\delta_C = 0$) with the transition-frequency of the unperturbed system **[50]**.

### 3.1.6. Stationary states and quasi-energies

It is a matter of straightforward evaluation to verify that if the TSS is launched (initiated) with amplitudes equal to the components of the "$P$" eigenvector of the rotated Hamiltonian of Eq. **(116)** ($C_1(0) = \xi_{P1}$ and $C_2(0) = \xi_{P2}$) then the parameters $\mathcal{A}$ and $\mathcal{C}$ of Eq. **(127)** each get the value of zero, the parameters $\mathcal{B}$ and $\mathcal{D}$ get the values of the "$P$" eigenstate components ($\mathcal{B} = C_1(0) = \xi_{P1}$ ; $\mathcal{D} = C_2(0) = \xi_{P2}$). The result is time-independent *magnitude* of the amplitudes ($|C_1(t)|^2 = |\xi_{P1}|^2$ ; $|C_2(t)|^2 = |\xi_{P2}|^2$).

---

[5] The author of (**[111]** Eq. 1.133) also derived a four-term expression for the wavefunction in the Schrödinger picture.

It follows from Eq. **(127)** that the TDH-TSS under such initial conditions is in a stationary state[6], characterized by *two* (quasi) energy levels $\left(\mathcal{E}_P(L)\,;\,\mathcal{E}_P(H)\right)$ as given by Eq. **(130)** below.

Similarly with the $\mathcal{A}$ and $\mathcal{C}$ parameters and the components of the "$N$" eigenvector $(C_1(0) = \xi_{N1}\,;\,C_2(0) = \xi_{N2} \Rightarrow \mathcal{A} = C_1(0) = \xi_{N1}\,;\,\mathcal{C} = C_2(0) = \xi_{N2}\,;\,\mathcal{B} = \mathcal{D} = 0)$ and in this case too, the TDH-TSS under the "$N$" initial conditions is in a stationary state, characterized by *two* (quasi) energy levels $\left(\mathcal{E}_N(L)\,;\,\mathcal{E}_N(H)\right)$ as given by Eq. **(130)**:

$$\mathcal{E}_P(L) \equiv \omega_P(L) = \left(\omega_0 - \frac{\omega_C}{2} + \Omega_{P,t}\right)\,;\,\mathcal{E}_P(H) \equiv \omega_P(H) = \left(\omega_0 + \frac{\omega_C}{2} + \Omega_{P,t}\right)$$
$$\mathcal{E}_N(L) \equiv \omega_N(L) = \left(\omega_0 - \frac{\omega_C}{2} - \Omega_{P,t}\right)\,;\,\mathcal{E}_N(H) \equiv \omega_N(H) = \left(\omega_0 + \frac{\omega_C}{2} - \Omega_{P,t}\right)$$
$$|\psi_{(\xi),P}(t)> = \left(\xi_{P1} \cdot e^{-j\cdot\omega_P(L)\cdot t}\right) \cdot |1> + \left(\xi_{P2} \cdot e^{-j\cdot\omega_P(H)\cdot t}\right) \cdot |2>$$
$$|\psi_{(\xi),N}(t)> = \left(\xi_{N1} \cdot e^{-j\cdot\omega_N(L)\cdot t}\right) \cdot |1> + \left(\xi_{N2} \cdot e^{-j\cdot\omega_N(H)\cdot t}\right) \cdot |2>$$

**(130)**

To summarize - if launched into the $P$-state, then the TDH-TSS will stay in the $P$-state with two quasi energies $\mathcal{E}_P(L), \mathcal{E}_P(H)$. If launched into the $N$-state, then the TDH-TSS will stay in the $N$-state with two quasi energies $\mathcal{E}_N(L), \mathcal{E}_N(H)$ (cf. *Figure 20* below). Namely:

$$\begin{pmatrix}C_1(0)\\C_2(0)\end{pmatrix}_P = \begin{pmatrix}\xi_{P1}\\\xi_{P2}\end{pmatrix} \Rightarrow \mathcal{P}_P(t) = 1\,;\,\mathcal{E}_P \Rightarrow \begin{pmatrix}\mathcal{E}_P(L)\\\mathcal{E}_P(H)\end{pmatrix} = \begin{pmatrix}\omega_0 - \frac{\omega_C}{2} + \Omega_{P,t}\\\omega_0 + \frac{\omega_C}{2} + \Omega_{P,t}\end{pmatrix}$$

$$\begin{pmatrix}C_1(0)\\C_2(0)\end{pmatrix}_N = \begin{pmatrix}\xi_{N1}\\\xi_{N2}\end{pmatrix} \Rightarrow \mathcal{P}_N(t) = 1\,;\,\mathcal{E}_N \Rightarrow \begin{pmatrix}\mathcal{E}_N(L)\\\mathcal{E}_N(H)\end{pmatrix} = \begin{pmatrix}\omega_0 - \frac{\omega_C}{2} - \Omega_{P,t}\\\omega_0 + \frac{\omega_C}{2} - \Omega_{P,t}\end{pmatrix}$$

**(131)**

These stationary states of quasi-energy levels (eq. **(131)**) **[61],[66],[109],[67]** are shown by the two panels of *Figure 17* below.

The state function describing the state of a given quantum mechanical system is in general made up of a sum of time-dependent amplitudes. An "Energy level" of a state (or "quasi-energy", cf. Eq. **(4)**) is identified with a constant magnitude of an amplitude and a constant phase rotational speed of that amplitude (constant rotational speed in the complex plane). In the case of near-resonance sinusoidally-driven TSSs (such that the RWA is justified **[50]**), two such energy levels are identified for the "$P$" initial conditions and two for the "$N$" initial conditions (Eq. **(131)**). Rotations of these constant-magnitude amplitudes in the complex plane are depicted schematically by the arrows of *Figure 18* below.

The spectral differences between the high-lying levels $\left(\mathcal{E}_P(H), \mathcal{E}_N(H)\right)$ and the low-lying levels $\left(\mathcal{E}_P(L), \mathcal{E}_N(L)\right)$ make a Mollow triplet (the center-positions of the triple Lorentzian lines, cf. *Figure 19* below):

---

[6] See "quasi-stationary" states discussed in relations to atom-light interaction in **[61]**.

$$\begin{aligned} Center&: \ \omega_C \\ Red&: \ \omega_C - 2 \cdot \Omega_{P,t} = \omega_C - \Omega_{GR,t} \\ Blue&: \ \omega_C + 2 \cdot \Omega_{P,t} = \omega_C + \Omega_{GR,t} \end{aligned}$$

(132)

The three spectral lines (center positions of [75],[33]) are shown by *Figure 19* below.

Like undamped *forced* mechanical vibrations, the TDH-TSS "vibrates" (radiates/absorbs) *in the frequency of the driving field* (and NOT in its "natural" vibration frequency). At resonance however, in the quantum mechanical case the amplitude of "vibrations" does not go to infinity [61]. And, unlike in the case of the mechanical system, in the case of the quantum mechanical TSS, two side lines at $\Omega_{GR,t}$ left and right distance from the centerline also appear[7].

If the TSS is initiated into the $|1>$ state ($C_1(0) = 1$ ; $C_2(0) = 0$) then, in the case of a near-resonance-driven TSS, state-occupation probabilities will oscillate (cf. *Figure 23* below for example). The state-occupation probability $|C_1(t)|^2$ is the system's (stimulated) absorption probability and $|C_2(t)|^2$ is the system's stimulated emission probability [48]. These oscillations of state-occupation probabilities (with frequency equals to the generalized Rabi "flopping frequency" [79]) represent then successive emission-absorption energy exchange [111],[96] (field's photons in the case of atom-light or molecule-light interaction) between the TDH-TSS and the (strong) driving field [61],[24]. If the TDH-TSS is launched into one of its two stationary states, then the emissions-absorptions are balanced and the state-occupation probabilities are time-independent.

### 3.1.7. Four quasi-energy levels following an equation by Meystre and Sargent

The existence of four quasi-energy level may be calculated through an equation from the book by Meystre and Sargent [48]. Considering atom-field interaction for two-level atoms, Meystre and Sargent write the wavefunction $\psi(\boldsymbol{r},t)$ for perturbed energy states $a$ and $b$ (in the Schrödinger picture) as ([48] Eq. 3.125):

$$\psi(\boldsymbol{r},t) = C_a(t) \cdot e^{-j \cdot \left(\omega_a - \frac{\delta}{2}\right) \cdot t} \cdot u_a(\boldsymbol{r}) + C_b(t) \cdot e^{-j \cdot \left(\omega_b + \frac{\delta}{2}\right) \cdot t} \cdot u_b(\boldsymbol{r})$$
$$\omega = \omega_a - \omega_b > 0 \ ; \ \delta = \omega - \nu$$

(133)

where $\nu$ is the frequency of the perturbing electric field (designated $\omega_C$ in our study here). Through Schrödinger's equation, solving for the probability amplitudes $(C_a(t), C_b(t))$ *in the rotating frame*, writing the solution in the $(\mathcal{A}, \mathcal{B}, \mathcal{C}, \mathcal{D})$ format *and inserting back* into Eq. (133) (the wavefunction in the Schrödinger picture) we get:

---

[7] At resonance ($\omega_C = \omega_A \to \delta_C = 0$), unlike the prediction of the classical theory for forced mechanical vibrations, a more accurate quantum-mechanical theory (*including the fast oscillating terms* that were neglected by adopting the RWA), predicts a *shift in the resonance frequency* of a perturbed TSS (an atom or otherwise) in the presence of a strong sinusoidal coupling field [41].

$$\psi(r,t) = \left( \mathcal{A} \cdot e^{-j\cdot\left(\omega_a - \frac{\delta}{2} + \Omega_{P,t}\right)\cdot t} + \mathcal{B} \cdot e^{-j\cdot\left(\omega_a - \frac{\delta}{2} - \Omega_{P,t}\right)\cdot t} \right) \cdot u_a(r)$$
$$+ \left( \mathcal{C} \cdot e^{-j\cdot\left(\omega_b + \frac{\delta}{2} + \Omega_{P,t}\right)\cdot t} + \mathcal{D} \cdot e^{-j\cdot\left(\omega_b + \frac{\delta}{2} - \Omega_{P,t}\right)\cdot t} \right) \cdot u_b(r)$$

(134)

We see then the *energy shift and split* of each of the original state-energies ($\omega_a$, $\omega_b$):

$$\omega_a \to \begin{cases} \omega_a^{(+)} = \left(\omega_a - \frac{\delta}{2}\right) + \Omega_{P,t} \\ \omega_a^{(-)} = \left(\omega_a - \frac{\delta}{2}\right) - \Omega_{P,t} \end{cases}$$

$$\omega_b \to \begin{cases} \omega_b^{(+)} = \left(\omega_b + \frac{\delta}{2}\right) + \Omega_{P,t} \\ \omega_b^{(-)} = \left(\omega_b + \frac{\delta}{2}\right) - \Omega_{P,t} \end{cases}$$

(135)

For a total of four quasi-energy levels. The resulting energy diagram is identical with the energy diagram of *Figure 19* below.

According to the quantum-mechanical dressed atom model, four "legitimate" energy levels are formed in a similar scenario. Considering two energy eigenstates of an unperturbed atom, if the atom is "dressed" by a "sea" of *near-resonance* photons, then each of the states splits in two [56] to form four eigenstates (dressed states) of the "dressed-atom Hamiltonian", say - ($|-,n>, |+,n>, |-,n+1>, |+,n+1>$) [50], [63],[85]. Each of the dressed states is associated with a distinct energy eigenvalue. The resulting four-energy diagram closely resembles the energy diagram of *Figure 19* below.

In their book on *Atoms in Strong Light Fields* [61], Delone and Krainov consider a two-level atom in a strong, *classical*, near-resonance electric field (*light is described classically and the atom quantum-mechanically*). Delone and Krainov show the split of each level in two, provide a list of four quasi-energies and plot an energy diagram that except for designations is identical to the energy diagram plotted in *Figure 19* below.

### 3.1.8. Graphical illustrations

The next four figures illustrate key characteristics of sinusoidally-driven TSSs.

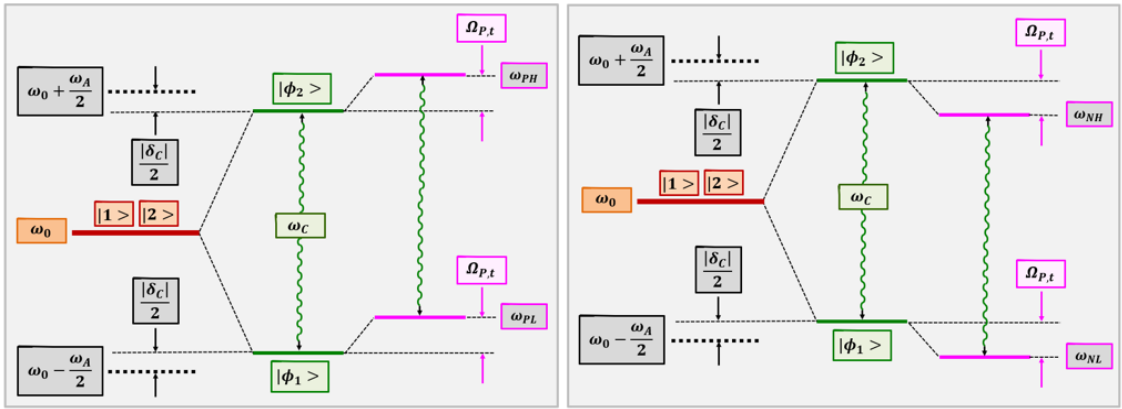

*Figure 17: Quasi-energy levels of TDH-TSSs excited into stationary states (cf. Eq. (130)). Left: Quasi-energy levels of stationary state $|\psi_{(\xi),P}(t)\rangle$. Right: Quasi-energy levels of stationary state $|\psi_{(\xi),N}(t)\rangle$. The states $|\phi_1(t)\rangle, |\phi_2(t)\rangle$ (framed green in the figure) are the $|1\rangle, |2\rangle$ basis states in the rotating frame (i.e. following the rotation of the basis states by the rotation matrix $(R_\omega(t))$ of Eq. (115)). The frequency-difference between each pair of quasi energies (left/right) is the frequency $(\omega_C)$ of the driving field and is smaller then $\omega_A$ for a red detuned driving field. In other words - the red detuned driving field "pulls" inwards the energy levels $\left(\omega_0 \pm \frac{\omega_A}{2}\right)$ of the unperturbed TSS. Like forced mechanical vibrations, the TSS "vibrates" (radiates/absorbs) in the frequency of the driving field (and NOT in its "natural" vibration frequency).*

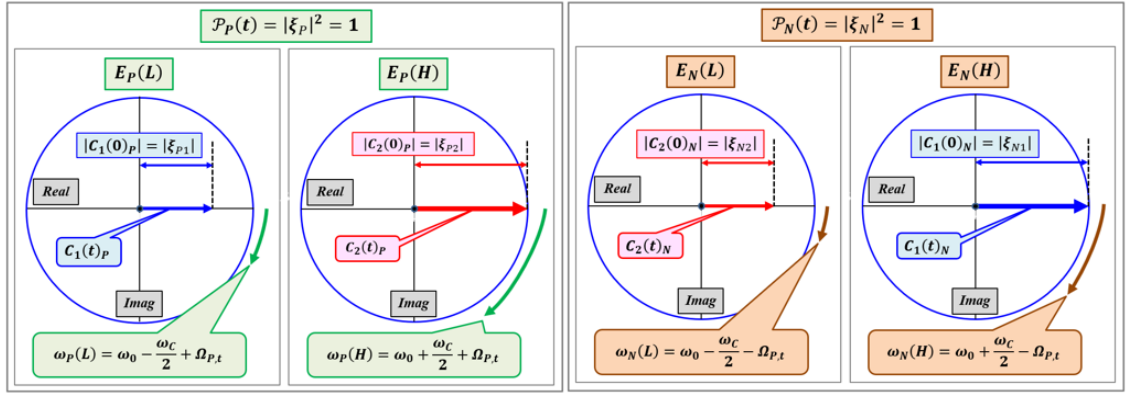

*Figure 18: Schematically-drawn phase rotational speeds of TSSs excited (initiated) into stationary states (cf. Eq. (130)). Left: Phase rotational speeds associated with stationary state $|\psi_{(\xi),P}(t)\rangle$ ($|C_1(0)| = |\xi_{P1}|, |C_2(0)| = |\xi_{P2}|$). Right: Phase rotational speeds associated with stationary state $|\psi_{(\xi),N}(t)\rangle$ ($|C_1(0)| = |\xi_{N1}|, |C_2(0)| = |\xi_{N2}|$). Compare the two pairs of phase rotational speeds shown here to the single pair of phase rotational speeds shown in **Figure 7**. A driven TSS launched into a (single) **stationary state** is characterized by **two** "definite energies" (quasi-energies).*

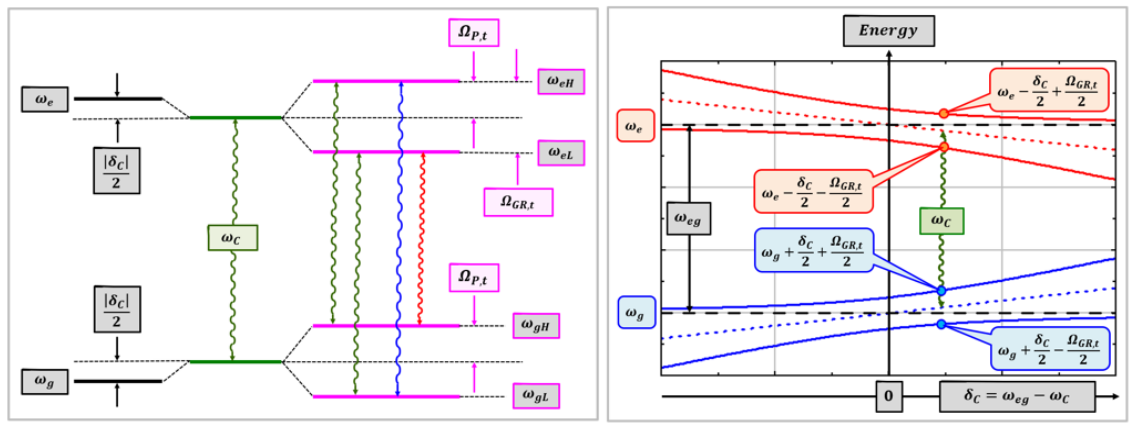

*Figure 19: Quasi-energy levels of a near-resonance sinusoidally-driven TSS. Left: Energy diagram. For brevity and clarity we designate here $\omega_0 - \frac{\omega_A}{2} \equiv \omega_g$ ; $\omega_0 + \frac{\omega_A}{2} \equiv \omega_e$. Note that on-resonance ($\delta_C = 0$) each split is narrower and symmetrical (w.r.t. the respective unperturbed level) **[64]**. The wavy arrows show the spectral centers of the various fluorescent lines that together make a Mollow triplet. Similar*

*diagram is obtained by the fully quantum-mechanical dressed-atom model [50],[63]. Right: Quasi-energies of a driven TSS vs. detuning of the driving field. Left half – blue detuning, right half – red detuning. Frequency spacing between each pair of solid lines (red / blue) is the generalized Rabi flopping frequency $(\Omega_{GR,t})$. Notice that even on red detuning of the driving field, the two extreme quasi-energy levels are "outside" the energy levels (dashed black lines) of the unperturbed TSS.*

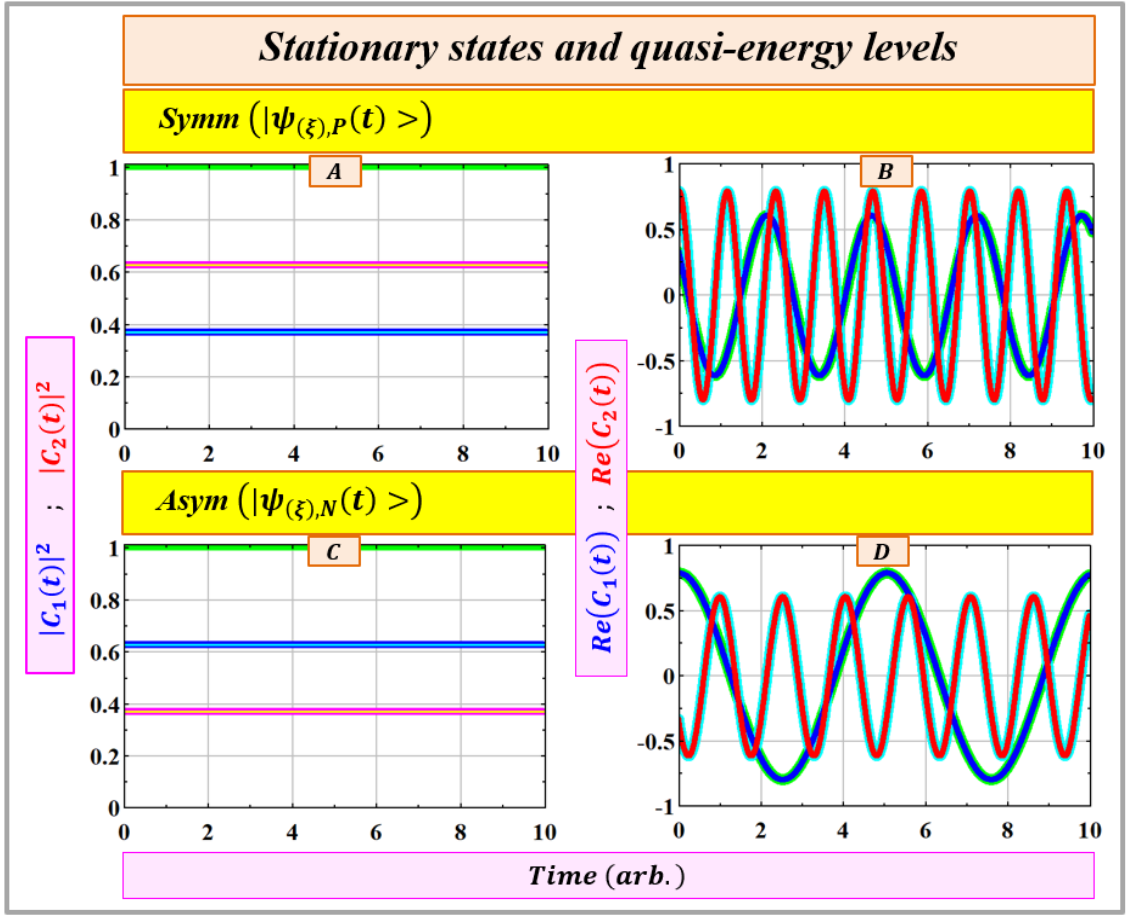

*Figure 20: Stationary states and quasi-energy levels of a sinusoidally-driven TSS. Left column (A,C): Stationary probabilities. A: System launched into the Symm $|\psi_{(\xi),P}(t)>$ eigenstate $(|C_1(0)| = |\xi_{P1}|, |C_2(0)| = |\xi_{P2}|)$. B: System launched into the Asym $|\psi_{(\xi),N}(t)$ eigenstate $(|C_1(0)| = |\xi_{N1}|, |C_2(0)| = |\xi_{N2}|$, cf. Eq. (130)). Right column (B,D): Real part of the probability amplitudes showing two pairs of phase rotational speeds. B: Time evolution of the probability amplitudes $(C_1(t), C_2(t))$ for a driven TSS launched into the **stationary** $|\psi_{(\xi),P}(t)>$ eigenstate. The time evolution of the probability amplitudes shows **two** distinct phase rotational speeds $\left(\omega_0 - \frac{\omega_c}{2} + \Omega_{P,t}; \omega_0 + \frac{\omega_c}{2} + \Omega_{P,t}\right)$. D: Same with **stationary** $|\psi_{(\xi),N}(t)>$ eigenstate $\left(\omega_0 - \frac{\omega_c}{2} - \Omega_{P,t}; \omega_0 + \frac{\omega_c}{2} - \Omega_{P,t}\right)$. Compare with the curves of **Figure 8** for a non-driven TSS showing two **degenerate** (same value) phase rotational speeds $((\omega_0 + \Omega_P) \text{ or } (\omega_0 - \Omega_P))$ for each of the stationary states.*

With the dynamics and some characteristics of a general abstract TDH-TSS at hand we proceed to examine three concrete driven TSSs.

## 3.2. A proton under a constant axial magnetic field and an harmonic transverse magnetic field

The first concrete TDH-TSS system we consider here is a proton in a constant axial magnetic field and an harmonic transverse magnetic field. The "driven-proton" TSS Hamiltonian fits the sinusoidally-driven TSS general Hamiltonian $\left(\left(\mathcal{H}_{TSS,t}\right)\right)$ of Eq. **(100)** (with $\omega_0 = 0$). Indeed, as stated in **[15]**, ***all** systems of two states can be made analogous to a spin one-half object precessing in a magnetic field* (including non-driven TSSs). Of course the driven-proton is at the heart of the highly successful MRI imaging technique **[4]**-**[9]**.

The two *independent* magnetic fields $(B_{const})$ and $(B_{osc}(t))$ are:

$$\boldsymbol{B}_{const} = (0,0,B_z) \; ; \; \boldsymbol{B}_{osc}(t) = (B_x \cdot cos(\omega_C \cdot t), 0, 0)$$

**(136)**

Magnetic field values taken for the simulation are the typical operating MRI value of the constant field (few Tesla) and $2 \cdot 10^5$ amplified typical value of the oscillating field (few micro-Tesla) to make the energy shifts "visible" **[93]**,**[6]**,**[5]**,**[7]**,**[9]**. The angular frequency of the driving field $(\omega_C)$ is selected (arbitrarily) to be close to the precession frequency ($2 \cdot \omega_{pr}$ - cf. Eq. **(52)** above). The calculated four quasi-energies are depicted by the panels of ***Figure 22***.

### 3.2.1. Basis states and time-dependent Hamiltonian

The canonical basis states for a proton in a constant axial magnetic field and in a sinusoidal transverse magnetic field ($|up>$ and $|dn>$) are:

$$|up> = \begin{pmatrix} 1 \\ 0 \end{pmatrix} \; ; \; |dn> = \begin{pmatrix} 0 \\ 1 \end{pmatrix}$$

**(137)**

The state vector $|\psi_M(t)>$ is then

$$|\psi_M(t)> = C_1(t) \cdot |up> + C_2(t) \cdot |dn>$$

**(138)**

Given the canonical basis, the Hamiltonian $\left(\left(\mathcal{H}_{\boldsymbol{B}(t)}\right)\right)$ for the proton in the magnetic fields (in angular frequency units) is **[15]**:

$$\left(\mathcal{H}_{\boldsymbol{B}(t)}\right) = \begin{pmatrix} -\frac{\mu_P \cdot B_z}{\hbar} & -\frac{\mu_P \cdot B_x \cdot cos(\omega_C \cdot t)}{\hbar} \\ -\frac{\mu_P \cdot B_x \cdot cos(\omega_C \cdot t)}{\hbar} & +\frac{\mu_P \cdot B_z}{\hbar} \end{pmatrix} \equiv \begin{pmatrix} -\frac{\omega_A}{2} & -\Omega_D \cdot cos(\omega_C \cdot t) \\ -\Omega_D \cdot cos(\omega_C \cdot t) & +\frac{\omega_A}{2} \end{pmatrix}$$

$$\omega_A \equiv \frac{2 \cdot \mu_P \cdot B_z}{\hbar} \cong 16 \cdot 10^8 \left(\frac{rad}{sec}\right)$$

$$B_x = G \cdot 3 \cdot 10^{-6} \, Tesla \; ; \; G = 2 \cdot 10^5 \, (set \, arbitrariliy)$$

$$\Rightarrow \Omega_D \equiv \frac{\mu_P \cdot B_x}{\hbar} = 0.8 \cdot 10^8 \left(\frac{rad}{sec}\right)$$

$$\delta_C = 0.06 \cdot \omega_A \Rightarrow \omega_C \cong 15.1 \cdot 10^8 \left(\frac{rad}{sec}\right)$$

(139)

Compared to the magnitude of the diagonal elements ($\omega_A/2$), the magnitude of the off-diagonal element is still rather small (smaller by a factor of $1/10$), despite our amplification of the amplitude of the typical transverse field by $2 \cdot 10^5$.

Given the Hamiltonian $\left(\mathcal{H}_{B(t)}\right)$ of Eq. **(139)** we have executed the rotate-solve-rotate back procedure of section **3.1** to find the values of the four quasi-energies for the driven-proton TSS (cf. ***Figure 22***).

### 3.2.2. Probability of the spin one-half particle to be in the $+x$ and or $-x$ state

The amplitude $A_+(t)$ to be in the $(+x)$ state at time $t$ is given by **[15]**:

$$A_+(t) = \frac{1}{\sqrt{2}} \cdot C_1(t) + \frac{1}{\sqrt{2}} \cdot C_2(t)$$

(140)

Similarly for the amplitude $A_-(t)$ to be in the $(-x)$ state at time $t$:

$$A_-(t) = \frac{1}{\sqrt{2}} \cdot C_1(t) - \frac{1}{\sqrt{2}} \cdot C_2(t)$$

(141)

where $\left(C_1(t); C_2(t)\right)$ are given by Eq. **(125)** or by Eq. **(127)**.

The probability $\left(\mathcal{P}_{(\pm x)}(t)\right)$ of the spin one-half particle to be in the $(\pm x)$ state, is then given explicitly in terms of the $(\mathcal{A}, \mathcal{B}, \mathcal{C}, \mathcal{D})$ parameters (cf. Eq. **(127)**) as:

$$U_1 \equiv \mathcal{A} \cdot \mathcal{B} + \mathcal{C} \cdot \mathcal{D}$$
$$U_{\pm 2} \equiv \pm(\mathcal{A} \cdot \mathcal{C} + \mathcal{B} \cdot \mathcal{D}) \; ; \; U_{\pm 3} \equiv \pm \mathcal{B} \cdot \mathcal{C} \; ; \; U_{\pm 4} \equiv \pm \mathcal{A} \cdot \mathcal{D}$$
$$cos_1 = cos(\Omega_{GR,t} \cdot t); \; cos_2 = cos(\omega_C \cdot t)$$
$$cos_3 = cos\left((\Omega_{GR,t} - \omega_C) \cdot t\right) \; ; \; cos_4 = cos\left((\Omega_{GR,t} + \omega_C) \cdot t\right)$$
$$\mathcal{P}_{(\pm x)}(t) = |A_\pm(t)|^2 = \frac{1}{2} + U_1 \cdot cos_1 + U_{\pm 2} \cdot cos_2 + U_{\pm 3} \cdot cos_3 + U_{\pm 4} \cdot cos_4$$

(142)

where $\Omega_{GR,t}$ is the generalized Rabi flopping frequency, given by Eq. **(120)**.

If the sinusoidally-driven TSS (of the spin one-half particle) is launched into one of its stationary states, then the probability expressions $\left(\mathcal{P}_{(\pm x)}(t)\right)$ are greatly simplified to

$$\left(\mathcal{P}_{(\pm x)}(t)\right)_P = \frac{1}{2} \pm \xi_{P1} \cdot \xi_{P2} \cdot cos(\omega_C \cdot t)$$

$$\left(\mathcal{P}_{(\pm x)}(t)\right)_N = \frac{1}{2} \pm \xi_{N1} \cdot \xi_{N2} \cdot cos(\omega_C \cdot t)$$

(143)

where $\xi_{P1}, \xi_{P2}, \xi_{N1}, \xi_{N2}$ are the components of the eigenvectors of the Hamiltonian in the rotating frame (cf. section **3.1.6**).

In another scenario, if the transverse field is a (strong) DC magnetic field ($\omega_C = 0$) then the probability $\left(\mathcal{P}_{(\pm x), \omega_C=0}(t)\right)$ of the spin one-half particle to be in the $(\pm x)$ state is given by the generalized Rabi flopping frequency as

$$\left(\mathcal{P}_{(\pm x), \omega_C=0}(t)\right) = \frac{1}{2} + U_{\pm 2} + (U_1 + U_{\pm 3} + U_{\pm 4}) \cdot cos(\Omega_{GR,t} \cdot t)$$

(144)

where the parameters of Eq. **(144)** are defined by Eq. **(142)** above.

A few comments are on order here.

- The "precession" of the spin one-half proton in a "driven-proton TSS" is not the familiar circular motion (making complete circles at a constant angle w.r.t. the $z$ axis). Rather, the "precession" is a *simultaneous* $x$ and $z$ "spiral" motion (cf. *Figure 21* below). Hence we enclose "precession" in quotation marks.

- The frequency of "precession" (of the spin one-half particle) is dominated by $\omega_C$ in the general initial conditions of the TSS (Eq. **(142)** and cf. *Figure 21* below), and is exactly $\omega_C$ if the TSS is launched into one of its stationary states ((Eq. **(143)** and cf. *Figure 21* below). The spin one-half particle's "precession" frequency then is *not* the gap energy ($\omega_A$) but rather (nearly) the frequency of the driving field ($\omega_C$). This "precession" frequency is another manifestation of the pull/push of the energy levels (of the unperturbed TSS) by the sinusoidal driving field.

- If the coupling transverse field is a constant field ($\omega_C = 0$) then the spin one-half particle's "precession" frequency is the generalized Rabi flopping frequency $(\Omega_{GR,t})$ in the general initial conditions case. If the TSS is launched into one of its eigenstates (and $\omega_C = 0$) then particle's precession ceases altogether.

- A Fourier transform of the two amplitudes involved $(C_1(t), C_2(t))$ will show the four quasi-energies of the driven TSS. Fourier transform of $cos_3, cos_4$ (Eq. **(142)** above) with half the argument will yield the four quasi-energies: $\pm \frac{1}{2} \cdot (\Omega_{GR,t} - \omega_C)$ and $\pm \frac{1}{2} \cdot (\Omega_{GR,t} + \omega_C)$.

The two panels on the right column of *Figure 21* show the probabilities of the (spin one-half) proton to be in the $(\pm x)$ state (Eqs. **(142)**,**(143)**).

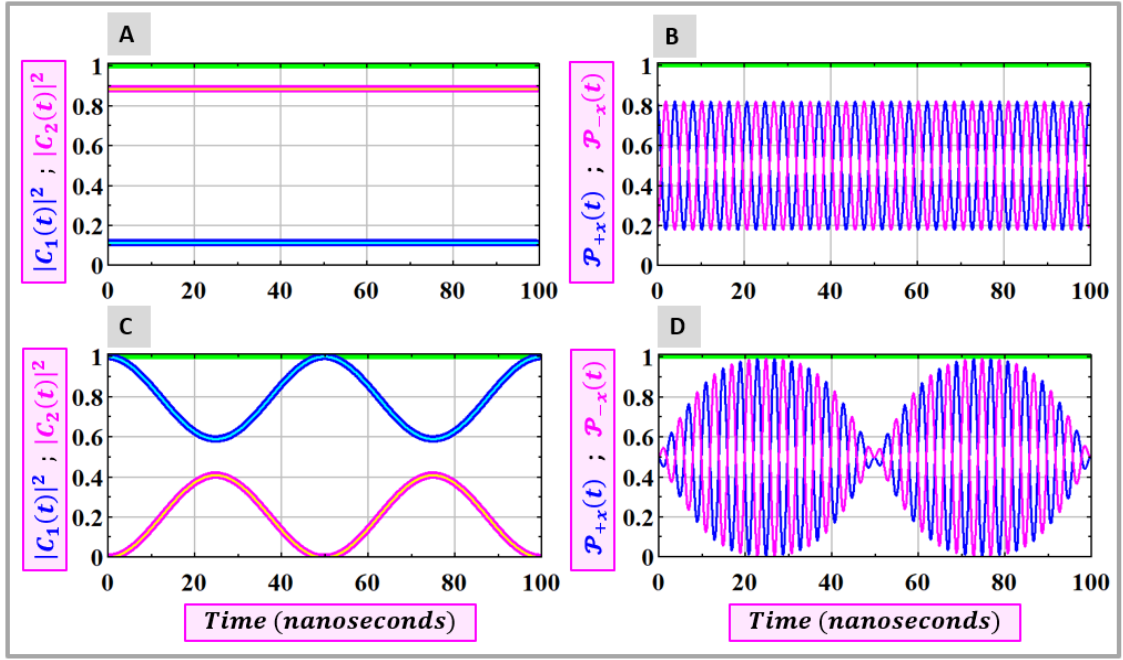

*Figure 21: Occupation probabilities of states of a driven spin one-half particle TSS. Left column: state-occupation probabilities in the canonical ("z") basis. Right column: state-occupation probabilities in the ($\pm x$) basis. A. Stationary state (Symm). Probabilities of finding the system in the "up" / "dn" states are independent of time. B. The ($\pm x$) probabilities $\left(\mathcal{P}_{(\pm x)}(t)\right)$ when the TSS is in the Symm stationary state. The two curves show "conical" precession at a fixed frequency equals to the frequency ($\omega_C$) of the driving field (and **not** to the frequency of the energy gap of the unperturbed system ($\omega_A$) – Eq. **(143)**). In a constant magnetic field ($\omega_C = 0$) precession frequency is (possibly slows down to) the generalized Rabi flopping frequency $\left(\Omega_{GR,t} \xrightarrow{\omega_C=0} \Omega_{GR}\right)$. C. General state of the TSS. The "up"/"dn" occupation probabilities oscillate at the generalized Rabi flopping frequency (note that the amplitude of the typical driving field is strongly amplified ($G = 2 \cdot 10^5$)). D. The ($\pm x$) probabilities $\left(\mathcal{P}_{(\pm x)}(t)\right)$ when the TSS is launched into a general ("z") state (as in panel C). "Precession" of the spin one-half particle is **nearly** at the frequency of the driving field ($\omega_C$) (Eq. **(142)**), but the peaks of probability amplitudes gradually change from $1/2$ (no precession) to nearly $1$ (the spin one-half particle is almost at the x-y plane). In a constant magnetic coupling field ($\omega_C = 0$), at all strengths down to zero ($\Omega_D \geq 0$), "precession" frequency is the generalized Rabi flopping frequency (Eq. **(142)**). The shown $\mathcal{P}_{(\pm x)}(t)$ curves (being the sum of **amplitudes** ($C_1(t)$ ; $C_2(t)$) prior to squaring) constitute another manifestation of the exitance of four quasi-energies in a near-resonance sinusoidally-driven TSS.*

Consulting ***Figure 21***, we see the difference between the "z" probabilities vs. the "x" probabilities in terms of the quantum picture. The "z" probabilities (($|C_1(t)|^2, |C_2(t)|^2$) - left column of ***Figure 21***) are independent of the back rotation. That is – their value in the rotating frame is equal to their value in the Schrödinger picture (the picture following the back-rotation). The "x" probabilities on the other hand ($\mathcal{P}_{(\pm x)}(t)$ - right column of ***Figure 21***) are seen to be radically different in each of the "frames". In the rotating frame (before the back rotation) the spin one-half particle's "precession" *is independent of the frequency* ($\omega_C$) *of the driving magnetic field*. As is the case with the four quasi-energies, it is only back in the Schrödinger picture (no frame rotation) that the complete TSS physics description (four quasi-energies as well as spin "spiral" precession) is realized.

### 3.2.3. Quasi-energies

Calculated quasi-energies for the driven-proton TSS (typical MRI parameters with $2 \cdot 10^5$ amplification of the oscillating transverse field) are depicted by the panels of *Figure 22*.

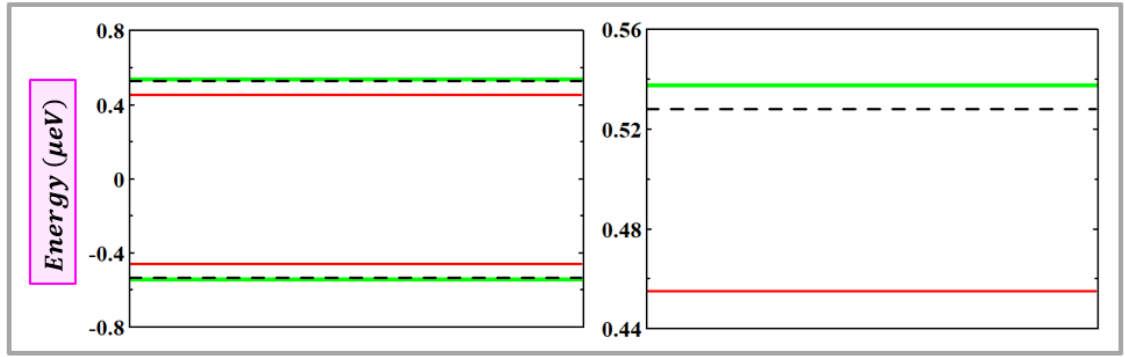

*Figure 22: Calculated four quasi-energies of a driven-proton under a constant magnetic field of* $3\ Tesla$ *and* $\delta_C = 0.06 \cdot \omega_A$. *The dashed black lines designate the energy levels of an un-driven proton (but still in a constant axial magnetic field). Left: Full view. The energy gap between the two pairs is about* $1\ \mu eV$. *Right: Zoom to the upper pair of energy levels. The shown doublet split of* $0.083\ \mu eV$ *(the gap between red and green) is somewhat exaggerated. The typical doublet split (no amplification of the transverse field) is* $\approx 0.063\ \mu eV$. ***On-resonance*** *the doublet splitting (for the assumed typical amplitude of the driving field) is* $2.6 \cdot 10^{-7}\ \mu eV$.

## 3.3. A cesium atom driven near resonance by an external harmonic magnetic field

The two quantum states of an atom to be considered here are the ground state hyperfine split $6^2S_{1/2}$ of the cesium atom. The resonance transition frequency of 9.192 631 770 GHz between these two hyperfine levels is the basis for the 1967 International System of Units definition of a second **[113]**. These two levels are taken to be coupled by a near-resonance sinusoidal radio frequency electromagnetic field. Both states have zero orbital angular momentum ($L = 0$), so transitions between them are not electric-dipole allowed. The magnetic dipole approximation (interaction energy $-\mu \cdot B$) is therefore relevant **[108]**. We refer to this TSS as a "cesium clock" TSS.

### 3.3.1. Basis states and time-dependent Hamiltonian

The canonical basis states for the two ground state hyperfine transition of a cesium atom ("cesium clock" transition) are designated $|g>$ and $|e>$:

$$|g> = \begin{pmatrix} 1 \\ 0 \end{pmatrix} \quad ; \quad |e> = \begin{pmatrix} 0 \\ 1 \end{pmatrix}$$

(145)

The state vector $|\psi_{Cs}(t)>$ is then

$$|\psi_{Cs}(t)> = C_1(t) \cdot |g> + C_2(t) \cdot |e>$$

(146)

The Hamiltonian for the cesium clock under microwave coupling field (in $\hbar$ units) reads:

$$(\mathcal{H}_{Cs}) = \begin{pmatrix} -\dfrac{\omega_{eg}}{2} & |\Omega_{eff}| \cdot e^{-j \cdot \phi_D} \cdot \cos(\omega_C \cdot t) \\ |\Omega_{eff}| \cdot e^{+j \cdot \phi_D} \cdot \cos(\omega_C \cdot t) & +\dfrac{\omega_{eg}}{2} \end{pmatrix}$$

$$\omega_{eg} \equiv \omega_e - \omega_g = 2 \cdot \pi \cdot 9.192631770 \cdot 10^9 \left(\dfrac{rad}{sec}\right)$$

$$|\Omega_{eff}| = G \cdot 2 \cdot \pi \cdot 5 \cdot 10^4 \left(\dfrac{rad}{sec}\right) \; ; \; G = 10^4 \text{ (set arbitrarily)} \; ; \; \phi_D = 0$$

$$\delta_C = 0.06 \cdot \omega_{eg} \Rightarrow \omega_C \cong 3.47 \cdot 10^9 \left(\dfrac{rad}{sec}\right)$$

(147)

Note that we have set the average energy level to zero $\left(\omega_0 \equiv \dfrac{\omega_e + \omega_g}{2} \to 0\right)$. The value of the diagonal element $(\omega_{eg})$ is the natural frequency of the cesium clock (in radians/sec). The value of the magnitude of off-diagonal element $(|\Omega_{eff}|)$ is the frequency of an operating cesium clock (in radians/sec) **[114]** multiplied by an arbitrary "amplifying" factor $(G)$ to make the energy shifts "visible". For the cesium clock simulation (***Figure 24***) we have set $G = 10^4$.

If the cesium clock TSS is initiated into the $|g>$ state $(C_1(0) = 1 \; ; \; C_2(0) = 0)$, then the state-occupation probabilities will oscillate as shown by the curves of ***Figure 23*** for an on-resonance driven TSS.

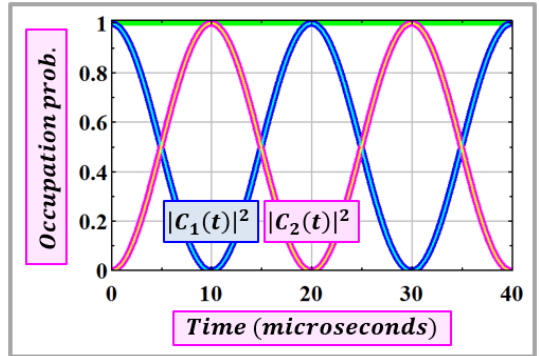

*Figure 23: Oscillations of the state-occupation probabilities for an on-resonance-driven cesium clock TSS. In the shown case the TSS was launched into the lower energy $|g>$ state $(C_1(0) = 1 \; ; \; C_2(0) = 0)$. In a casual phrasing – "clock's state populations 'slowly' oscillate". The state-occupation probabilities can be viewed as the time-dependent probabilities of consecutive absorption and stimulated emission of energy from the driving field and back to the driving field by the two states* **[48]**.

But if the TSS is launched into one of its two stationary states, then the state-occupation probabilities are time-independent.

### 3.3.2. Quasi-energies

Calculated quasi-energies for a strongly driven ($G = 10^4$) cesium clock TSS are depicted by the panels of *Figure 24*.

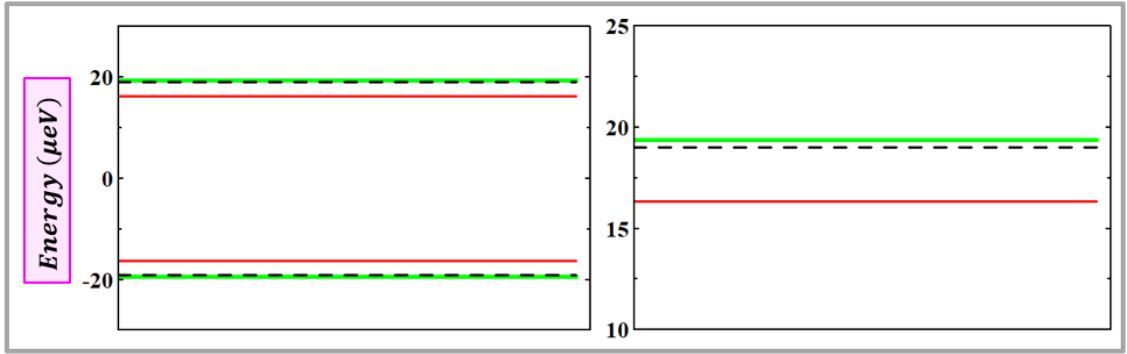

*Figure 24:* Calculated four quasi-energies for a strongly driven ($G = 10^4$) cesium clock TSS with $\delta_C = 0.06 \cdot \omega_{eg}$. The dashed black lines designate the energy levels of an un-driven cesium clock TSS. Left: Full view. The energy gap between the two pairs is about $38 \, \mu eV$. Right: Zoom to the upper pair of energy levels. The shown doublet split of $3.1 \, \mu eV$ (the gap between red and green) is rather exaggerated. The typical doublet split (no amplification of the transverse field) is $\approx 2.3 \, \mu eV$. **On-resonance** the doublet splitting (for the assumed typical amplitude of the driving field) is $2.1 \cdot 10^{-4} \, \mu eV$.

### 3.4. An ammonia molecule in an harmonic electric field

The two quantum states of an ammonia molecule ($NH_3$) to be considered here are the two quantum states associated with the energy levels of a symmetric (low) and antisymmetric (high) *superposition eigenstate*s of the molecule [98]. The energy gap ($2 \cdot A$) between these two states is (cf. Eq. (69)) $2 \cdot A = 23.786 \, GHz \rightarrow \cong 98.4 \, \mu eV$. These two basis states, the states of the free ammonia molecule, are taken to be coupled by a sinusoidal electric field (as is the case of the ammonia maser). Typical magnitude of the amplitude of the coupling sinusoidal electric field ($E_0$) is $E_0 = 2.36 \cdot 10^{-2} \, \frac{Volt}{meter}$ leading to an interaction energy amplitude ($\omega_D \equiv \mu_E \cdot E_0$) of $\omega_D = 2.82 \cdot 10^3 \, \frac{radians}{second} \rightarrow \cong 1.86 \cdot 10^{-12} \, eV$ [20]. Interaction time is assumed to be very long (to allow for a large number of Rabi cycles).

The interaction energy of the driving field with the electric dipole moment of $NH_3$ appears as the diagonal elements of the TDH-TSS Hamiltonian [15],[20]. Therefore, the solution to the dynamics of the *driven-ammonia-molecule* TSS requires a preparation step to bring the Hamiltonian to the general form as in Eq. (113) (interaction energy as the off-diagonal elements).

Below we perform the preparation step (switch to the eigenstates basis) and proceed according the recipe of section 3.1 to calculate the four quasi-energy levels of the driven-ammonia-molecule TSS.

### 3.4.1. Basis states and time-dependent Hamiltonian

The canonical basis states for the driven-ammonia-molecule TSS are $|1>$ and $|2>$:

$$|1> = \begin{pmatrix}1\\0\end{pmatrix} \; ; \; |2> = \begin{pmatrix}0\\1\end{pmatrix}$$

(148)

The state vector $|\psi_{12,free}(t)>$ is then

$$|\psi_{12,free}(t)> = C_{1,free}(t) \cdot |1> + C_{2,free}(t) \cdot |2>$$

(149)

The Hamiltonian for the *free* ammonia molecule in the $|1>, |2>$ basis is (in $\hbar$ units):

$$(\mathcal{H}_{12,free}) = \omega_0 \cdot (I) + \begin{pmatrix} 0 & -\frac{\omega_A}{2} \\ -\frac{\omega_A}{2} & 0 \end{pmatrix}$$

$$\omega_A = 2 \cdot \pi \cdot 23.786 \cdot 10^9 \left(\frac{rad}{sec}\right) \; ; \; \omega_0 = 1.2 \cdot \frac{\omega_A}{2} \text{ (set arbitrarily)}$$

(150)

Eigenvectors of the $(\mathcal{H}_{12,free})$ Hamiltonian are:

$$\boldsymbol{\eta_P} = \frac{1}{\sqrt{2}} \cdot \begin{pmatrix}-1\\1\end{pmatrix} \; ; \; \boldsymbol{\eta_N} = \frac{1}{\sqrt{2}} \cdot \begin{pmatrix}1\\1\end{pmatrix} \; ; \; (\eta) = (\eta)^{-1} = (\boldsymbol{\eta_P}, \boldsymbol{\eta_N})$$

(151)

The state vector $|\psi_{\eta,free}(t)> \; (= |\psi_{12,free}(t)>)$ is

$$|\psi_{\eta,free}(t)> = \eta_{P,free}(t) \cdot |\boldsymbol{\eta_P}> + \eta_{N,free}(t) \cdot |\boldsymbol{\eta_N}>$$

(152)

The relations between the two amplitude sets is (cf. Eq. (32)):

$$\begin{pmatrix}\eta_{P,free}(t)\\\eta_{N,free}(t)\end{pmatrix} = (\eta)^{-1} \cdot \begin{pmatrix}C_{1,free}(t)\\C_{2,free}(t)\end{pmatrix} = \begin{pmatrix}-C_{1,free}(t) + C_{2,free}(t)\\C_{1,free}(t) + C_{2,free}(t)\end{pmatrix}$$

(153)

Note that $\eta_{P,free}(t)$ is an antisymmetric superposition of the $\boldsymbol{C}_{free}(t)$ amplitudes, associated with the higher eigenvalue of the $(\mathcal{H}_{12,free})$ Hamiltonian, i.e. associated with the upper energy level of the ammonia molecule. The $\eta_{N,free}(t)$ is a symmetric superposition of the $\boldsymbol{C}(t)$ amplitudes, associated with the lower eigenvalue of the $(\mathcal{H}_{12,free})$ Hamiltonian, i.e. associated with the lower energy level of the ammonia molecule.

The Hamiltonian for the *driven*-ammonia-molecule TSS in the $|1>, |2>$ basis is:

$$\left(\mathcal{H}_{12,AC}(t)\right) = \omega_0 \cdot (I) + \begin{pmatrix} -G \cdot \omega_D \cdot \cos(\omega_C \cdot t) & -\frac{\omega_A}{2} \\ -\frac{\omega_A}{2} & +G \cdot \omega_D \cdot \cos(\omega_C \cdot t) \end{pmatrix} \; ; \; \omega_D \in \mathbb{R}; \; \omega_D \geq 0$$

$$\omega_D \equiv \mu_E \cdot E_0 = 2.82 \cdot 10^3 \left(\frac{rad}{sec}\right) \; ; \; G = 2 \cdot 10^6 \text{ (set arbitrarily)}$$

(154)

The value of the ($\omega_D$) parameter in the diagonal element of the Hamiltonian **(154)** is the product of the known electric dipole moment of $NH_3$ ($\mu_E$) and an estimated magnitude of the amplitude of the coupling sinusoidal electric field ($E_0$) inside an $NH_3$ maser cavity **[20]**. The parameter $G$ is an arbitrary "amplifying" factor to make the energy splits "visible". For the driven-ammonia simulation (*Figure 25*) we have set $G = 2 \cdot 10^6$. The value of (twice) the off-diagonal element ($\omega_A$) (cf. Eq. **(150)**) is the tabulated angular frequency of the antisymmetric-symmetric transition of the ammonia molecule (cf. Eq. **(69)**).

The Hamiltonian for the *driven-ammonia-molecule* in the ($\boldsymbol{\eta}_P, \boldsymbol{\eta}_N$) basis is:

$$\left(\mathcal{H}_{\eta,AC}(t)\right) = (\eta)^{-1} \cdot \left(\mathcal{H}_{12,AC}(t)\right) \cdot (\eta)$$

$$= \omega_0 \cdot (I) + \begin{pmatrix} +\frac{\omega_A}{2} & -G \cdot \omega_D \cdot cos(\omega_C \cdot t) \\ -G \cdot \omega_D \cdot cos(\omega_C \cdot t) & -\frac{\omega_A}{2} \end{pmatrix}$$

(155)

The $\delta_C$ parameter in the rotation matrix $\left(R_\omega(t)\right)$ of Eq. **(115)** is taken to be

$$\delta_C = 0.06 \cdot \omega_A$$

(156)

Given the Hamiltonian $\left(\mathcal{H}_{\eta,AC}(t)\right)$ of Eq. **(155)** we have executed the rotate-solve-rotate back procedure of section **3.1** to find the values of the four quasi-energies for the driven-ammonia-molecule TSS (cf. *Figure 25*).

### 3.4.2. Quasi-energies

Calculated quasi-energies for a strongly-driven ammonia-molecule TSS are depicted by the three panels of *Figure 25*.

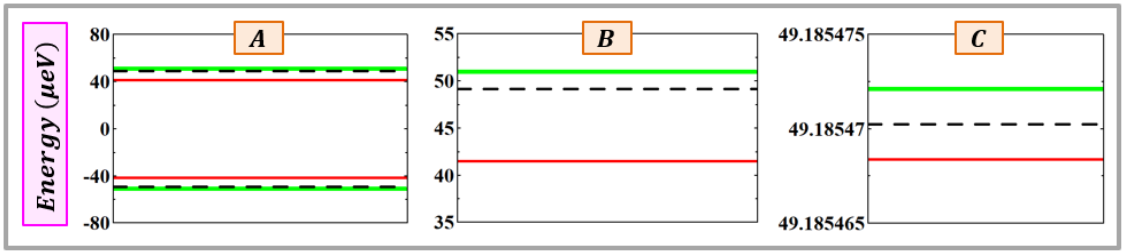

*Figure 25:* Calculated four quasi-energies for a strongly-driven ammonia-molecule TSS ($G = 2 \cdot 10^6$) with $\delta_C = 0.06 \cdot \omega_A$. The dashed black lines designate the energy levels of an un-driven ammonia molecule. A: Full view. The energy gap between the two pairs is about $98.4 \ \mu eV$. B: Zoom to the upper pair of energy levels. The shown doublet split of $9.5 \ \mu eV$ (the gap between red and green) is greatly exaggerated. The typical doublet split (no amplification of the transverse field) is $\approx 5.9 \ \mu eV$. C: **On-resonance** doublet splitting (for the assumed typical amplitude of the driving field) is $3.7 \cdot 10^{-6} \ \mu eV$.

With characterization of the driven-ammonia-molecule TSS our discussion of both non-driven and driven TSSs comes to an end. To summarize these two chapters – for the case of non-driven TSSs we have shown the existence of two stationary states each associated with a single definite energy. Similarly, we have shown two stationary states in the case of driven-TSSs, but this time each stationary states is associated with two definite "quasi-energies".

Next, in an important section of our present contribution, we provide through a set of two coupling-probe computer simulations, an "experimental" proof to the two-level shift and split into four quasi-energies for an atom TSS driven by near-resonance coherent light.

## 4. Coupling-probe computer simulations

The present "coupling-probe computer simulations" section is dedicated to analyzing the state-occupation dynamics of specific three energy states of a single static isolated atom (experimental realizations possible **[61]**) in the presence of two coupling lasers. The objective of the coupling-probe simulations is better understanding, and actually providing an "experimental" proof, of the shift and doublet-splitting of each of two specific energy levels, once the atom is exposed to coherent light oscillating with frequency near the two-levels' resonance.

We consider, as already stated, three non-degenerate energy eigenstates of an isolated atom. The three energy levels are the lower ("ground") state ($|g>$), a mid-level ("excited") state ($|e>$) and an upper near-ionization ("Rydberg") state ($|r>$). The energies of each of these levels in units of angular frequencies are $(\omega_g; \omega_e; \omega_r)$ respectively (cf. *Figure 26*).

The atom is assumed to be simultaneously exposed to two monochromatic lasers, each radiating electric (and magnetic) field in the form of an ideal planewave (constant and infinite amplitude in space and time). The first of the two lasers is a strong "coupling laser" characterized by an amplitude ($E_{0C}$) and angular frequency ($\omega_C$) near the resonance of the two lower levels ($|g>$ and $|e>$). The second of the two lasers is a weak "probing laser" characterized by an amplitude ($E_{0PE}$) or ($E_{0PG}$) and angular frequency ($\omega_{PE}$) or ($\omega_{PG}$). The effect of spontaneous emission is not included in the presented semiclassical treatment of the atom-field interaction **[55]**.

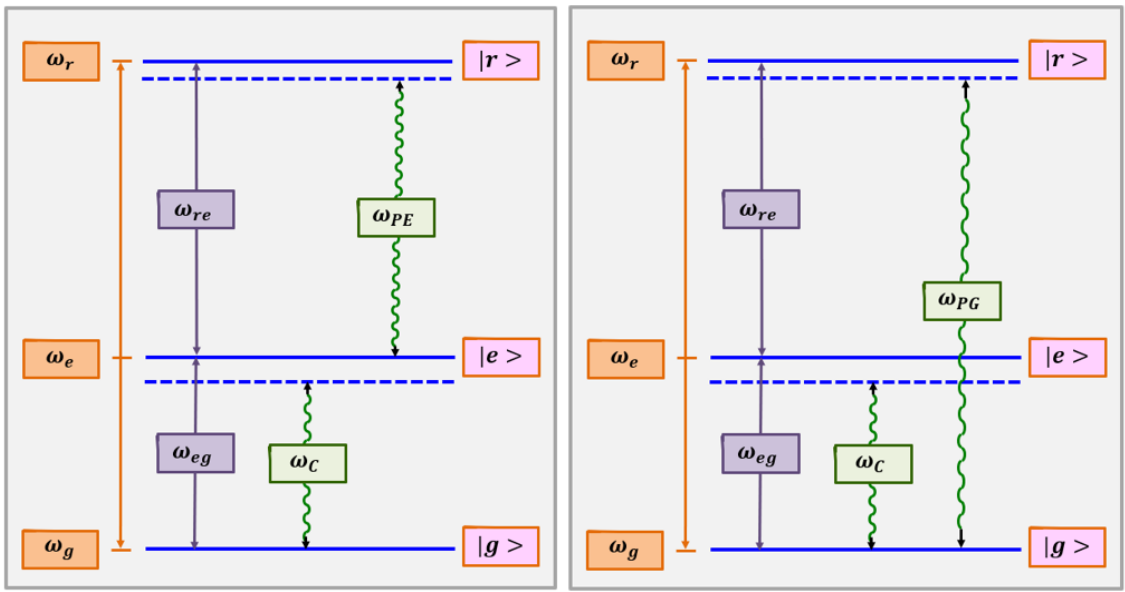

*Figure 26: Coupling-probe simulations. Left: Probing the mid-level $|e>$ state. Right: Probing the lower-level $|g>$ state. The two lower levels, creating a TSS, are coupled near-resonance by a strong monochromatic electromagnetic planewave.*

We simulate two probing scenarios:

- **Probing mid-level ($|e>$) to top level**
- **Probing low level ($|g>$) to top level**

Explicitly, the electromagnetic planewaves participating in each of the two scenarios are:

$$E_C(t) = E_{0C} \cdot cos(\omega_C \cdot t) \ ; \ E_{PE}(t) = E_{0PE} \cdot cos(\omega_{PE} \cdot t)$$
$$E_C(t) = E_{0C} \cdot cos(\omega_C \cdot t) \ ; \ E_{PG}(t) = E_{0PG} \cdot cos(\omega_{PG} \cdot t)$$

(157)

The resulting off-diagonal dipole-coupling elements are $\mathcal{D}_C \cdot cos(\omega_C \cdot t)$ coupling the $(g,e)$ levels, $\mathcal{D}_{PE} \cdot cos(\omega_{PE} \cdot t)$ probing the $(e,r)$ levels, and $\mathcal{D}_{PG} \cdot cos(\omega_{PG} \cdot t)$ probing the $(g,r)$ levels. The two scenarios are illustrated by the panels of ***Figure 26***.

### 4.1. Probing mid-level ($|e>$) to top level

Let's start the TSS dynamics simulation by probing the $|e>$ to $|r>$ transition (cf. the left panel of ***Figure 26***).

In the canonical basis representation (the eigenstates of the unperturbed Hamiltonian) the unperturbed Hamiltonian operator $((\mathcal{H}_0))$ reads:

$$(\mathcal{H}_0) = \begin{pmatrix} \omega_g & 0 & 0 \\ 0 & \omega_e & 0 \\ 0 & 0 & \omega_r \end{pmatrix}$$

(158)

The interaction energy operator $((\mathcal{V})_{er})$ holds off-diagonal elements only:

$$(\mathcal{V})_{er} = \begin{pmatrix} 0 & \mathcal{D}_C \cdot cos(\omega_C \cdot t) & 0 \\ \mathcal{D}_C^* \cdot cos(\omega_C \cdot t) & 0 & \mathcal{D}_{PE} \cdot cos(\omega_{PE} \cdot t) \\ 0 & \mathcal{D}_{PE}^* \cdot cos(\omega_{PE} \cdot t) & 0 \end{pmatrix}$$

(159)

The perturbed-system Hamiltonian operator $((\mathcal{H}_{PRT})_{er})$ is the sum $(\mathcal{H}_{PRT})_{er} = (\mathcal{H}_0) + (\mathcal{V})_{er}$:

$$(\mathcal{H}_{PRT})_{er} = \begin{pmatrix} \omega_g & \mathcal{D}_C \cdot cos(\omega_C \cdot t) & 0 \\ \mathcal{D}_C^* \cdot cos(\omega_C \cdot t) & \omega_e & \mathcal{D}_{PE} \cdot cos(\omega_{PE} \cdot t) \\ 0 & \mathcal{D}_{PE}^* \cdot cos(\omega_{PE} \cdot t) & \omega_r \end{pmatrix}$$

(160)

Define now a unitary rotation operator $(R_{GER}(t))$ to rotate the frame (and aim to obtain a TIH):

$$(R_{GER}(t)) = \begin{pmatrix} e^{-\mathbb{j} \cdot \omega_G \cdot t} & 0 & 0 \\ 0 & e^{-\mathbb{j} \cdot \omega_E \cdot t} & 0 \\ 0 & 0 & e^{-\mathbb{j} \cdot \omega_R \cdot t} \end{pmatrix} ; (R_{GER}(t))^\dagger = \begin{pmatrix} e^{+\mathbb{j} \cdot \omega_G \cdot t} & 0 & 0 \\ 0 & e^{+\mathbb{j} \cdot \omega_E \cdot t} & 0 \\ 0 & 0 & e^{+\mathbb{j} \cdot \omega_R \cdot t} \end{pmatrix}$$

(161)

Note that $\omega_G, \omega_E, \omega_R$ are at this stage general phase rotational speeds and thus $R_{GER}(t)$ is a general rotation operator and is *not* the time-evolution operator for the unperturbed Hamiltonian (leading to a rotating frame but *not* to the interaction picture **[110]**). The values of the three rotational speeds $(\omega_G, \omega_E, \omega_R)$ are to be set later-on.

Anticipating the adoption of the rotating wave approximation we designate an Hamiltonian operator in the rotating frame as $(\mathcal{H}^{er}_{ROT,RWA})$ and write **[110]**:

$$(\mathcal{H}^{er}_{ROT,RWA}) = (R_{GER}(t))^\dagger \cdot (\mathcal{H}_{PRT}(t))_{er} \cdot (R_{GER}(t)) - \mathbb{j} \cdot (R_{GER}(t))^\dagger \cdot \frac{d(R_{GER}(t))}{dt}$$

**(162)**

The origin of the energy ladder is set midway between $\omega_g$ and $\omega_e$ so that

$$\omega_g = -\frac{\omega_{eg}}{2} \quad ; \quad \omega_e = +\frac{\omega_{eg}}{2} \quad ; \quad \omega_r = \frac{\omega_{eg}}{2} + \omega_{re}$$

**(163)**

Executing Eq. **(162)**, adopting the RWA, and setting the values of the three rotational speeds $(\omega_G, \omega_E, \omega_R)$ as -

$$\omega_E = +\frac{\omega_C}{2} \quad ; \quad \omega_E - \omega_G = \omega_C \quad ; \quad \omega_R - \omega_E = \omega_{PE}$$

**(164)**

we arrive at a TIH in the rotating frame:

$$(\mathcal{H}^{er}_{ROT,RWA}) = \begin{pmatrix} -\frac{\delta_C}{2} & \frac{1}{2} \cdot \mathcal{D}_C & 0 \\ \frac{1}{2} \cdot \mathcal{D}_C^* & \frac{\delta_C}{2} & \frac{1}{2} \cdot \mathcal{D}_{PE} \\ 0 & \frac{1}{2} \cdot \mathcal{D}_{PE}^* & \delta_{Pe} + \frac{\delta_C}{2} \end{pmatrix}$$

$$\delta_C \equiv \omega_{eg} - \omega_C \quad ; \quad \delta_{Pe} \equiv \omega_{re} - \omega_{PE}$$

**(165)**

In the "experiment", it is possible to vary the entries of the rotating-frame Hamiltonian of Eq. **(165)** by varying either one of four parameters. The first parameter-pair is the two coupling field amplitudes - $E_{0C}$ and $E_{0PE}$ (cf. Eq. **(157)**) effecting $\mathcal{D}_C$ and $\mathcal{D}_{PE}$ respectively. The second parameter-pair is the two frequencies of the coupling fields - $\omega_C$ and $\omega_{PE}$ effecting the three diagonal elements.

Due to the coupling fields, the states are no longer the eigenstates of the unperturbed (TIH) Hamiltonian. The states *exchange energies with the driving fields* **[61]** and are thus not well energy-defined. Some researchers refer to the energies of these states in the presence of oscillating electric fields as *quasi-energies* **[65],[61],[50],[66],[67]**.

Designating the states of the unperturbed Hamiltonian (Eq. **(158)**) as $|g>, |e>, |r>$, the time-dependent state function $(|\psi_{ROT}(t)>)$ is given by the superposition:

$$|\psi_{ROT}(t)> = C_g(t) \cdot |g> + C_e(t) \cdot |e> + C_r(t) \cdot |r>$$

**(166)**

Defining a column vector $\boldsymbol{C}_R(t) = [C_g(t), C_e(t), C_r(t)]^T$, the Schrödinger equation for the *time-dependent* probability amplitudes reads ($\hbar = 1$):

$$\frac{d\boldsymbol{C}_R(t)}{dt} = -\mathbb{j} \cdot \left(\mathcal{H}^{er}_{ROT,RWA}\right) \cdot \boldsymbol{C}_R(t)$$

(167)

The standard matrix-multiplication solution **[89],[90]** to the coupled equations **(167)** reads:

$$\boldsymbol{C}_R(t) = (\boldsymbol{\xi}_R) \cdot \left(I_{\Lambda R}(t)\right) \cdot (\boldsymbol{\xi}_R)^{-1} \cdot (\boldsymbol{C}_R(0))$$

(168)

where $(\boldsymbol{\xi}_R)$ is the "eigenvectors matrix" (eigenvectors of the $\left(\mathcal{H}^{er}_{ROT,RWA}\right)$ Hamiltonian operator – Eq. **(165)**) and $\left(I_{\Lambda R}(t)\right)$ is the "eigenvalues matrix" (phase-rotation exponentials of).

In the "experiment", the three probability amplitudes $\left(\boldsymbol{C}_R(t)\right)$ can be affected by **eight independent parameters** – by the four parameters of the Hamiltonian **(165)** as explained above, and by additional four parameters dictating the *initial* state-occupation values of each of the three states (two independent absolute values and two relative phases).

### 4.2. Probing low level ($|g>$) to top level

Next, through a similar procedure, we study TSS dynamics by probing the $|g>$ to $|r>$ transition (cf. the right panel of ***Figure 26***).

In this second probing scenario the perturbed Hamiltonian reads:

$$(\mathcal{H}_{PRT})_{gr} = \begin{pmatrix} \omega_g & \mathcal{D}_C \cdot cos(\omega_C \cdot t) & \mathcal{D}_{PG} \cdot cos(\omega_{PG} \cdot t) \\ \mathcal{D}_C^* \cdot cos(\omega_C \cdot t) & \omega_e & 0 \\ \mathcal{D}_{PG}^* \cdot cos(\omega_{PG} \cdot t) & 0 & \omega_r \end{pmatrix}$$

(169)

Executing Eq. **(162)** with $(\mathcal{H}_{PRT})_{gr}$, adopting the RWA, and setting the values of the three phase rotational speeds $(\omega_G, \omega_E, \omega_R)$ as -

$$\omega_G = -\frac{\omega_C}{2} \; ; \; \omega_E - \omega_G = \omega_C \; ; \; \omega_R - \omega_G = \omega_{PG}$$

(170)

we arrive at a TIH in the rotating frame:

$$\left(\mathcal{H}^{gr}_{ROT,RWA}\right) = \begin{pmatrix} -\frac{\delta_C}{2} & \frac{1}{2} \cdot \mathcal{D}_C & \frac{1}{2} \cdot \mathcal{D}_{PG} \\ \frac{1}{2} \cdot \mathcal{D}_C^* & \frac{\delta_C}{2} & 0 \\ \frac{1}{2} \cdot \mathcal{D}_{PG}^* & 0 & \delta_{Pg} - \frac{\delta_C}{2} \end{pmatrix}$$

$$\delta_C \equiv \omega_{eg} - \omega_C \; ; \; \delta_{Pg} \equiv \left(\omega_{eg} + \omega_{re}\right) - \omega_{PG}$$

(171)

Notice the *plus* sign of $\frac{\delta_C}{2}$ in $(\mathcal{H}^{er}_{ROT,RWA})[3,3]$ of Eq. **(165)** probing the $|e>$ state vs. the *minus* sign of $\frac{\delta_C}{2}$ in $(\mathcal{H}^{gr}_{ROT,RWA})[3,3]$ of Eq. **(171)** probing the $|g>$ state.

### 4.3. TSS level-occupation dynamics

We proceed now, as planned, to studying the level occupation dynamics of the lower two states (the TSS, ignoring all other states of the atom **[40]**) as altered by the presence of the strong, near resonance coupling field $E_C(t) = E_{0C} \cdot cos(\omega_C \cdot t)$. We consider the following three different initial system settings (for each of the two probing scenarios):

A. **Eigenstates-initiated TSS; Coupling field at resonance ($\boldsymbol{\delta_C = 0}$)** (two independent simulation runs).

B. **Eigenstates-initiated TSS; Red/Blue detuning of the coupling field ($\boldsymbol{\pm|\delta_C| \neq 0}$)** (four independent simulation runs).

C. **General initial state of the TSS; Red detuning of the coupling field ($\boldsymbol{\delta_C > 0}$)** (a single simulation run).

In addition, in a separate sub-section, we briefly discuss

D. **Upper level line-widths**

Let's start with the on-resonance TSS-coupling simulation runs.

A. **Eigenstates-initiated TSS; Coupling field at resonance ($\boldsymbol{\delta_C = 0}$).**

In the rotating frame, the $2x2$ Hamiltonian operator $\left((\mathcal{H}^{2x2}_{ROT,RWA})\right)$ for both probing scenarios reads (cf. Eqs. **(165)** and **(171)**):

$$(\mathcal{H}^{2x2}_{ROT,RWA}) = \begin{pmatrix} -\frac{\delta_C}{2} & \frac{1}{2} \cdot \mathcal{D}_C \\ \frac{1}{2} \cdot \mathcal{D}^*_C & \frac{\delta_C}{2} \end{pmatrix}$$

**(172)**

The generalized Rabi flopping frequency $(\Omega^{2x2}_{GR})$ for the $2x2$ Hamiltonian of Eq. **(172)** is –

$$\Omega^{2x2}_{GR} = \sqrt{\delta^2_C + |\mathcal{D}_C|^2}$$

**(173)**

In the case of on-resonance TSS coupling ($\delta_C = 0$), the two eigenvectors of $(\mathcal{H}^{2x2}_{ROT,RWA})$ are perfectly symmetric and antisymmetric:

$$\xi_P \equiv \xi_{Symm} = \frac{1}{\sqrt{2}} \cdot \begin{pmatrix} 1 \\ 1 \end{pmatrix} \quad ; \quad \xi_N \equiv \xi_{Asym} = \frac{1}{\sqrt{2}} \cdot \begin{pmatrix} -1 \\ 1 \end{pmatrix}$$

**(174)**

In the four consecutive computer-simulation runs shown by the panels of ***Figure 27*** the TSS is initiated with "50:50" probability amplitudes - equal to the components of the $\left(\mathcal{H}_{ROT,RWA}^{2x2}\right)$ eigenvectors (i.e. either Symm or Asym as marked). As we have seen already, these initial conditions would have set an *isolated* driven TSS into stationary states. In all simulation runs, the upper "probe-target" state ($|r>$) *starts* unpopulated. (Here and in the text below we use the casual "population" to stand for the rigorous "state-occupation probability" **[55],[50]**).

The panels of *Figure 27* show *maximum* population of the upper level as a function of the *probe* detuning (left column) along with the time evolution of the populations (center and right columns) of all three involved-states at the peak (w.r.t. probe detuning) of the upper-level population.

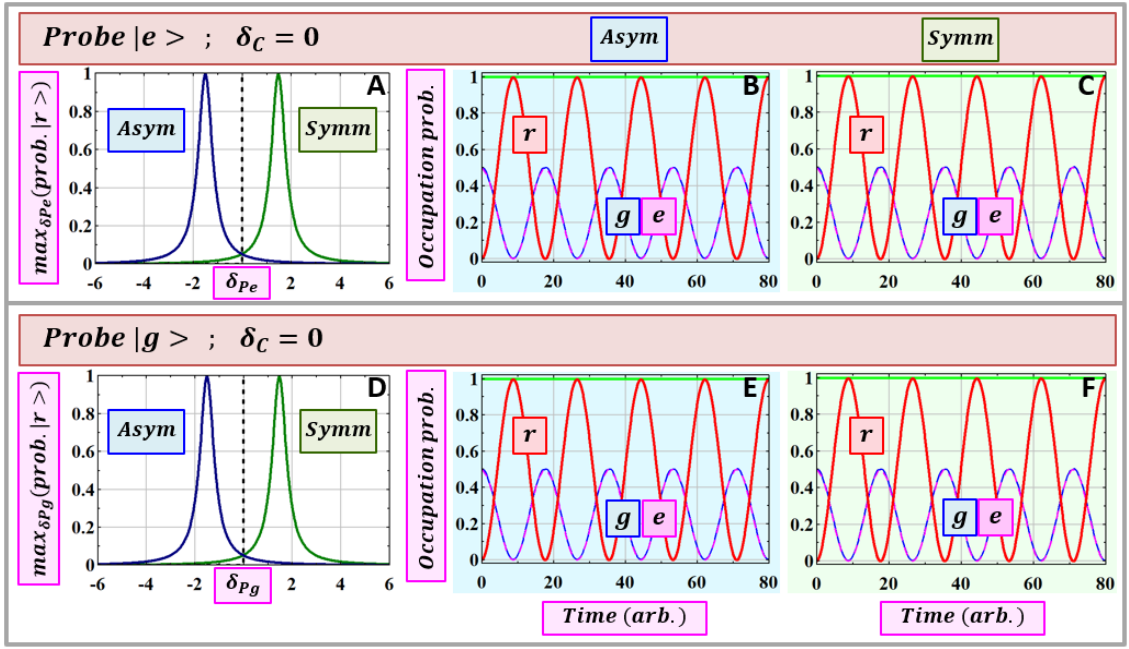

*Figure 27: Dynamics of state occupations in coupling-probe numerical "experiments". TSS coupling is **on-resonance** ($\delta_C = 0$). Top row: Probe $|e>$. Bottom row: Probe $|g>$. A: Maximum population of the $|r>$ state vs. probe detuning in **two consecutive runs** - Asym (blue)/Symm (green) eigenstate-initiation of the TSS. The two curves are each of a Lorentzian shape and are spectrally separated by $\Omega_{GR}^{2x2} = |\mathcal{D}_C|$ (Eq. (173)). B,C: Time-evolution of the populations at the peak (vs. probe detuning) of the upper level population. The populations (blue and magenta) of the two lower-energy states (the TSS) are simultaneously reduced **to zero** while the population of the upper level (red) increases **to unity**. Then a reversed half cycle starts, and so on. The horizontal green line at a level "1" is the sum of the three occupation probabilities. D,E,F: Another set of two separate runs, this time probing the lower $|g>$ level. Curves of the Lorentzian lines (D) and curves of population evolution (E,F) are unchanged (vs. the respective curves in A,B,C). Together, the four Lorentzian lines in A and D indicate (prove the existence of) two $\Omega_{GR}^{2x2}$-separated quasi-energy doublets of an on-resonance sinusoidally-driven TSS of a free atom (cf. Figure 19).*

### B. Eigenstates-initiated TSS; Large detuning of the coupling field ($\pm|\delta_C| \neq 0$)

The curves of the next two figures (*Figure 28* and *Figure 29*) are similar to the curves of *Figure 27*, only this time the frequency of the coupling field is taken to be detuned from the resonance frequency of the two lower-energy states (the TSS). The eigenvectors of the $2x2$ Hamiltonian (Eq. **(172)**) are still symmetric and antisymmetric (in the soft sense of equal/opposite components' sign) but the magnitudes of the components of each eigenvector are different. The TSS is again consecutively launched (initiated) into its eigenstates (initial probability amplitudes equal the components of the Symm/Asym eigenvector of the detuned-coupled TSS).

Several changes in the shown curves (vs. the $\delta_C = 0$ curves of **Figure 27**) immediately catch the eye:

- Shift of the Lorentzian doublet lines – "inwards" for a red detuned ($\delta_C > 0$) coupling field (and, not shown, "outwards" for a blue detuned ($\delta_C < 0$) coupling field) - cf. **Figure 17** and **Figure 19**. (The doublet splitting in each of the shown four probing scenarios is still $\Omega_{GR}^{2x2}$).

- As shown by the curves in each of the left four panels, the pair of Lorentzian lines are of different width (to be discussed in a dedicated section below).

- Oscillations frequency of the states' probabilities are very different in the Asym/Symm cases (high oscillation frequency in the wide-Lorentzian case and low oscillation frequency for the narrow-Lorentzian case).

In all eight cases shown in the two figures (**Figure 28** and **Figure 29**), at a certain detuning of the probe field, the populations, initially occupying the lower two states only, are time-periodically *fully transferred to the upper state* (periodical *unity occupation probability* of the $|r>$ state).

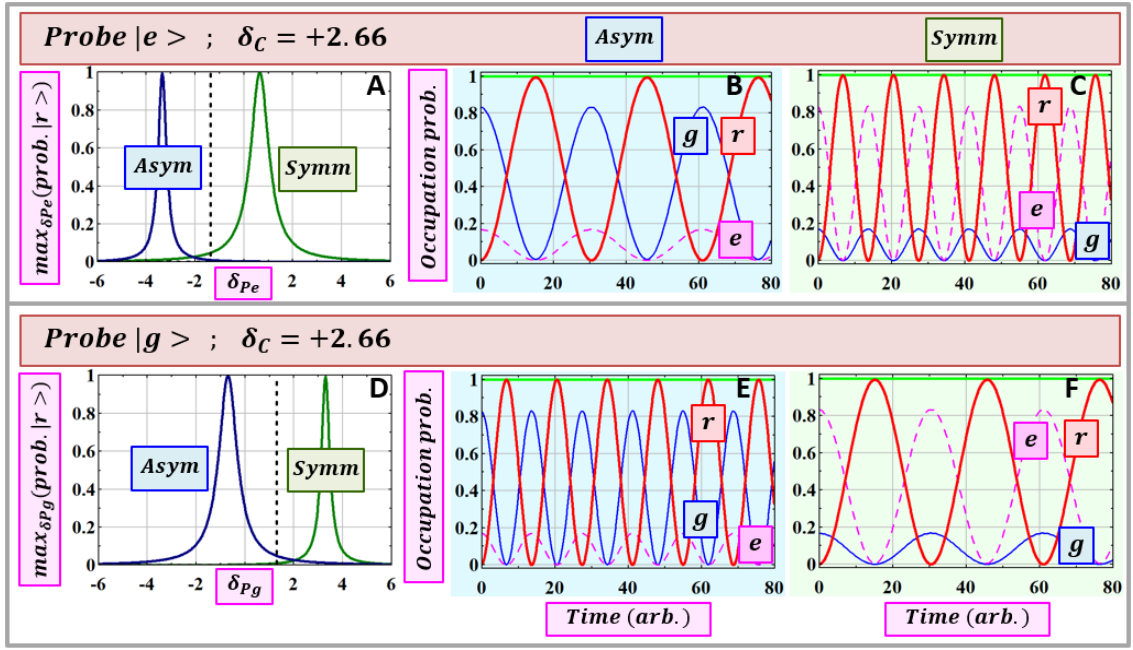

*Figure 28:* Same as *Figure 27* except that the coupling field is red detuned ($\delta_C > 0$). A: Probe $|e>$. For $\delta_C > 0$ the doublet-center (dashed black line) is shifted "downwards". The centers of the shown two Lorentzian lines correspond to $\mathcal{E}_P(H)$ (Symm) and $\mathcal{E}_N(H)$ (Asym) of Eq. **(130)**. D: Probe $|g>$. For $\delta_C > 0$, the doublet-center (dashed black line) is shifted "upwards". The shown two Lorentzian lines correspond to $\mathcal{E}_P(L)$ (Symm) and $\mathcal{E}_N(L)$ (Asym) of Eq. **(130)**. The difference in line widths is discussed in a dedicated section below. B,C,E,F: Time oscillations of the *states' populations at the peak-population (vs. probe detuning) of the $|r>$ state. Populations of the lower-energy states simultaneously go down to zero while the $|r>$ population increases to unity. Population-oscillation periods in the Asym/Symm cases are found to be very different (high population-oscillations frequency for a corresponding wide Lorentzian line).*

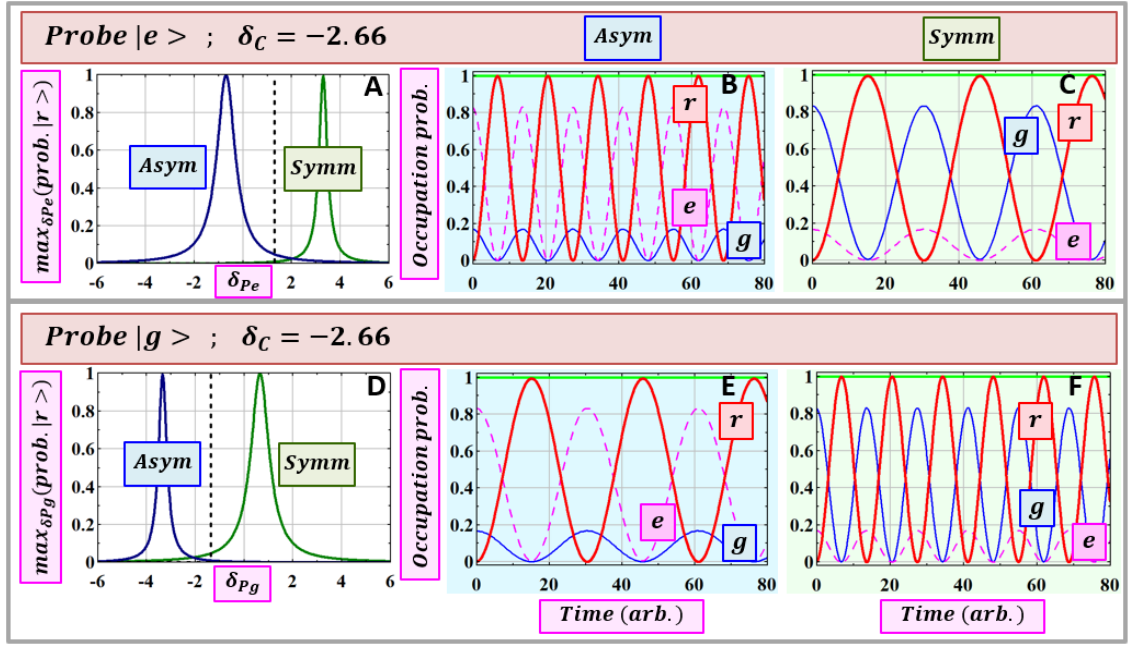

*Figure 29: Same as Figure 28, this time with a blue-detuned coupling field ($\delta_C < 0$). In the blue detuning case, as indicated by the dotted vertical black lines, the doublet center of the perturbed $|e>$ state moves "upwards" (relative to the energy level of the unperturbed state) whereas the doublet center of the perturbed $|g>$ state moves "downwards". Overall the energy gap of a TSS driven by a blue detuned coherent field **expands** slightly (to match the energy of the slightly too "fat" photons of the driving field). And here again, at a specific detuning value of the probing field the maximum population of the $|r>$ state is (periodically) at a value of unity (complete depletion of the two lower states).*

### C. General initial state of the TSS; Red detuning of the coupling field ($\delta_C > 0$).

The curves of *Figure 30* show two run-results, each for a TSS launched into a general state (arbitrary initial probability amplitudes for $|g>$ and for $|e>$).

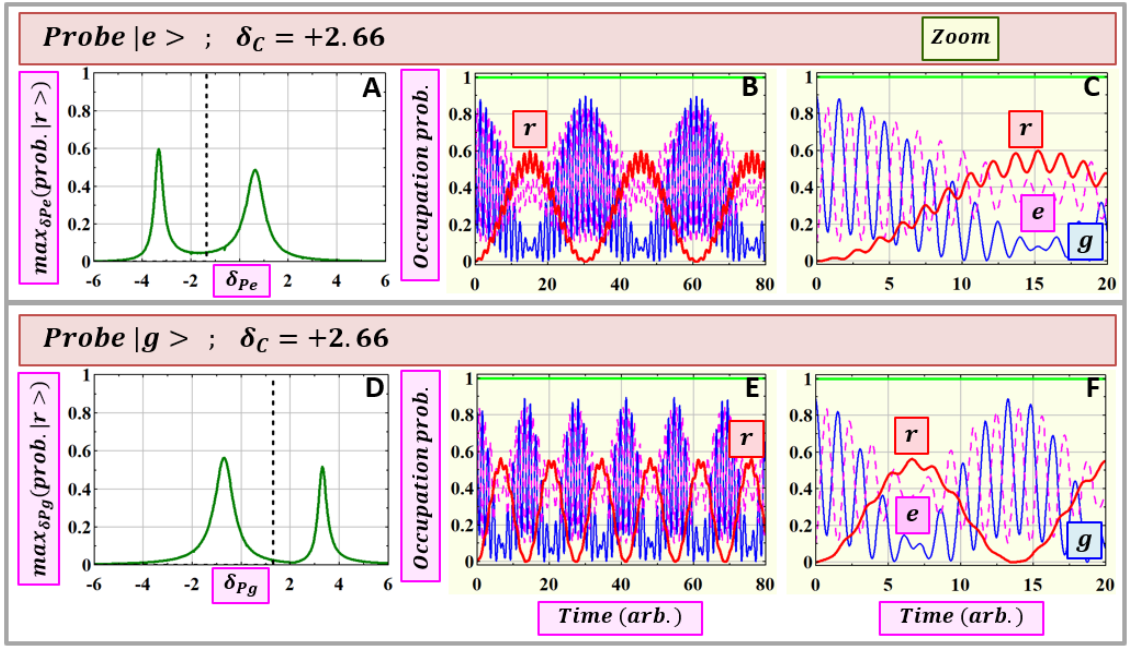

*Figure 30: Each row shows the results of a **single** simulation run. In each run the TSS is launched (initiated) into a general state (general initial value of the TSS probability amplitudes). The TSS coupling field is red detuned ($\delta_C > 0$). A,D: Two peaks (vs. probe frequency), $\Omega_{GR}^{2x2}$ apart, are recorded. The doublet centers (dotted vertical black lines) shift such that The overall TSS energy gap slightly shrinks. B,E: evolution of state populations at the probe frequency that resulted in the higher of the two $|r>$ population peaks (the left peak in A,D as shown). In these TSS initial conditions the $|r>$ population does NOT reach unity. C,F: Zoom-in. Lower levels' populations are seen to rapidly oscillate and the TSS is never fully depleted.*

Probe-absorption curves similar to the curves shown in the left column of *Figure 30* are presented in (**[48]** figure 18.3.), in the context of cavity quantum electrodynamics.

### D. Upper level line-widths

Looking at the left columns of *Figure 27* to *Figure 29* the difference in width of the shown Lorentzian lines attracts viewer's attention. In this "upper level line-widths" sub-section we show that indeed the lines are all of a Lorentzian shape and briefly discuss the reasons for the difference in their widths.

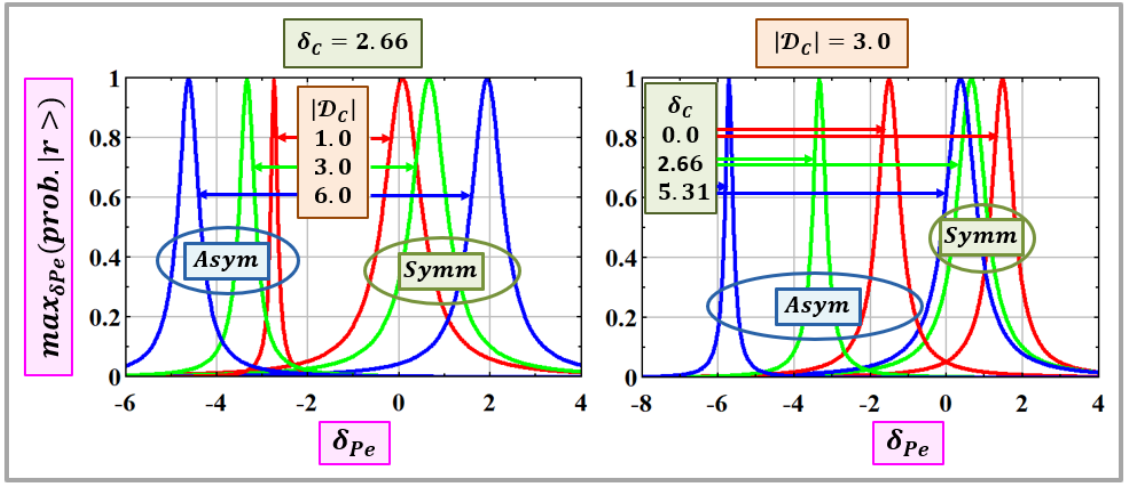

*Figure 31: Lorentzian lines showing the population of the $|r>$ state vs. probe detuning resulted during probing the upper $|e>$ state of the TSS. Left – vs. TSS coupling strength ($|\mathcal{D}_C|$). Stronger coupling – narrower Symm lines, wider Asym lines, and wider Symm-Asym spectral distance. Right: Vs. TSS coupling-field detuning ($\delta_C$). Further detuning – wider Symm lines, narrower Asym lines and again larger Symm-Asym spectral distances. Consult the curves of **Figure 3** and of **Figure 4** for the correlations with the components of the TSS eigenvectors.*

The top-left 2x2 elements of the $\left(\mathcal{H}_{ROT,RWA}^{er}\right)$ Hamiltonian (Eq. **(165)**) and of the $\left(\mathcal{H}_{ROT,RWA}^{gr}\right)$ Hamiltonian (Eq. **(171)**) both include only two parameters – $\delta_C$ and $\mathcal{D}_c$. In the panels of **Figure 31** we show how $\delta_C$ and $|\mathcal{D}_C|$ affect the line shapes of the top-level populations (vs. the change in frequency of the coherent probe light-field). Looking at the figure, two effects stand out – the width-difference between Symm/Asym initial conditions and the changes in the Symm-Asym spectral distances with the different values of the two parameters $(\delta_C; |\mathcal{D}_C|)$. *Both effects can be traced down to the components of the 2x2 Hamiltonian eigenvectors.*

The change in the components of the 2x2 Hamiltonian eigenvectors with each of the two parameters is shown by the curves of **Figure 3** and of **Figure 4**. The vertical dotted black lines in these two figures designate the components of the eigenvectors computed for **Figure 27** to **Figure 29** above. Consulting **Figure 31** and consulting **Figure 27** to **Figure 29** in relations to **Figure 3** and **Figure 4** the following conclusion is drawn – for the probe $|e>$ simulation, the Lorentzian line width is proportional to the *initial* probability amplitude of the $|e>$ state (which is one of the eigenvector's components). *Larger **initial** probability amplitude results in a wider Lorentzian line shape*.

But the Lorentzian line width depends also on the probe field coupling strength ($\mathcal{D}_{PE}$ for the probe $|e>$ simulation). So the Lorentzian line shape can be written by some "half-width" function $Q_e$ of the three parameters $\left(Q_e(\delta_C, |D_C|, |D_{PE}|)\right)$ as:

$$FIT_{e,Symm}(\delta_{Pe}; \delta_C, |D_C|, |D_{PE}|) = \frac{[Q_e(\delta_C, |D_C|, |D_{PE}|)]^2}{[Q_e(\delta_C, |D_C|, |D_{PE}|)]^2 + \left[\delta_{Pe} + \frac{\delta_C}{2} - \lambda_P\right]^2}$$

$$FIT_{e,Asym}(\delta_{Pe}; \delta_C, |D_C|, |D_{PE}|) = \frac{[Q_e(\delta_C, |D_C|, |D_{PE}|)]^2}{[Q_e(\delta_C, |D_C|, |D_{PE}|)]^2 + \left[\delta_{Pe} + \frac{\delta_C}{2} - \lambda_N\right]^2}$$

**(175)**

where $(\lambda_P, \lambda_N)$ are the eigenvalues of the 2x2 Hamiltonian operator (Eq. **(172)**). And a similar set of two fit equations for the probe $|g>$ simulation.

The functional dependence of the three-parameter $Q_e$ function $\left(Q_e(\delta_C, |D_C|, |D_{PE}|)\right)$ is yet to be worked-out.

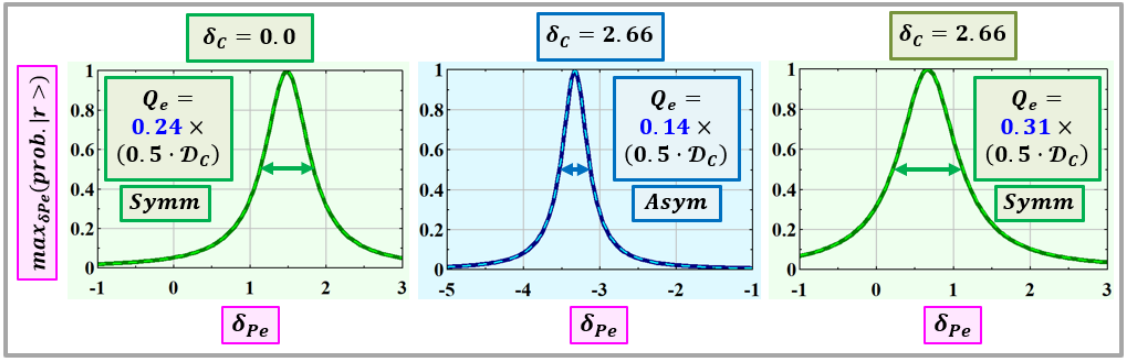

*Figure 32: Curve fits (Eqs. **(175)**) to the $|r>$ population curves in the probe $|e>$ simulation (cf. panel A of **Figure 27** and of **Figure 28**). The fits indicate general much narrower widths vs. the modulation-depth width $(0.5 \cdot \mathcal{D}_C)$ of an isolated driven TSS (Eq. **(46)**). The spectral positions of the centers of the simulated Lorentzian lines $\left(\frac{\delta_C}{2} - \lambda_P ; \frac{\delta_C}{2} - \lambda_N ; \text{Eqs. } \mathbf{(175)}\right)$ confirm the analytical predictions (cf. Eq. **(130)** and **Figure 17**).*

The panels of **Figure 32** show three fit-curves as prescribed by Eqs. **(175)**. The width of the modulation-depth curve (cf. Eq. **(46)**) of an isolated driven TSS with the 2x2 Hamiltonian of Eq. **(172)** would have been $\mathcal{D}_C/2$. As indicated by the legend's numbers (blue fonts), the three curves of **Figure 32** resulted during probing of the $|e>$ state by a *weak* field (small $|\mathcal{D}_{PE}|$ value) are much narrower. And as indicated by the fit curves (Eqs. **(175)**), the spectral positions of the centers of the simulated Lorentzian lines $\left(\frac{\delta_C}{2} - \lambda_P ; \frac{\delta_C}{2} - \lambda_N\right)$ confirm the analytical predictions (cf. Eq. **(130)** and **Figure 17**).

## 5. Summary

Many TSS studies are focused on state-occupation probabilities. Once the state amplitudes are squared (absolute value of) phase-evolution information is partly lost. This phase-evolution information loss is particularly severe in the case of driven TSSs, where the state-occupation *probabilities* are (correctly) calculated in the rotating frame. Thus, the relevant phase-evolution information exposed through (the unexecuted) back rotation operation is never even reached. However, it is through the phase-evolution information included in the expressions for the state *amplitudes* that the TSS *energies* are identified (cf. Eqs **(1)**-**(4)**).

Expressions for the state amplitudes are arrived-at through the solution of a suitable Schrödinger equation. Often the solution is found through the standard method of matrix multiplication **[89]**,**[90]**. The phase-evolution information in the expression obtained by matrix multiplication appears in an exponential factor but also in trigonometric functions. Identification of the TSS energies through such expressions is not straight forward. We have suggested here an ($\mathcal{A}, \mathcal{B}, \mathcal{C}, \mathcal{D}$) solution format in which time-evolution of each state amplitude is given solely by exponential factors (the product of "$\mathbb{j}$", a phase rotational speed, and time in each exponent's argument) such that identification of the TSS energies is far more attainable (cf. Eqs **(1)**-**(4)**).

Following, in a way of summary, is a set of statements related to energies of two-state systems:

- A non-driven TSS (characterized by a TIH) may be in one of two stationary states. Each stationary state (state-occupation probabilities independent of time) is also a state of a definite energy (a single phase rotational speed in the argument of a single phase-evolution factor). The value of a high/low definite energy equals the high/low eigenvalue of the corresponding Hamiltonian (section **2.1.7**).

- Given a driven TSS (characterized by a TDH), we are facing a unique state of affairs: the TSS can still be in one of only two stationary states (state-occupation probabilities independent of time) but each stationary state is a state of *two* definite energies (referred-to in some studies as "quasi-energies" – sections **3.1.6** and **3.1.7**).

- A spin one-half particle in a constant axial magnetic field will precess (make complete circles at a constant angle w.r.t. the $z$ axis) at the generalized Rabi flopping frequency (section **2.2**).

  A spin one-half particle in a constant axial magnetic field and a *constant* transverse magnetic field (the "level-coupling" field) will "precess" (*simultaneous $x$ and $z$* "spiral"-type motion) at the generalized Rabi flopping frequency. If the TSS is launched into one of its eigenstates then particle's precession ceases altogether. (sections **2.2** and **3.2**).

  If the transverse coupling field is a near-resonance sinusoidal field, then the spin one-half particle will "precess" at nearly the frequency of the transverse sinusoidal driving field (Eq. **(142)** and section **3.2**).

- The energy separation between doublets of all TSS types (non-driven TSSs as well as driven TSSs), in terms of angular frequency units, is the generalized Rabi flopping frequency ($\Omega_{GR}$ – Eq. **(14)** ; $\Omega_{GR,t}$ Eq. **(120)**).

- Average energy of a non-driven TSS depends on the TSS initial conditions and its value is between the two eigenvalues of the corresponding Hamiltonian (section **2.1.9**).

- In a set of two coupling-probe computer simulations, the existence of four quasi-energy levels of a driven TSS is proved "experimentally" (section **4**). In each of the probe simulations, an Autler-Townes doublet can be identified (section **4**). In a fluorescence experiment with a driven-atom TSS, a Mollow triplet will be detected.

The peak frequency of the center line of the triplet is the frequency of the driving field. Two side lines, separated by the generalized Rabi flopping frequency from the center line will also be detected (section **4** and ***Figure 19***).

- If a driven TSS is launched into a stationary state, then upon scanning the frequency of a probe field (in a coupling-probe computer simulation) the curve of maximum population of the "target" state is of a Lorentzian shape. The width of such Lorentzian line will vary with TSS and probe parameters (section **4**).

Two-level quantum systems called "qubits" are the building blocks of concrete realizations of systems executing bit-operations in the "hot" fields of quantum computing and quantum information processing **[24]**,**[115]**. Energies of these TLSs (as TSSs are frequently referred-to), and in particular energy changes in driven TLSs play a significant role in the designs and runs of these concrete systems. Energy changes are involved not only in the control of the qubits' states but also in the control of circuit environment.

One such recently proposed environment-control example is the reduction of dielectric loss in superconducting resonators soaked in a bath of TLSs with broad distribution of energies. The loss, attributed to the interaction of the resonator and the surrounding near-resonance TLSs is proved to be significantly reduced by *fast frequency sweeping of a bias electric field* forcing the TLSs in and out of resonance with the resonator **[116]**. Energy shifts of near-resonance sinusoidally-driven TSSs are proved analytically and numerically in the present work.

**Acknowledgements**

I would like to thank Yotam Shapira for his guidance in performing the coupling-probe computer simulations.

I would like to thank David Mukamel and Ziv Meir for their valuable comments and constructive suggestions during preparation of the manuscript.